\pdfoutput=1
\documentclass[11pt,letterpaper]{article}

\usepackage[letterpaper,margin=1in]{geometry}
\usepackage[T1]{fontenc}
\usepackage[utf8]{inputenc}
\usepackage{lmodern}
\usepackage{microtype}
\usepackage{amsmath,amssymb,amsthm}
\usepackage{paralist}
\usepackage{dsfont}
\usepackage{mathtools}
\usepackage{graphicx}
\usepackage[dvipsnames]{xcolor}
\usepackage{tcolorbox}
\usepackage{booktabs}
\usepackage{makecell}
\usepackage{todonotes}
\usetikzlibrary{arrows.meta}

\usepackage{titlesec}
\titleformat{\subparagraph}[runin]{\normalfont\normalsize\bfseries}{}{0pt}{}
\titlespacing*{\subparagraph}{0pt}{\smallskipamount}{\wordsep}
\titlespacing*{\section}{0pt}{2.5ex plus .5ex minus .2ex}{1.2ex plus .2ex}
\titlespacing*{\subsection}{0pt}{2ex plus .5ex minus .2ex}{.8ex plus .2ex}
\titleclass{\subsubparagraph}{straight}[\subparagraph]
\newcounter{subsubparagraph}
\renewcommand{\thesubsubparagraph}{\Alph{subsubparagraph}}
\titleformat{\subsubparagraph}[runin]{\normalfont\normalsize\itshape}{\thesubsubparagraph}{1em}{}
\titlespacing*{\subsubparagraph}{0pt}{3.25ex plus 1ex minus .2ex}{1em}
\makeatletter
\newcommand*{\l@subsubparagraph}{\@dottedtocline{6}{14em}{6em}}
\newcommand*{\toclevel@subsubparagraph}{6}
\makeatother
\usepackage[
  colorlinks=true,
  linkcolor=MidnightBlue,
  citecolor=OliveGreen,
  urlcolor=MidnightBlue,
  bookmarksnumbered=true,
  bookmarksdepth=2,
]{hyperref}
\usepackage{xurl}

\usepackage{etoolbox}
\patchcmd{\abstract}{\small}{\normalsize}{}{\PackageWarning{main}{abstract patch failed}}

\usepackage{algorithm}
\usepackage[noend]{algpseudocode}

\usepackage[capitalise,noabbrev,nameinlink]{cleveref}

\theoremstyle{plain}
\newtheorem{theorem}{Theorem}[section]
\newtheorem{lemma}[theorem]{Lemma}

\theoremstyle{definition}

\newtheorem{example}[theorem]{Example}
\theoremstyle{remark}
\newtheorem{remark}[theorem]{Remark}
\crefname{claim}{Claim}{Claims}

\newcommand{\eps}{\varepsilon}
\newcommand{\defas}{\coloneqq}
\newcommand{\pr}{\mathbb{P}}
\newcommand{\NN}{\mathbb{N}}
\newcommand{\RR}{\mathbb{R}}
\newcommand{\ER}{\exists\mathbb{R}}
\newcommand{\coER}{\text{co-}\exists\mathbb{R}}
\newcommand{\1}{\mathds{1}}
\newcommand{\support}{\text{supp}}
\newcommand{\pati}{\text{Patience}}
\newcommand{\val}{\text{\textup{val}}}

\newcommand{\reach}{\text{Reach}}

\newcommand{\prob}{\delta}
\newcommand{\Actions}{\mathcal{A}}
\newcommand{\A}{\Actions}
\newcommand{\action}{a}
\newcommand{\Bactions}{\mathcal{B}}
\newcommand{\baction}{b}
\newcommand{\States}{S}
\newcommand{\state}{s}
\newcommand{\Plays}{\Omega}
\newcommand{\play}{\omega}
\newcommand{\Target}{T}
\newcommand{\strategyone}{\sigma}
\newcommand{\strategytwo}{\tau}
\newcommand{\Strategyone}{\Sigma}
\newcommand{\Strategytwo}{\Gamma}
\newcommand{\low}{L}
\newcommand{\up}{U}
\newcommand{\CSGD}{\textnormal{\textsc{CsgDecision}}}
\newcommand{\CSGGV}{\textnormal{\textsc{CsgGapValue}}}
\newcommand{\CSGAV}{\textnormal{\textsc{CsgApproxValue}}}
\newcommand{\F}{\mathcal{F}}
\newcommand{\D}{\mathcal{D}}

\newcommand{\rel}{\text{rel}}
\newcommand{\trunc}{\text{Trunc}}

\DeclareMathOperator*{\op}{op}
\newcommand{\liminfavg}{\text{LimAvg}}

\newcommand{\disc}{\text{Disc}}
\newcommand{\Binp}{B_{\mathrm{inp}}}
\newcommand{\Beps}{B_{\eps}}
\newcommand{\Blam}{B_{\lambda}}
\newcommand{\Bsep}{B_{\mathrm{sep}}}

\title{The Computational Complexity of Concurrent Stochastic Games:\\ Exact Value and Approximation}

\author{
  Ali Asadi\\
  Institute of Science and Technology Austria (ISTA)\\
  \texttt{ali.asadi@ist.ac.at}
  \and
  Krishnendu Chatterjee\\
  Institute of Science and Technology Austria (ISTA)\\
  \texttt{krishnendu.chatterjee@ist.ac.at}
}

\date{}

\hypersetup{
  pdftitle={The Computational Complexity of Concurrent Stochastic Games: Exact Value and Approximation},
  pdfauthor={Ali Asadi, Krishnendu Chatterjee},
}

\begin{document}

\pagenumbering{roman}
\maketitle
\thispagestyle{empty}

\begin{abstract}
We consider two-player zero-sum concurrent stochastic games played over a finite state space for an infinite number of stages.
In every stage, both players simultaneously and independently choose an action.
Then, given the current state and chosen actions, the successor state is selected according to a stochastic transition function.
We consider limit-average objectives, where Player~1 maximizes the long-run average of rewards assigned to every transition.
The value from an initial state is the maximal expectation that Player~1 can ensure.
We study the two basic computational problems: (a)~the approximation problem of computing the values within an arbitrary additive error, and (b)~the exact problem of deciding whether the value at a state is at least a given threshold.
The previous results in the literature establish: (a)~the approximation problem belongs to TFNP[NP], and 
(b)~the exact problem belongs to PSPACE, via the theory of the reals, and is SQRT-SUM hard.

Our contributions are as follows.
(a)~Our main result shows that the approximation problem is in TFNP, and that the associated gap or promise problem
is in NP$\cap$coNP.
(b)~We also show that the exact problem is in $\ER \cap \coER$, i.e., it reduces to both the existential and the universal theory of the reals, and that it is PosSLP-hard.
\end{abstract}

\clearpage

\tableofcontents
\clearpage

\pagenumbering{arabic}

\section{Introduction}\label{sec:intro}
In this work, we present improved complexity results for the value computation and approximation 
of concurrent stochastic games with limit-average objectives.
We informally present the model, the objectives,
the computational problems, previous results, and our contributions.

\subparagraph{Concurrent stochastic games.}
Concurrent stochastic games~\cite{shapley1953stochastic, gillette1957StochasticGamesZero, everett1957RecursiveGames} are two-player zero-sum games played on finite-state graphs
for an infinite number of stages.
In every stage, each player simultaneously and independently of the other player chooses an action.
The current state and the actions chosen by the players give a probability distribution
over the successor state.
An infinite number of such stages results in a play which is an infinite sequence of states and actions.
Concurrent stochastic games are one of the most fundamental models in game theory.
They were introduced by Shapley~\cite{shapley1953stochastic} as stopping games, or equivalently discounted-sum games, which halt with probability~$1$.
The fundamental extension to games of infinite duration was studied by Gillette~\cite{gillette1957StochasticGamesZero} and Everett~\cite{everett1957RecursiveGames}.
Concurrent stochastic games have been widely studied from the mathematical perspective~\cite{shapley1953stochastic, gillette1957StochasticGamesZero, everett1957RecursiveGames, mertens1981StochasticGames, martin1998DeterminacyBlackwellGames, filar2012competitive},
and from the algorithmic and computational complexity perspective, including
the complexity of reachability objectives~\cite{dealfaro1998ConcurrentReachabilityGames, etessami2008RecursiveConcurrentStochastic, hansen2009WinningConcurrentReachability, frederiksen2013ApproximatingValueConcurrent},
algorithms and complexity for limit-average objectives~\cite{chatterjee2008LimitAverageEXPTIME, hansen2011ExactAlgorithms, oliubarton2021NewAlgorithms, bose2024BoundedMemory},
the qualitative and quantitative analysis of $\omega$-regular objectives~\cite{dealfaro2000ConcurrentOmegaregularGames, dealfaro2001QuantitativeSolutionOmegaregular, chatterjee2006ComplexityQuantitativeConcurrent, asadi2024StatefulDiscounted},
and temporal logics~\cite{alur2002alternating}.
In particular, in analysis of reactive systems, concurrent games provide the appropriate model for
reactive systems with components that interact synchronously~\cite{alur2002alternating,de2000control,de2001control}.

\subparagraph{Objectives.}
An objective assigns a payoff to every play. Player~1 aims to maximize its expectation and Player~2 is adversarial.
We consider the following classical objectives.
In \emph{limit-average} objectives, also known as mean-payoff objectives, every transition is assigned a reward and the payoff of a play is the long-run average of the rewards~\cite{gillette1957StochasticGamesZero,mertens1981StochasticGames}.
These objectives generalize \emph{reachability} objectives, where the goal of Player~1 is to eventually reach a target state.
For our technical results, we use \emph{discounted-sum} objectives, where the payoff is the discounted sum of the rewards~\cite{shapley1953stochastic}.

\subparagraph{Strategies, values, and the computational problems.}
Strategies are recipes that define the choice of actions of the players.
They are functions that, given the sequence of states visited and actions chosen so far, return a probability distribution over actions.
Given a concurrent stochastic game and an objective, the value of Player~1 at a state $s$, $\val(s)$, is
the maximal expectation of the objective that Player~1 can guarantee
against all strategies of Player~2.
Given a concurrent stochastic game and an objective, the main computational problems are as follows:
(a)~the \emph{value-decision} problem given a state $s$ and threshold $\alpha$ asks whether $\val(s) \geq \alpha$;
(b)~the \emph{value-gap} problem given a state $s$, a threshold $\alpha$, and an error $\eps \in (0, 1)$ asks whether $\val(s) \geq \alpha + \eps$ or $\val(s) \leq \alpha - \eps$, where any answer is correct if $\val(s)$ lies strictly between $\alpha - \eps$ and $\alpha + \eps$; and
(c)~the \emph{value-approximation} problem given an error $\eps \in (0, 1)$ asks to compute the values of all states within an additive error $\eps$.

\subparagraph{Previous results.} We discuss the previous results for the exact and approximation 
questions.

\smallskip\noindent{\em Value decision problem.}
The value-decision problem for concurrent stochastic games is in $\coER$, i.e., it reduces to the universal theory of the reals, for reachability objectives~\cite{etessami2008RecursiveConcurrentStochastic} and in PSPACE for limit-average objectives~\cite{chatterjee2008LimitAverageEXPTIME, chatterjee2015Value1}, via the theory of the reals.
Moreover, it is SQRT-SUM hard for reachability objectives~\cite{etessami2008RecursiveConcurrentStochastic},
and the hardness also holds for limit-average objectives since they generalize reachability objectives.
The SQRT-SUM problem is a classical problem in computational geometry~\cite{garey1976NPCompleteGeometric} and whether SQRT-SUM belongs to NP has been a long-standing open problem~\cite{allender2009ComplexityNumericalAnalysis}.
An interesting open question is whether the upper and lower bounds can be improved for concurrent stochastic games with limit-average and reachability objectives.

\smallskip\noindent{\em Approximation problems.}
The complexity of the value-approximation problem for concurrent stochastic games has been an intriguing and challenging open problem, even for reachability objectives.
The classical approach in the study of stochastic games is to guess approximately optimal strategies and then verify the guess.
The complexity of this approach depends on the size of the guess and the complexity of the verification procedure.
For reachability objectives, it was claimed in~\cite{chatterjee2004NashEquilibriaStochastic}, using the above classical approach, that for every fixed error $\eps \in (0, 1)$, the value-gap problem belongs to NP $\cap$ coNP.
However, it was pointed out in~\cite{etessami2008RecursiveConcurrentStochastic} that the proof of this claim in~\cite{chatterjee2004NashEquilibriaStochastic} is incorrect.
Later~\cite{hansen2009WinningConcurrentReachability,chatterjee2017StrategyComplexityConcurrent} show that approximately optimal strategies for both players require double exponentially small
numbers, so their explicit representations require exponential space.
This rules out the classical approaches of guessing strategies explicitly and verifying.
The next approach that was adopted is to guess the strategies symbolically.
It was shown that there exists a
symbolic polynomial-size representation of approximately optimal strategies for which the verification procedure requires an NP oracle~\cite{frederiksen2013ApproximatingValueConcurrent}.
This leads to a TFNP[NP] upper bound for the value-approximation problem for reachability objectives.
For limit-average objectives, approximately optimal strategies may require infinite memory~\cite{blackwell1968BigMatch}, so they cannot be guessed directly.
Instead, the value of limit-average objectives is the limit of the values of discounted-sum objectives as the discount factor tends to one~\cite{mertens1981StochasticGames}.
Moreover, these values are approximated by the values of discounted-sum objectives with discount factors that are double exponentially close to one~\cite{oliubarton2021NewAlgorithms}.
The TFNP[NP] upper bound was extended to limit-average objectives~\cite{bose2024BoundedMemory}.
For the subclass of ergodic games, in which all states are visited infinitely often with probability one, the value-approximation problem for limit-average objectives is in FNP~\cite{chatterjee2014ErgodicMeanPayoff}.
Whether a TFNP bound can be obtained for general games has remained an important open question~\cite{frederiksen2013ApproximatingValueConcurrent, chatterjee2017StrategyComplexityConcurrent, asadi2024StatefulDiscounted, grobelna2025StoppingCriteria}, even for reachability objectives.
Note that even for special cases of concurrent stochastic games, such as turn-based stochastic games, aka simple stochastic games (SSGs)~\cite{condon1992complexity},
the value-approximation problem for reachability objectives is in TFNP and the existence of a polynomial-time algorithm is a major long-standing open problem.
We summarize the previous complexity results in \Cref{Table: Previous results}.

\begin{table}[ht]
  \centering
  \begin{tabular}{llll}
    \toprule
    Objective & \CSGD{} & \CSGAV{} & \CSGGV{} \\
    \midrule
    Reachability
      & \makecell[tl]{$\coER$~\cite{etessami2008RecursiveConcurrentStochastic} \\ SQRT-SUM hard~\cite{etessami2008RecursiveConcurrentStochastic}}
      & \makecell[tl]{TFNP[NP]~\cite{frederiksen2013ApproximatingValueConcurrent} \\ SSG-hard~\cite{condon1992complexity}}
      & \makecell[tl]{NP[NP] $\cap$ coNP[NP]$^\dagger$ \\ SSG-hard~\cite{condon1992complexity}} \\
    \addlinespace
    Limit-average
      & \makecell[tl]{PSPACE~\cite{chatterjee2008LimitAverageEXPTIME, chatterjee2015Value1} \\ SQRT-SUM hard~\cite{etessami2008RecursiveConcurrentStochastic}}
      & \makecell[tl]{TFNP[NP]~\cite{bose2024BoundedMemory} \\ SSG-hard~\cite{condon1992complexity}}
      & \makecell[tl]{NP[NP] $\cap$ coNP[NP]$^\dagger$ \\ SSG-hard~\cite{condon1992complexity}} \\
    \bottomrule
  \end{tabular}
  \caption{Previous complexity bounds for concurrent stochastic games.
    In each cell, the first line is the upper bound and the second line is the lower bound.
    SSG-hard means at least as hard as solving simple stochastic games.
    $^\dagger$Immediate consequence of the upper bound for \CSGAV{}.}
  \label{Table: Previous results}
\end{table}

\subparagraph{Our contributions.}
In this work, we resolve the open questions. 
\begin{compactitem}
\item {\em Approximation problems.} Our main result shows that the value-approximation problem for concurrent stochastic games with limit-average and reachability objectives is in TFNP, see \cref{Result: CSG in TFNP}.
Moreover, the value-gap problem is in NP $\cap$ coNP for all these objectives.
Along the way, we obtain the same results for discounted-sum objectives with discount factors that are double exponentially close to one, see \cref{Result: Discounted in TFNP}, which may be of independent interest.

   \item {\em Value decision problem.} 
We also show that the value-decision problem for limit-average and reachability objectives is PosSLP-hard and is in $\ER \cap \coER$, i.e., it reduces to both the existential and the universal theory of the reals, see \cref{Result: Value decision}.
\end{compactitem}

We summarize our results in \Cref{Table: Our results}. On one hand our results improve long-standing computational complexity bounds for the basic computational problems for a fundamental game-theoretic model, and on the other hand, we discuss further extensions of our results (e.g., for $\omega$-regular objectives) in \Cref{Section: Extensions}.

\begin{table}[ht]
  \centering
  \begin{tabular}{llll}
    \toprule
    Objective & \CSGD{} & \CSGAV{} & \CSGGV{} \\
    \midrule
    Reachability
      & \makecell[tl]{$\pmb{\ER \cap \coER}$ \\ \textbf{PosSLP-hard}}
      & \makecell[tl]{\textbf{TFNP} \\ SSG-hard~\cite{condon1992complexity}}
      & \makecell[tl]{\textbf{NP $\cap$ coNP} \\ SSG-hard~\cite{condon1992complexity}} \\
    \addlinespace
    Limit-average
      & \makecell[tl]{$\pmb{\ER \cap \coER}$ \\ \textbf{PosSLP-hard}}
      & \makecell[tl]{\textbf{TFNP} \\ SSG-hard~\cite{condon1992complexity}}
      & \makecell[tl]{\textbf{NP $\cap$ coNP} \\ SSG-hard~\cite{condon1992complexity}} \\
    \bottomrule
  \end{tabular}
  \caption{Complexity bounds for concurrent stochastic games, including our results in bold (\Cref{Result: CSG in TFNP} and \Cref{Result: Value decision}), with the conventions of \Cref{Table: Previous results}.}
  \label{Table: Our results}
\end{table}

\section{Preliminaries}
\label{Section: Preliminaries}

We present standard notations and definitions for concurrent stochastic games, similar to~\cite{chatterjee2017StrategyComplexityConcurrent, etessami2008RecursiveConcurrentStochastic, asadi2024StatefulDiscounted}.

\subparagraph{Basic notations.} Given a finite set $X$, a probability distribution over $X$ is a function $\mu \colon X \to [0, 1]$ such that $\sum_{x \in X} \mu(x) = 1$.
The set of all probability distributions over $X$ is denoted by $\Delta(X)$.
For $\mu \in \Delta(X)$, the support of $\mu$, i.e., the set $\{x \mid \mu(x) > 0\}$, is denoted by $\support(\mu)$.
For two functions $f, g\colon X \to \RR$, we say $f \preceq g$ if for all $x \in X$ we have $f(x) \le g(x)$.
For a function $\rho \colon X \to \RR$, we denote its $L_1$ norm and $L_\infty$ norm by $\|\rho\|_1 \defas \sum_{x \in X} |\rho(x)|$ and $\|\rho\|_\infty \defas \max_{x \in X} |\rho(x)|$, respectively.
For functions $f \colon X \to Y$ and $g \colon Y \to Z$, we denote by $g \circ f$ their composition, i.e., $(g \circ f)(x) \defas g(f(x))$ for all $x \in X$.
We denote by $\NN$ the set of positive integers and by $\NN_0 \defas \NN \cup \{0\}$ the set of non-negative integers.
For $k \in \NN$, we denote $[k] \defas \{i \in \NN \mid i \le k\}$.

\subparagraph{Concurrent stochastic games.}
A concurrent stochastic game (CSG) is a two-player finite game
$G = (\States, \Actions, \Bactions, \prob)$ consisting of
\begin{compactitem}
    \item the set of states $\States$, of size $n$;
    \item the action sets $\Actions$ and $\Bactions$ of Players~1 and~2, respectively, each with at most $m$ actions; and
    \item the stochastic transition function $\prob \colon \States \times \Actions \times \Bactions \to \Delta(\States)$.
\end{compactitem}

\subparagraph{Stages.}
Given an initial state $\state_0 \in \States$, the game proceeds as follows.
In each stage, both players choose an action simultaneously and independently, $\action \in \Actions$ and $\baction \in \Bactions$.
Based on both actions $(\action, \baction)$ and current state $\state$, the next state $\state' \in \States$ is drawn
according to the probability distribution $\prob(\state, \action, \baction)$.

\subparagraph{Histories and plays.}
A history in a CSG is a finite sequence $\langle \state_0, \action_0, \baction_0, \state_1, \action_1, \baction_1, \ldots, \state_k \rangle$ of states and actions, where $\action_t \in \Actions$ and $\baction_t \in \Bactions$,
such that, for any stage $0 \le t < k$, we have $\state_{t+1} \in \support(\prob(\state_t, \action_t, \baction_t))$.
The set of all histories is denoted by $\mathcal{H}$.
Similarly, a play in a CSG is an infinite sequence $\play = \langle \state_0, \action_0, \baction_0, \state_1, \action_1, \baction_1, \ldots \rangle$
such that, for any stage $t \ge 0$, we have $\state_{t+1} \in \support(\prob(\state_t, \action_t, \baction_t))$.
The set of all plays is denoted by $\Plays$.
For any state $\state$, the set of all plays starting at $\state$, i.e., $\play = \langle \state_0, \action_0, \baction_0, \state_1, \ldots \rangle$ where $\state_0 = \state$, is denoted by $\Plays_\state$.

\subparagraph{Objectives.} An objective is a measurable function $\phi \colon \Plays \to [0, 1]$ that assigns a payoff to every play.
Player~1 aims to maximize the expectation of the objective, while Player~2 aims to minimize it.
In this work, we consider the following objectives, where $\play = \langle \state_0, \action_0, \baction_0, \state_1, \action_1, \baction_1, \ldots \rangle$ denotes a play.
\begin{compactitem}
  \item \emph{(Reachability)} Given a target set $\Target \subseteq \States$, the reachability objective consists in eventually reaching $\Target$, i.e., $\reach(\Target)(\play) \defas 1$ if $\state_t \in \Target$ for some $t \ge 0$, and $\reach(\Target)(\play) \defas 0$ otherwise.
  \item \emph{(Discounted-sum)} Given a reward function $r \colon \States \times \Actions \times \Bactions \to [0, 1]$ and a discount factor $\lambda \in [0, 1)$, the discounted-sum objective is the discounted sum of the rewards, i.e.,
    \[
      \disc(r, \lambda)(\play) \defas (1 - \lambda) \sum_{t \ge 0} \lambda^t \, r(\state_t, \action_t, \baction_t) \,.
    \]
  \item \emph{(Limit-average)} Given a reward function $r \colon \States \times \Actions \times \Bactions \to [0, 1]$, the limit-average objective is the long-run average of the rewards, i.e.,
    \[
      \liminfavg(r)(\play) \defas \liminf_{k \to \infty} \frac{1}{k} \sum_{t = 0}^{k - 1} r(\state_t, \action_t, \baction_t) \,.
    \]
    
\end{compactitem}
Note that reachability objectives are special cases of limit-average objectives, see \Cref{Section: Liminf-average}.

\subparagraph{Strategies.}
A strategy for Player~1 is a function $\strategyone \colon \mathcal{H} \to \Delta(\Actions)$ that assigns a probability distribution over actions to every history.
Similarly, a strategy for Player~2 is a function $\strategytwo \colon \mathcal{H} \to \Delta(\Bactions)$.
Given strategies $\strategyone$ and $\strategytwo$, the game proceeds as follows.
At each stage, given the current history $h$, Player~1 (resp. Player~2) chooses an action according to the distribution $\strategyone(h)$ (resp. $\strategytwo(h)$), independently of the other player.
The set of all strategies for Player~1 and Player~2 is denoted by $\Strategyone$ and $\Strategytwo$ respectively.
A stationary strategy depends on the history only through the current state, i.e., it is a function that assigns a probability distribution over actions to every state and is denoted by $\strategyone \colon \States \to \Delta(\Actions)$ for Player~1 (resp. $\strategytwo \colon \States \to \Delta(\Bactions)$ for Player~2).
The set of all stationary strategies for Player~1 and Player~2 is denoted by $\Strategyone^\mathcal{S}$ and $\Strategytwo^\mathcal{S}$ respectively.
A stationary strategy is pure if it assigns a Dirac distribution to every state.

\subparagraph{Markov decision processes and Markov chains.}
For $\mathfrak{p} \in \{1, 2\}$, a player-$\mathfrak{p}$ Markov decision process (player-$\mathfrak{p}$ MDP) is a special class of CSGs
where the other player has only one action and is denoted by $P = (\States, \A, \prob \colon \States \times \Actions \to \Delta(\States))$.
A Markov chain (MC) is a special class of MDPs where both players have only one action and is denoted by $C = (\States, \prob \colon \States \to \Delta(\States))$.

\subparagraph{Probability space.}
An initial state $\state \in \States$ and a pair of strategies $(\strategyone, \strategytwo)$ induce a unique probability measure over $\Plays_\state$, endowed with the standard sigma-algebra generated by the finite histories.
We denote by $\pr_\state^{\strategyone, \strategytwo}$ and $\mathbb{E}_\state^{\strategyone, \strategytwo}$ the probability and the expectation, respectively.
We state the fundamental determinacy for CSGs with the objectives above.

\begin{theorem}[Determinacy~{\cite{everett1957RecursiveGames, shapley1953stochastic, mertens1981StochasticGames}}]
    For all CSGs, states $\state$, and reachability, discounted-sum, or limit-average objectives $\phi$, we have
    \[
        \sup_{\strategyone \in \Strategyone} \mathop{\vphantom{\sup}\inf}_{\strategytwo \in \Strategytwo} \mathbb{E}_{\state}^{\strategyone, \strategytwo}[\phi] =
        \mathop{\vphantom{\sup}\inf}_{\strategytwo \in \Strategytwo} \sup_{\strategyone \in \Strategyone} \mathbb{E}_{\state}^{\strategyone, \strategytwo}[\phi] \,.
    \]
\end{theorem}

\subparagraph{Values, and $\eps$-optimal strategies.}
The above determinacy result implies that switching the quantification order of strategies does not make
a difference and leads to the unique notion of value, defined for a state $\state$ as
\[
    \val_\phi^G(\state) \defas \sup_{\strategyone \in \Strategyone} \mathop{\vphantom{\sup}\inf}_{\strategytwo \in \Strategytwo} \mathbb{E}_{\state}^{\strategyone, \strategytwo}[\phi] \,.
\]
We drop $G$ if the context is clear.
Moreover, while we define limit-average as lim-inf average, the values do not change if $\liminf$ is replaced by $\limsup$ in this definition~\cite{mertens1981StochasticGames}.
For the objectives above, we write $\val_\Target$, $\val_\lambda$, and $\val$ for the values of $\reach(\Target)$, $\disc(r, \lambda)$, and $\liminfavg(r)$, respectively, where the reward function $r$ is clear from the context.
For $\eps \geq 0$, a strategy $\strategyone$ for Player~$1$ is \emph{$\eps$-optimal} if, for all states $\state \in \States$, we have
\[
    \inf_{\strategytwo \in \Strategytwo} \mathbb{E}_\state^{\strategyone, \strategytwo}[\phi] \ge \val_\phi(\state) - \eps \,.
\]
We say the strategy is \emph{optimal} if $\eps = 0$.
The notion of $\eps$-optimal strategies for Player $2$ is defined analogously.
For reachability objectives, for every $\eps \in (0, 1)$, there exists a stationary $\eps$-optimal strategy~\cite{everett1957RecursiveGames} for each player.
Moreover, Player~2 has optimal stationary strategies~\cite{himmelberg1976ExistencePEquilibriumOptimal, parthasarathy1971DiscountedPositiveStochastic}.
For discounted-sum objectives, both players have optimal stationary strategies~\cite{shapley1953stochastic}.
In contrast, for limit-average objectives, $\eps$-optimal strategies require infinite memory in general~\cite{blackwell1968BigMatch}.

\subparagraph{MDPs obtained by fixing a strategy.}
Given a stationary strategy $\strategyone$ for Player~1 in a game $G$ with reward function $r$, by fixing $\strategyone$, we obtain the player-2 MDP $G_\strategyone = (\States, \Bactions, \prob_\strategyone)$ with reward function $r_\strategyone$, where $\prob_\strategyone(\state, \baction)(\state') \defas \sum_{\action \in \Actions} \strategyone(\state)(\action) \cdot \prob(\state, \action, \baction)(\state')$ and $r_\strategyone(\state, \baction) \defas \sum_{\action \in \Actions} \strategyone(\state)(\action) \cdot r(\state, \action, \baction)$ for all states $\state, \state' \in \States$ and actions $\baction \in \Bactions$.
Analogously, we obtain the player-1 MDP $G_\strategytwo$ by fixing a stationary strategy $\strategytwo$ for Player~2.

\subparagraph{Bellman operator $\Psi$.}
For discounted-sum objectives, the values of MDPs are characterized by the Bellman operator.
For $\mathfrak{p} \in \{1, 2\}$, consider a player-$\mathfrak{p}$ MDP $P$ with reward function $r$ and discount factor $\lambda$.
The \emph{Bellman operator} of $P$ maps every function $\rho \colon \States \to \RR$ to the function $\Psi_P(\rho) \colon \States \to \RR$ defined by
\[
    \Psi_P(\rho)(\state) \defas \op_{\action \in \Actions} \left( (1 - \lambda) \, r(\state, \action) + \lambda \sum_{\state' \in \States} \prob( \state, \action)(\state') \rho(\state') \right) \,,
\]
where $\op$ is $\max$ (resp. $\min$) for player-1 (resp. player-2) MDPs.
Since the discount factor is less than one, the Bellman operator is a contraction, and its unique fixed point is $\val_\lambda^P$~\cite{shapley1953stochastic}.

\subparagraph{Computational problems.}
An instance of the computational problems consists of a CSG $G$ with transition probabilities that are rational numbers defined using at most $\Binp$ bits and an objective $\phi$,
which is a reachability objective given by a set of states, or a limit-average objective given by a reward function whose values are rational numbers defined using at most $\Binp$ bits.
The additive error $\eps = 2^{-\Beps} \in (0, 1)$, for an integer $\Beps \ge 1$, is given in binary, i.e., with $\Beps + 1$ bits.
We consider the following three computational problems.
	\begin{tcolorbox}
	    \subparagraph{\CSGD.} Given an instance $(G, \phi)$, a state $s$, and a rational threshold $\alpha$ defined using at most $\Binp$ bits,
	    decide whether $\val_\phi(s) \ge \alpha$.
	\end{tcolorbox}

	\begin{tcolorbox}
	    \subparagraph{\CSGGV.} Given an instance $(G, \phi)$, a state $s$, a rational threshold $\alpha$ defined using at most $\Binp$ bits, and $\eps = 2^{-\Beps}$ in binary,
	    decide whether $\val_\phi(s) \ge \alpha + \eps$ or $\val_\phi(s) \le \alpha - \eps$, where any answer is correct if $\val_\phi(s)$ lies strictly between $\alpha - \eps$ and $\alpha + \eps$.
	\end{tcolorbox}

	\begin{tcolorbox}
	    \subparagraph{\CSGAV.} Given an instance $(G, \phi)$ and $\eps = 2^{-\Beps}$ in binary,
	    compute a function $\hat{v} \colon \States \to \mathbb{Q}$ such that $|\hat{v}(\state) - \val_\phi(\state)| \le \eps$ for all states $\state \in \States$.
	\end{tcolorbox}

Complexity statements for the \CSGGV{} problem assume the promise that $\val_\phi(s)$ does not lie strictly between $\alpha - \eps$ and $\alpha + \eps$.

\section{Overview of Results and Techniques: Value Approximation}
\label{Section: Overview}

In this section, we focus on the \CSGAV{} and \CSGGV{} problems, which are the main subject of this paper.
We recall the relevant previous results, state our results, and present the techniques that we use to obtain them.

\subsection{Previous Results}
\label{Section: Previous Results}

For reachability objectives, since $\eps$-optimal stationary strategies always exist, a natural approach is to guess such strategies and verify their $\eps$-optimality.
Thus, the computational complexity crucially depends on the complexity of $\eps$-optimal stationary strategies.
The mathematical definition to capture the complexity of distributions associated with stationary
strategies is the notion of patience.

\subparagraph{Patience of strategies.}
For a probability distribution $\mu \in \Delta(X)$, its \emph{patience} is the inverse of minimum non-zero probability, i.e.,
$
    \pati(\mu) \defas \max \left\{ \frac{1}{\mu(x)} \mid x \in \support(\mu) \right\} \,.
$
For a stationary strategy $\pi$ of either player, the patience is the maximum of patience over the probability distributions of $\pi$ in all states $\state \in \States$,
i.e., $\pati(\pi) \defas \max \left\{ \pati(\pi(\state)) \mid \state \in \States \right\}$.
For reachability objectives, double exponential patience is necessary~\cite{hansen2009WinningConcurrentReachability, chatterjee2017StrategyComplexityConcurrent} and sufficient~\cite[full version, Corollary~42]{hansen2011ExactAlgorithms} for $\eps$-optimal stationary strategies of both players.
Hence, explicitly guessing an $\eps$-optimal strategy requires exponential space, and a different representation of stationary strategies is necessary.

\subparagraph{Symbolic representation.}
The work of \cite{frederiksen2013ApproximatingValueConcurrent} presents a symbolic representation of
strategies called floating-point representation.
They show that $\eps$-optimal strategies, for $\eps \in (0, 1)$, for the reachability player have polynomial-size
floating-point representation.
However, the verification procedure given such a guessed strategy requires an NP oracle.
Indeed, verifying that a guessed strategy $\strategyone$ is $\eps$-optimal reduces to approximating the value of the MDP obtained by fixing $\strategyone$,
and the verification quantifies over the pure stationary strategies of the opponent in this MDP.
Hence, they obtain a TFNP[NP] upper bound for the \CSGAV{} problem for reachability objectives.
An open problem is to obtain a polynomial-time verification procedure.

\subparagraph{Generalization to limit-average objectives.}
For limit-average objectives, $\eps$-optimal strategies may require infinite memory, so they cannot be guessed directly.
Instead, their values are approximated by the values of discounted-sum objectives with a discount factor that is double exponentially close to one~\cite{oliubarton2021NewAlgorithms},
for which both players have optimal stationary strategies.
The TFNP[NP] upper bound of~\cite{frederiksen2013ApproximatingValueConcurrent} was extended to limit-average objectives in the work of~\cite{bose2024BoundedMemory}.

\subsection{Our Results and Techniques}
\label{Section: Our Results}

Our main result improves the upper bounds in \Cref{Table: Previous results} from TFNP[NP] to TFNP.

\begin{theorem}[Value approximation]
\label{Result: CSG in TFNP}
  For limit-average and reachability objectives, (i)~the \CSGAV{} problem is in TFNP; and (ii)~the \CSGGV{} problem is in NP $\cap$ coNP.
\end{theorem}

We now present our techniques and compare them with previous approaches.

\subparagraph{Reduction to discounted-sum objectives.}
We approximate limit-average values by discounted-sum values.
The required discount factor is double exponentially close to one, but it can be chosen as $\lambda = 1 - 2^{-2^{\Blam}}$, where $\Blam \in \NN_0$ is polynomial in the input size~\cite[Theorem~3]{oliubarton2021NewAlgorithms}.
We call such a discount factor \emph{succinct}.
Therefore, \Cref{Result: CSG in TFNP} follows from the following result.

\begin{theorem}[Value approximation for discounted-sum objectives]
\label{Result: Discounted in TFNP}
    For discounted-sum objectives with a succinct discount factor, (i)~the \CSGAV{} problem is in TFNP; and (ii)~the \CSGGV{} problem is in NP $\cap$ coNP.
\end{theorem}

\begin{remark}[Discount factors given in binary]
  \label{Remark: Binary discount factors}
  For discount factors given in binary, the \CSGAV{} problem is in UEOPL~\cite{batziou2025MonotoneContractions}, a subclass of TFNP.
  This result does not apply to succinct discount factors, since their binary representations have exponentially many bits, 
  and an approximate fixed point $x$ of the Bellman operator $\Psi$ of the game with $\|x - \Psi(x)\|_\infty \le \eps$ only guarantees $\|x - \val_\lambda\|_\infty \le \frac{\eps}{1 - \lambda}$, so approximating the value requires the precision $\eps (1 - \lambda)$, which also has exponentially many bits.
\end{remark}

To prove \Cref{Result: Discounted in TFNP}, we guess a witness that consists of stationary strategies for both players and interval bounds for the values, and we verify it in polynomial time.
The probabilities of $\eps$-optimal strategies may be double exponentially small~\cite{hansen2009WinningConcurrentReachability, chatterjee2017StrategyComplexityConcurrent}, and the succinct discount factor is double exponentially close to one.
Hence, we first need a representation of numbers that supports exact arithmetic with such numbers.

\subparagraph{Power-series representation.}
The floating-point representation of~\cite{frederiksen2013ApproximatingValueConcurrent} represents such numbers with polynomially many bits, but it does not support exact arithmetic, since the result of an operation may require exponentially many bits.
For example, $2^{-1}$ and $2^{-2^n}$ have floating-point representations with polynomially many bits, whereas their difference $2^{-1} - 2^{-2^n}$ has $2^n - 1$ significant binary digits.
We present a new symbolic representation, called power-series representation (\Cref{Section: Power Series Representation}), where a number is a sum of polynomially many signed powers of two whose exponents are written in binary.
This representation supports exact addition, subtraction, multiplication, and comparison in polynomial time (\Cref{lemm:power_arithmetic}),
and it represents the succinct discount factor exactly (\Cref{Section: Discounted-sum}).

\subparagraph{Witness and verification.}
We guess $\eps$-optimal stationary strategies $\strategyone$ and $\strategytwo$ for both players, which are now power-series representable.
The difficulty lies in verifying their $\eps$-optimality.
Previous approaches approximate the values of the MDPs $G_\strategyone$ and $G_\strategytwo$ obtained by fixing these strategies, which quantifies over the strategies of the opponent and requires an NP oracle~\cite{frederiksen2013ApproximatingValueConcurrent}.
Our key idea is to include certificates for these values in the witness.
\begin{compactitem}
  \item \emph{(Witness)}
    Besides the strategies, the witness contains interval bounds $(\low_\strategyone, \up_\strategyone)$ and $(\low_\strategytwo, \up_\strategytwo)$ for the values of $G_\strategyone$ and $G_\strategytwo$, which are required to be $\frac{\eps}{2}$-close, i.e., the intervals have length at most $\frac{\eps}{2}$, and shrinking, i.e., $\low_\strategyone \preceq \Psi_{G_\strategyone}(\low_\strategyone)$ and $\Psi_{G_\strategyone}(\up_\strategyone) \preceq \up_\strategyone$, and similarly for $G_\strategytwo$.
  \item \emph{(Soundness)}
    Since the Bellman operator is monotone and its unique fixed point is the value, shrinking interval bounds contain the value (\Cref{Result: Properties of Bellman operator}).
    Hence, $\low_\strategyone$ and $\up_\strategytwo$ certify the $\eps$-optimality of $\strategyone$ and $\strategytwo$ (\Cref{Result: Soundness of witnesses}).
  \item \emph{(Polynomial-time verification)}
    Our verifier (\Cref{algorithm: polynomial verifier}) checks that the interval bounds are shrinking by applying the Bellman operator once, for every state and action, which takes deterministic polynomial time (\Cref{Result: Polynomial-time verifier}).
\end{compactitem}
The key challenge is the completeness of witnesses, i.e., the existence of polynomial-size witnesses (\Cref{Result: Completeness of witnesses}), which has two components:
(a)~polynomial-size strategies, which are obtained by standard techniques; and 
(b)~polynomial-size $\eps$-close shrinking interval bounds, which is the main challenge since the values may be double exponentially close to each other.

\subparagraph{Existence of polynomial-size strategies.}
Similar to~\cite{frederiksen2013ApproximatingValueConcurrent, asadi2024StatefulDiscounted}, we obtain $\eps$-optimal stationary strategies of polynomial size in power-series representation from an optimal stationary strategy in two steps.
First, we remove the probabilities that are smaller than a threshold depending on $\eps$ and the discount factor, which bounds the exponents in the power-series representation and yields an $\eps$-optimal strategy (\Cref{Result: Bounded patience,Result: Continuity}).
Second, we truncate every remaining probability to polynomially many significant binary digits (\Cref{Result: Distribution representation in power series}).
To bound the resulting error independently of the discount factor, we interpret discounted-sum objectives as reachability objectives (\Cref{Result: Stopping MDP})
and apply the continuity result of~\cite{solan2003ContinuityValueCompetitive} for relative perturbations of transition probabilities (\Cref{Result: Relative continuity,Result: Existence of Floating-Point Strategies}).

\begin{figure}[t]
  \centering
  \begin{minipage}[c]{0.38\textwidth}
    \centering
    \begin{tikzpicture}[scale=3, >={Stealth[length=1.5mm]}]
      \draw[->] (0, 0) -- (1.1, 0) node[right] {$x$};
      \draw[->] (0, 0) -- (0, 1.1);
      \draw[dashed] (0, 0) -- (1, 1);
      \foreach \x/\y/\i in {0.3/0.18/1, 0.45/0.285/2, 0.8/0.5825/3, 1/0.7725/4} {
        \draw[dotted] (\x, 0) -- (\x, \y);
        \fill (\x, \y) circle (0.25pt);
        \node[below, font=\footnotesize] at (\x, 0) {$v_{\i}$};
      }
      \node[below, font=\footnotesize] at (0, 0) {$v_0$};
      \draw[thick] (0, 0) -- (0.3, 0.18) -- (0.45, 0.285) -- (0.8, 0.5825) -- (1, 0.7725) node[right] {$f(x)$};
      \draw[densely dotted] (0.45, 0.285) -- (0.8, 0.285);
      \draw[<->] (0.45, 0.25) -- node[below, font=\footnotesize] {$g_3$} (0.8, 0.25);
      \draw[<->] (0.84, 0.285) -- node[right, font=\footnotesize] {$d_3$} (0.84, 0.5825);
    \end{tikzpicture}
  \end{minipage}\hfill
  \begin{minipage}[c]{0.58\textwidth}
    \caption{A convex piecewise-linear function $f$ with breakpoints at the values $v_0 < \dots < v_4$, below the identity (dashed).
      The slope $d_j / g_j$ of $f$ between $v_{j-1}$ and $v_j$ is the ratio of the increment $d_j$ and the gap $g_j$, and it increases in $j$.
      The lower bound of a state with the value $v_j$ is $f(v_j)$.}
    \label{Figure: Convex transformation}
  \end{minipage}
\end{figure}
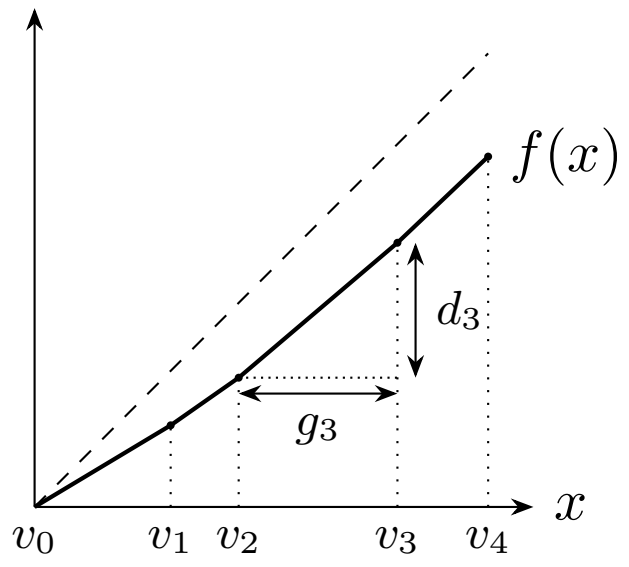

\subparagraph{Existence of polynomial-size interval bounds.}
We now construct $\eps$-close shrinking interval bounds of polynomial size for the MDPs $G_\strategyone$ and $G_\strategytwo$ (\Cref{Result: Shrinking Bounds in MDP}).
This is the main technical challenge of our approach.
Consider an MDP $P$ with the value function $\val_\lambda$.
On the one hand, the values are in general not power-series numbers, e.g., $\frac{1}{3}$, and hence the bounds must deviate from the values.
On the other hand, the bounds must approximate the values, and hence they must be close to the values.
Moreover, distinct values may be double exponentially close to each other.
Since the inequalities of shrinking bounds compare the bound of every state with a $\lambda$-weighted average of the bounds of its successors, where $\lambda$ is double exponentially close to one, these challenges make many natural candidates fail (\Cref{Section: Failed candidates}).
Instead, we construct the bounds as follows.
\begin{compactitem}
  \item \emph{(Convex transformations)}
    We obtain the lower bound $\low \defas f \circ \val_\lambda$ by applying a non-decreasing convex function $f \colon [0, 1] \to \RR$ such that $f(x) \le x$ to the value function.
    For every state $\state$, let $\action$ be an optimal action at $\state$ if $P$ is a player-1 MDP, and an arbitrary action otherwise.
    Then, we have
    \begin{align*}
      f(\val_\lambda(\state))
        &\le f \Big( (1 - \lambda) \, r(\state, \action) + \lambda \sum_{\state'} \prob(\state, \action)(\state') \, \val_\lambda(\state') \Big)
          && (f \text{ is non-decreasing}) \\
        &\le (1 - \lambda) \, r(\state, \action) + \lambda \sum_{\state'} \prob(\state, \action)(\state') \, f(\val_\lambda(\state'))
          && (f \text{ is convex},\ f(x) \le x) \,.
    \end{align*}
    The right-hand side is the term of the action $\action$ in the definition of $\Psi_P(\low)(\state)$.
    For player-1 MDPs, $\Psi_P(\low)(\state)$ is the maximum of these terms over all actions, and hence it is at least the right-hand side.
    For player-2 MDPs, the inequality holds for every action, so it also holds for the minimum over all actions, which is $\Psi_P(\low)(\state)$.
    Therefore, we have $\low \preceq \Psi_P(\low)$ for every such function $f$ (\Cref{Result: Convex concave transformations}), and it remains to choose $f$ such that $\low$ has polynomial size.
  \item \emph{(Choice of the function)}
    Let $0 = v_0 < v_1 < \dots < v_J = 1$ be the distinct values of the states of $P$ together with $0$ and $1$.
    We choose $f$ piecewise linear with $f(0) = 0$ and breakpoints at $v_0, \dots, v_J$ (\Cref{Figure: Convex transformation}).
    Its increment $d_j \defas f(v_j) - f(v_{j-1})$ is the gap $g_j \defas v_j - v_{j-1}$, scaled by a factor slightly smaller than one and truncated to polynomially many significant binary digits.
    Truncation decreases every slope $d_j / g_j$ slightly, but the factors increase from left to right by more than this decrease, and hence the slopes increase and $f$ is convex.
    Moreover, all slopes are at most one, which implies $f(x) \le x$, and at least $1 - \frac{\eps}{2}$, which implies $\val_\lambda(\state) - \frac{\eps}{2} \le \low(\state) \le \val_\lambda(\state)$ for all states $\state$ (see the proof of \Cref{Result: Shrinking Bounds in MDP}).
  \item \emph{(Polynomial size)}
    Every bound is the sum of at most $n + 1$ increments, and truncation to polynomially many significant binary digits requires few terms in power-series representation, no matter how small the gaps are (\Cref{Result: Number representation in power series}).
    Moreover, the gaps are at least double exponentially small by Cramer's rule and Hadamard's inequality (\Cref{Result: Value separation}), and hence the exponents of the power-series representation have polynomially many bits.
  \item \emph{(Upper bound)}
    We obtain the upper bound $\up$ by applying the same construction to the MDP with the rewards $1 - r$ and the opposite player, whose values are $1 - \val_\lambda(\state)$.
    Equivalently, $\up = g \circ \val_\lambda$ for a non-decreasing concave function $g$.
\end{compactitem}

\subparagraph{Novelty.}
The main novelty of our approach is the existence of polynomial-size certificates for the values of MDPs with succinct discount factors, which are verified in deterministic polynomial time.
Previous approaches obtain polynomial-size representations of $\eps$-optimal strategies, but verifying their $\eps$-optimality requires an NP oracle~\cite{frederiksen2013ApproximatingValueConcurrent, bose2024BoundedMemory}.
Although inequalities involving the Bellman operator certify bounds on the values of MDPs~\cite{chatterjee2025FixedPointCertificates}, the main challenge in our setting is to show that such certificates have polynomial size even if the discount factor is double exponentially close to one.
We overcome this challenge with the following two ingredients.
\begin{compactitem}
  \item \emph{(Power-series representation)}
    We represent numbers as sums of polynomially many signed powers of two whose exponents are written in binary.
    In contrast to floating-point numbers~\cite{frederiksen2013ApproximatingValueConcurrent}, this representation supports exact addition, subtraction, multiplication, and comparison with double exponentially small numbers in polynomial time.
  \item \emph{(Polynomial-size $\eps$-close shrinking interval bounds)}
    Our main technical contribution is the construction of lower and upper bounds of polynomial size that satisfy the inequalities involving the Bellman operator exactly.
    We obtain them by applying convex and concave piecewise-linear transformations to the value function.
    Their increments are the gaps between consecutive values, scaled by increasing factors and truncated with a fixed relative precision, which preserves convexity and yields polynomial-size representations even if distinct values are double exponentially close to each other.
    Several natural candidates for such bounds fail (\Cref{Section: Failed candidates}), and the main technical results appear as \Cref{Result: Convex concave transformations}, \Cref{Result: Value separation}, and \Cref{Result: Shrinking Bounds in MDP} (\Cref{Section: Interval Bounds in MDPs}).
\end{compactitem}
Together with polynomial-size representations of $\eps$-optimal strategies, these bounds eliminate the need for an NP oracle and yield the TFNP upper bound for the \CSGAV{} problem.
Moreover, the same witnesses yield the NP $\cap$ coNP upper bound for the \CSGGV{} problem, since a witness with $\low_\strategyone(\state) \ge \alpha$ (resp. $\up_\strategytwo(\state) \le \alpha$) certifies that $\val_\lambda(\state) \ge \alpha$ (resp. $\val_\lambda(\state) \le \alpha$), and such a witness exists if the value is at least $\alpha + \eps$ (resp. at most $\alpha - \eps$).

\section{Overview of Results and Techniques: Value Decision}
\label{Section: Exact problem}

In this section, we focus on the \CSGD{} problem.
We recall the relevant previous results, state our results, and present the techniques that we use to obtain them.

\subsection{Previous Results}
\label{Section: Decision previous results}

\subparagraph{Existential and universal theories of the reals.}
The \emph{existential theory of the reals} (ETR) is the problem of deciding whether a sentence of the form $\exists x_1 \cdots \exists x_k \, \varphi(x_1, \ldots, x_k)$ holds over the reals, where $\varphi$ is a quantifier-free Boolean combination of polynomial equations and inequalities with integer coefficients.
Analogously, the \emph{universal theory of the reals} (UTR) is the problem of deciding whether a sentence of the form $\forall x_1 \cdots \forall x_k \, \varphi(x_1, \ldots, x_k)$ holds over the reals.
The complexity classes $\ER$ and $\coER$ consist of the problems that reduce in polynomial time to ETR and UTR, respectively~\cite{schaefer2009GeometricTopological, schaefer2017FixedPoints}.
We have NP $\subseteq \ER \subseteq$ PSPACE and coNP $\subseteq \coER \subseteq$ PSPACE~\cite{canny1988AlgebraicPSPACE}.

\subparagraph{Straight-line programs.}
A \emph{straight-line program} is a sequence of gates $1, \ldots, k$, where every gate $i$ either is a constant $z_i \in \{0, 1\}$, or computes $z_i \defas z_j \circ z_{j'}$ for gates $j, j' < i$ and an operation $\circ \in \{+, -, \times\}$.
The program computes the integer $z_k$, which may have exponentially many bits.
The \emph{PosSLP} problem asks, given a straight-line program, whether the integer that it computes is positive~\cite{allender2009ComplexityNumericalAnalysis}.
The SQRT-SUM problem reduces to PosSLP under polynomial-time Turing reductions, and PosSLP is in the counting hierarchy~\cite{allender2009ComplexityNumericalAnalysis}.

\subparagraph{Lower bounds.}
The \CSGD{} problem is SQRT-SUM-hard for reachability objectives~\cite{etessami2008RecursiveConcurrentStochastic}, and hence also for limit-average objectives.
PosSLP-hardness was previously known for recursive Markov chains~\cite{etessami2009RecursiveMarkovChains} and for the cost problem in Markov chains~\cite{haase2015OddsStayingBudget}, but, to the best of our knowledge, not for concurrent stochastic games with finitely many states.

\subparagraph{Upper bounds.}
For reachability objectives, the \CSGD{} problem is in PSPACE, and in fact in $\coER$, since~\cite{etessami2008RecursiveConcurrentStochastic} reduces the question whether the value is less than a threshold to the existential theory of the reals.
For discounted-sum objectives, the question whether the value is greater than a threshold reduces to the existential theory of the reals, using a stationary strategy together with a bound on the value function as a witness~\cite{chatterjee2007Thesis, chatterjee2008LimitAverageEXPTIME}.
For limit-average objectives, \cite{chatterjee2008LimitAverageEXPTIME} reduces the \CSGD{} problem to the theory of the reals with a constant number of quantifier alternations, and hence the problem is in PSPACE~\cite{chatterjee2015Value1}.

\subsection{Our Results and Techniques}
\label{Section: Decision our results}

For the \CSGD{} problem, we show PosSLP-hardness, which implies SQRT-SUM hardness under polynomial-time Turing reductions~\cite{allender2009ComplexityNumericalAnalysis}, and we improve the PSPACE upper bound to $\ER \cap \coER$.

\begin{theorem}[Value decision]
\label{Result: Value decision}
  For limit-average and reachability objectives, the following assertions hold.
  \begin{compactenum}
    \item \emph{(Lower bound)} The \CSGD{} problem is PosSLP-hard under polynomial-time many-one reductions, even for the threshold $\frac{1}{2}$ and for games in which every transition is deterministic, both players have two actions, and every play reaches one of two absorbing states.
    \item \emph{(Upper bound)} The \CSGD{} problem is in $\ER \cap \coER$.
  \end{compactenum}
\end{theorem}

\subparagraph{Lower bound.}
To show PosSLP-hardness, we simulate straight-line programs by games with reachability objectives (\Cref{Section: Decision lower bound}).
We consider games with two absorbing states $\top$ and $\bot$, one of which every play reaches, and the target set $\Target \defas \{\top\}$.
For every state $\state$, we use its \emph{centered value} $2 \val_\Target(\state) - 1 \in [-1, 1]$, i.e., the probability of reaching $\top$ minus the probability of reaching $\bot$ when both players follow optimal strategies.
We construct three gadgets: (i)~the dual of a state negates its centered value, similar to the complementary components in the reduction of~\cite{etessami2009RecursiveMarkovChains} for recursive Markov chains; (ii)~an averaging state averages two centered values; and (iii)~a multiplication state, which is our main new construction and in which both players randomize, multiplies two centered values up to the factor $\frac{1}{8}$.
The integers computed by the gates may be double exponentially large, whereas centered values are in $[-1, 1]$.
Hence, for every gate, we construct a numerator state and a denominator state such that the integer computed by the gate is the ratio of their centered values.
In contrast to the reduction of~\cite{etessami2009RecursiveMarkovChains}, which scales all integers at the same depth by the same factor, the denominator states allow a different factor for every gate.
Lastly, the sign of the output is determined by comparing the value of the numerator state of the output gate with the threshold $\frac{1}{2}$.
Since all transition probabilities are integer multiples of $\frac{1}{4}$, we replace stochastic transitions by deterministic transitions without changing the values~\cite{etessami2008RecursiveConcurrentStochastic, chatterjee2015RandomnessForFree}.

\subparagraph{Upper bound.}
The upper bound combines standard techniques with the separation of the values from rational thresholds.
Strict comparisons between the values of discounted-sum objectives and real thresholds are witnessed by a stationary strategy together with a lower or an upper bound on the value function~\cite{chatterjee2007Thesis, chatterjee2008LimitAverageEXPTIME} (\Cref{Result: Real witnesses}).
The value of a limit-average objective is an algebraic number whose degree and coefficients are bounded~\cite{oliubarton2021NewAlgorithms}, and hence, if it differs from a rational threshold $\alpha$, then it differs from $\alpha$ by at least a double exponentially small amount $\eta$ (\Cref{Result: Algebraic values,Result: Separation from thresholds}).
Since discounted-sum values with a succinct discount factor approximate limit-average values up to $\frac{\eta}{4}$ (\Cref{Result: Liminf-average approximation}), we compare them strictly with the shifted threshold $\alpha - \frac{\eta}{2}$, and we define the double exponentially small constants by repeated squaring~\cite{ibsenjensen2023ConcurrentGames, bose2024BoundedMemory}.
This reduces both comparisons $\val(\state) \ge \alpha$ and $\val(\state) < \alpha$ to the existential theory of the reals.

\subparagraph{Novelty.}
The main novelty of our value-decision results is the lower bound.
To the best of our knowledge, it is the first PosSLP-hardness result for concurrent stochastic games, and it strengthens the SQRT-SUM-hardness of~\cite{etessami2008RecursiveConcurrentStochastic}, even for reachability objectives, deterministic transitions, and two actions per player.
In contrast to the PosSLP-hardness for recursive Markov chains~\cite{etessami2009RecursiveMarkovChains}, which relies on recursion, we simulate multiplication with finitely many states by a new multiplication state in which both players randomize, together with numerator and denominator states that scale the integer of every gate separately.

\section{TFNP Algorithm for \texorpdfstring{\CSGAV}{CsgApproxValue}}
\label{Section: TFNP Algorithm}

In this section, we prove \Cref{Result: Discounted in TFNP}, where the input consists of a game, a reward function whose values are rational numbers defined using at most $\Binp$ bits, and a succinct discount factor.
To do so, we first introduce the power-series representation of numbers (\Cref{Section: Power Series Representation}).
Then, we define witnesses, which consist of power-series representable strategies and $\frac{\eps}{2}$-close shrinking interval bounds, and we present a polynomial-time verifier for them (\Cref{Section: Witnesses and Verifier}).
We defer the proof of the completeness of witnesses, i.e., that polynomial-size witnesses exist (\Cref{Result: Completeness of witnesses}), to \Cref{Section: Existence of Witness}.

\subsection{Power Series Representation and Properties}
\label{Section: Power Series Representation}

In this subsection, we introduce the power-series representation of numbers and present its properties.
We call an object \emph{power-series representable} if the numbers that specify it are represented in power series, namely,
(a)~the entries of an interval bound;
(b)~the probabilities of a distribution (see $\D(\ell, B)$ below); and
(c)~the probabilities of the distributions that a stationary strategy assigns to the states.

\subparagraph{Power series representation.}
Let $\F(\ell, B)$ be a subset of rationals consisting of the sum of at most $\ell$ signed powers of $2$, where the exponent is non-positive and greater than $-2^{B}$, i.e., its absolute value is written in binary with $B$ bits.
Formally,
    \begin{align*}
      \F(\ell, B)
        \defas \left\{ \sum_{i = 1}^k \pm 2^{-a_i} \quad \mid \quad k \le \ell, \quad \forall i \le k \quad a_i \in \NN_0, \quad a_i < 2^{B} \right\} \,.
    \end{align*}
The power series representation of an element $x = \pm 2^{-a_1} \pm \dots \pm 2^{-a_k} \in \F(\ell, B)$ uses at most $\ell (B + 1)$ bits, namely, a sign bit and $B$ bits for the exponent of every term.
Note that the exponents may be exponentially large in the size of the input, while the representation remains polynomial, since the exponents are written in binary.
We denote non-negative power-series representable numbers by $\F_+(\ell, B) \defas \F(\ell, B) \cap [0, \infty)$.

\subparagraph{Comparison to floating-point.}
A floating-point number $(m, x)$ represents $m2^x$.
The addition/subtraction of two floating-point numbers is not necessarily polynomial-size representable floating-point number, e.g.,
two floating-point numbers $2^{-1}$ and $2^{-2^n}$ can be represented in floating-point with $O(n)$ bits,
but their subtraction $2^{-1} - 2^{-2^n}$ requires $2^n - 1$ bits. In contrast, power-series representation supports polynomial-time addition, subtraction, multiplication, and comparison.

\subparagraph{Arithmetic operations.} Basic operations over power-series numbers have the following properties.
\begin{lemma}
  \label{lemm:power_arithmetic}
  If $x \in \F(\ell_1, B_1)$ and $y \in \F(\ell_2, B_2)$, then
  \begin{compactitem}
      \item Addition and Subtraction: $x + y, x - y \in \F(\ell_1 + \ell_2, \max\{B_1, B_2\})$
      \item Multiplication: $x \cdot y \in \F(\ell_1 \cdot \ell_2, \max\{B_1, B_2\} + 1)$
  \end{compactitem}
  and all operations, together with the comparison $x < y$, can be performed in polynomial time with respect to the input size.
\end{lemma}
\begin{proof}
  Consider $x = \sum_{i = 1}^{k} (-1)^{b_i} \cdot 2^{-a_i}$ and $y = \sum_{i=1}^{\tilde{k}} (-1)^{\tilde{b}_i} \cdot 2^{-\tilde{a}_i}$, and denote $B \defas \max\{B_1, B_2\}$.
  \begin{compactitem}
    \item \emph{(Addition)} The value
      \[
        x + y = \sum_{i=1}^{k} (-1)^{b_i} \cdot 2^{-a_i} + \sum_{i=1}^{\tilde{k}} (-1)^{\tilde{b}_i} \cdot 2^{-\tilde{a}_i} \in \F(\ell_1 + \ell_2, B) \,,
      \]
      and can be computed in $\mathcal{O}\left ((\ell_1 + \ell_2) (B + 1) \right )$.
    \item \emph{(Subtraction)} The value
      \[
        x - y = \sum_{i=1}^{k} (-1)^{b_i} \cdot 2^{-a_i} + \sum_{i=1}^{\tilde{k}} (-1)^{1 - \tilde{b}_i} \cdot 2^{-\tilde{a}_i} \in \F(\ell_1 + \ell_2, B) \,,
      \]
      and can be computed in $\mathcal{O}\left ((\ell_1 + \ell_2) (B + 1) \right )$.
    \item \emph{(Multiplication)} The value
      \[
        x \cdot y = \sum_{i=1}^{k} \sum_{j=1}^{\tilde{k}} (-1)^{b_i + \tilde{b}_j} \cdot 2^{-(a_i + \tilde{a}_j)} \in \F(\ell_1 \cdot \ell_2, B + 1) \,,
      \]
      since $a_i + \tilde{a}_j < 2^{B_1} + 2^{B_2} \le 2^{B + 1}$, and can be computed in $\mathcal{O}\left (\ell_1 \ell_2 (B + 1) \right )$.
    \item \emph{(Comparison)} It suffices to decide the sign of $z \defas x - y$.
    By subtraction, we have $z = \sum_{i=1}^{\hat{k}} (-1)^{\hat{b}_i} 2^{-\hat{a}_i}$ where $\hat{k} \le \ell_1 + \ell_2$.
    First, we rewrite this sum, without changing its value, so that all exponents are pairwise distinct.
    As long as two terms have the same exponent $a$, we replace them as follows:
    if they have opposite signs, we remove both, since $2^{-a} - 2^{-a} = 0$;
    otherwise, they have the same sign $(-1)^{b}$, and we replace them by the single term $(-1)^{b} 2^{-(a - 1)}$, since $2^{-a} + 2^{-a} = 2^{-(a-1)}$.
    Every replacement decreases the number of terms, so there are at most $\hat{k}$ replacements.
    Moreover, every replacement decreases an exponent by at most one, so all exponents remain integers of absolute value at most $\hat{k} + 2^{B}$.
    If no term remains, then $x = y$.
    Otherwise, we obtain $z = \sum_{i=1}^{k^*} (-1)^{b^*_i} 2^{-a^*_i}$ with distinct exponents $a^*_1 < a^*_2 < \dots < a^*_{k^*}$.
    Second, the first term dominates the others, since the exponents $a^*_2, \ldots, a^*_{k^*}$ are distinct integers larger than $a^*_1$, and therefore
    \[
      \sum_{i = 2}^{k^*} 2^{-a^*_i} < \sum_{j > a^*_1} 2^{-j} = 2^{-a^*_1} \,.
    \]
    Hence, the sign of $z$ is the sign $(-1)^{b^*_1}$ of its first term.
    Lastly, keeping the terms sorted by exponent, every replacement takes $\mathcal{O}\left( \hat{k} (\log \hat{k} + B + 1) \right)$ time,
    so the procedure takes $\mathcal{O}\left( (\ell_1 + \ell_2)^2 (\log (\ell_1 + \ell_2) + B + 1) \right)$ time.
  \end{compactitem}
  The desired result follows.
\end{proof}

\subparagraph{Rational coefficients.}
The rewards and transition probabilities of the input are rational numbers that are in general not power-series numbers, e.g., $\frac{1}{3}$, and power-series numbers are not closed under division.
Nevertheless, the following result shows that the sign of a linear combination of power-series numbers with rational coefficients is decided exactly in polynomial time.

\begin{lemma}[Rational coefficients]
  \label{Result: Rational coefficients}
  Let $x_1, \ldots, x_k \in \F(\ell, B)$, and let $c_1, \ldots, c_k$ be rational numbers defined using at most $\Binp$ bits.
  Then, the sign of $\sum_{i = 1}^{k} c_i x_i$ is decided in time polynomial in $k$, $\ell$, $B$, and $\Binp$.
\end{lemma}

\begin{proof}
  Let $Q$ be the product of the denominators of $c_1, \ldots, c_k$, and let $B_Q \defas k \Binp$.
  Since $Q \cdot 2^{-B_Q} > 0$, the sum $\sum_{i = 1}^{k} c_i x_i$ has the same sign as
  \[
    Q \cdot 2^{-B_Q} \sum_{i = 1}^{k} c_i x_i = \sum_{i = 1}^{k} \big( Q c_i 2^{-B_Q} \big) \, x_i \,.
  \]
  For every $i \in [k]$, the number $Q c_i$ is the product of the numerator of $c_i$ and the denominators of the other coefficients, and hence it is an integer with $|Q c_i| < 2^{B_Q}$, which is computed in polynomial time.
  Let $|Q c_i| = \sum_{j = 0}^{B_Q - 1} z_j 2^{j}$ be the binary representation of $|Q c_i|$, where $z_j \in \{0, 1\}$.
  Then, we have
  \[
    Q c_i 2^{-B_Q} = \pm \sum_{j = 0}^{B_Q - 1} z_j 2^{-(B_Q - j)} \in \F(B_Q, B_Q) \,,
  \]
  where the sign is the sign of $c_i$, and the membership holds because $1 \le B_Q - j \le B_Q < 2^{B_Q}$.
  Hence, by \Cref{lemm:power_arithmetic}, we have $(Q c_i 2^{-B_Q}) \, x_i \in \F(B_Q \ell, \max\{B, B_Q\} + 1)$ for every $i \in [k]$, and the right-hand side is in $\F(k B_Q \ell, \max\{B, B_Q\} + 1)$.
  Lastly, since $B_Q = k \Binp$, by \Cref{lemm:power_arithmetic}, the right-hand side is computed and compared with $0$ in time polynomial in $k$, $\ell$, $B$, and $\Binp$.
\end{proof}

\subparagraph{Approximation.}
We define the relative distance between two numbers $x, x' \in (0, 1]$ by
\[
  \rel(x, x')
    \defas \max \left\{ \frac{x}{x'}, \frac{x'}{x} \right\} - 1
    = \inf \{ \gamma \ge 0 \enspace \mid \enspace x \le (1 + \gamma) x' , \quad x' \le (1 + \gamma) x \}\,,
\]
i.e., $x$ and $x'$ are equal up to a multiplicative factor $1 + \rel(x, x')$.
Moreover, we set $\rel(0, 0) \defas 0$ and $\rel(x, 0) = \rel(0, x) \defas \infty$ for $x \in (0, 1]$.
The set $\F(\ell, B)$ approximates the interval $(0, 1]$ in the following way.
\begin{lemma}[Truncation]
  \label{Result: Number representation in power series}
  Let $2^{B} \ge \ell \ge 1$, and let $x \in (2^{\ell - 2^{B}}, 1]$.
  Denote by $\trunc_\ell(x)$ the number obtained from the binary expansion of $x$ by keeping its first $\ell$ digits, starting from the first non-zero digit, where we use the finite binary expansion of $x$ if it exists.
  Then, $\trunc_\ell(x) \in \F_+(\ell, B)$, $\trunc_\ell(x) \le x$, and $\rel(x, \trunc_\ell(x)) \le 2^{1-\ell}$.
  In particular, $(1 - 2^{1 - \ell})\, x \le \trunc_\ell(x) \le x$.
\end{lemma}

\begin{proof}
  Consider a binary sequence $(x_i)_{i \ge 1}$ such that $x = \sum_{i \ge 1} 2^{-i} x_i$ (if $x$ has two such sequences, we choose the finite one), and denote by $i_0 \defas \min \{ i \mid x_i = 1 \}$ the index of its first nonzero element.
  Note that, since $2^{\ell - 2^{B}} < x \le \sum_{i \ge i_0} 2^{-i} = 2^{1 - i_0}$, the index $i_0$ is bounded from above as follows
  \[
    i_0 < 2^{B} - \ell + 1 \,.
  \]
  Define the approximation $x'$ as the following truncated binary expansion starting at $i_0$ and containing at most $\ell$ terms
  \[
    x' \defas \trunc_\ell(x) \defas \sum_{i = i_0}^{i_0 + \ell - 1} 2^{-i} x_i \in \F_+(\ell, B) \,,
  \]
  where the membership holds because $i_0 + \ell - 1 < 2^{B}$.
  Clearly, $x' \le x$.
  Then, their distance is bounded as follows
  \begin{align*}
    \rel(x, x')
      &= \frac{x}{x'} - 1
        &(x' \le x) \\
      &= \frac{1}{x'} (x - x') \\
      &\le 2^{i_0} \left( \sum_{i \ge i_0 + \ell} 2^{-i} x_i \right)
        &(x' \ge 2^{-i_0})\\
      &\le 2^{i_0} 2^{-i_0 - \ell + 1} \\
      &= 2^{1 - \ell} \,.
  \end{align*}
  Lastly, $\rel(x, x') \le 2^{1 - \ell}$ implies $x' \ge x / (1 + 2^{1 - \ell}) \ge (1 - 2^{1 - \ell}) x$, which concludes the proof.
\end{proof}

\subparagraph{Power-series probability distribution representation.}
We denote the set of probability distributions whose probabilities are power-series numbers by
\[
    \D(\ell, B)
      \defas \left\{ \mu \in \Delta([t]) \mid \mu(i) \in \F_+(\ell, B) \text{ for all } i \in [t] \right\} \,.
\]

Extending the relative distance $\rel$ to probability distributions by
\[
    \rel (\mu, \mu') \defas \max \{ \rel(\mu(i), \mu'(i)) \mid i \in [t]\} \,,
\]
we have the following result.

\begin{lemma}[Distribution approximation]
    \label{Result: Distribution representation in power series}
    Let $t \ge 1$ and $2^{B} \ge \ell \ge 2$, and let $\mu \in \Delta([t])$ be a distribution with $\pati(\mu) < 2^{2^{B} - \ell}$.
    Then, there exists a distribution $\mu' \in \D(t \ell, B)$ such that $\support(\mu') = \support(\mu)$ and $\rel(\mu, \mu') \le t \cdot 2^{1 - \ell}$.
\end{lemma}

\begin{proof}
    Let $i^* \in [t]$ be such that $\mu(i^*)$ is maximal. Therefore, we have $\mu(i^*) \ge 1/t$ and hence $1 - \mu(i^*) \le (t - 1) \mu(i^*)$.
    For every $i \in [t]$, we define
    \[
      \mu'(i) \defas
        \begin{cases}
          1 - \sum_{j \ne i^*} \mu'(j) & i = i^* \,, \\
          \trunc_\ell(\mu(i)) & i \ne i^* \text{ and } \mu(i) > 0 \,, \\
          0 & \text{otherwise.}
        \end{cases}
    \]
    For all $i \in \support(\mu)$ such that $i \neq i^*$, since $\mu(i) \ge 1 / \pati(\mu) > 2^{\ell - 2^{B}}$, \Cref{Result: Number representation in power series} yields $\mu'(i) \in \F_+(\ell, B)$, $\mu'(i) \le \mu(i)$, and $\rel(\mu(i), \mu'(i)) \le 2^{1 - \ell}$.
    By definition, we have $\sum_{i \in [t]} \mu'(i) = 1$ and, by \Cref{lemm:power_arithmetic}, $\mu'(i^*) \in \F(t \ell, B)$.
    Moreover, we have
    \[
      0 \le \mu'(i^*) - \mu(i^*) = \sum_{i \ne i^*} \big( \mu(i) - \mu'(i) \big) \le 2^{1 - \ell} \big( 1 - \mu(i^*) \big) \le (t - 1) \, 2^{1 - \ell} \mu(i^*) \,.
    \]
    Therefore, we have $\mu'(i^*) > 0$ and $\rel(\mu(i^*), \mu'(i^*)) \le (t - 1) \, 2^{1 - \ell}$, which completes the proof.
\end{proof}

\subparagraph{Succinct discount factors.}
\label{Section: Discounted-sum}
A succinct discount factor $\lambda = 1 - 2^{-2^{\Blam}}$ is represented exactly in power-series representation, namely, $\lambda \in \F(2, \Blam + 1)$ and $1 - \lambda \in \F(1, \Blam + 1)$.

\subsection{Witnesses and Verifier}
\label{Section: Witnesses and Verifier}

In this subsection, we first show properties of the Bellman operator (\Cref{Result: Properties of Bellman operator}) and define $\eps$-close shrinking interval bounds.
Then, we define witnesses and show their soundness (\Cref{Result: Soundness of witnesses}) and completeness (\Cref{Result: Completeness of witnesses}), whose proof we defer to \Cref{Section: Existence of Witness}.
Next, we present a verifier for witnesses and show that it runs in polynomial time (\Cref{Result: Polynomial-time verifier}).
Lastly, we prove \Cref{Result: Discounted in TFNP}.

\begin{lemma}[Properties of the Bellman operator]
  \label{Result: Properties of Bellman operator}
  For $\mathfrak{p} \in \{1, 2\}$, let $P$ be a player-$\mathfrak{p}$ MDP with a reward function and a discount factor $\lambda$, and let $\rho, \rho' \colon \States \to \RR$.
  Then, the following assertions hold.
  \begin{compactenum}
    \item \emph{(Monotonicity)} If $\rho \preceq \rho'$, then $\Psi_P(\rho) \preceq \Psi_P(\rho')$.
    \item \emph{(Lower bound)} If $\rho \preceq \Psi_P(\rho)$, then $\rho \preceq \val_\lambda$.
    \item \emph{(Upper bound)} If $\Psi_P(\rho) \preceq \rho$, then $\val_\lambda \preceq \rho$.
  \end{compactenum}
\end{lemma}

\begin{proof}
  We present the proof of the three items.
  \begin{compactitem}
    \item \emph{(Monotonicity)}
      For all functions $\rho \preceq \rho'$, since the coefficients of $\rho$ in the definition of $\Psi_P(\rho)$ are non-negative and $\op$ is monotone, we have $\Psi_P(\rho) \preceq \Psi_P(\rho')$.
    \item \emph{(Lower bound)}
      First, we show that the Bellman operator is a contraction with respect to the $L_\infty$ norm.
      For all functions $\rho, \rho' \colon \States \to \RR$ and states $\state \in \States$, we have
      \begin{align*}
        \big| \Psi_P(\rho)(\state) - \Psi_P(\rho')(\state) \big|
          &\le \max_{\action \in \Actions} \Big| \lambda \sum_{\state' \in \States} \prob(\state, \action)(\state') \big( \rho(\state') - \rho'(\state') \big) \Big|
            && (\op \text{ is } 1\text{-Lipschitz}) \\
          &\le \lambda \| \rho - \rho' \|_\infty
            && (\prob(\state, \action) \in \Delta(\States)) \,.
      \end{align*}
      Taking the maximum over all states $\state$, we have $\| \Psi_P(\rho) - \Psi_P(\rho') \|_\infty \le \lambda \| \rho - \rho' \|_\infty$, where $\lambda < 1$.
      Therefore, by the Banach fixed-point theorem, $\Psi_P$ has a unique fixed point, which is $\val_\lambda$~\cite{shapley1953stochastic}, and, for every function $\rho$, the iterates $\Psi_P^k(\rho)$ converge to $\val_\lambda$ as $k \to \infty$.
      Second, suppose that $\rho \preceq \Psi_P(\rho)$.
      We show by induction that $\rho \preceq \Psi_P^k(\rho)$ for all $k \ge 1$.
      The base case $k = 1$ holds by assumption.
      If $\rho \preceq \Psi_P^k(\rho)$, then, by monotonicity, we have $\Psi_P(\rho) \preceq \Psi_P^{k+1}(\rho)$, and therefore $\rho \preceq \Psi_P(\rho) \preceq \Psi_P^{k+1}(\rho)$.
      Lastly, since $\preceq$ is preserved under pointwise limits, taking the limit $k \to \infty$ yields $\rho \preceq \val_\lambda$.
    \item \emph{(Upper bound)}
      Suppose that $\Psi_P(\rho) \preceq \rho$.
      Analogously to the lower bound, by induction and monotonicity, we have $\Psi_P^k(\rho) \preceq \rho$ for all $k \ge 1$.
      Taking the limit $k \to \infty$, we have $\val_\lambda \preceq \rho$, which completes the proof.
  \end{compactitem}
\end{proof}

\subparagraph{Interval bounds.}
An \emph{interval bound} is a pair $I = (\low, \up)$ of functions $\low, \up \colon \States \to \RR$ such that $\low \preceq \up$.
For interval bounds $I = (\low, \up)$ and $I' = (\low', \up')$, we write $I' \subseteq I$ if $[\low'(\state), \up'(\state)] \subseteq [\low(\state), \up(\state)]$ for all states $\state \in \States$, i.e., $\low \preceq \low'$ and $\up' \preceq \up$.
For $\mathfrak{p} \in \{1, 2\}$ and a player-$\mathfrak{p}$ MDP $P$, we say that $I$ is \emph{shrinking} if $(\Psi_P(\low), \Psi_P(\up)) \subseteq I$.
For $\eps \in (0, 1)$, we say that $I$ is \emph{$\eps$-close} if $\up(\state) - \low(\state) \le \eps$ for all states $\state \in \States$.

\begin{remark}[Shrinking interval bounds]
  \label{Remark: Shrinking interval bounds}
  In words, an interval bound is shrinking if applying the Bellman operator once to the lower and upper bounds yields intervals that are contained in the original ones.
  Similar interval bounds appear in interval value iteration for MDPs~\cite{haddad2018Interval, hartmanns2020Optimistic} and concurrent stochastic games~\cite{grobelna2025StoppingCriteria}, where they serve as a sound stopping criterion, and in certificates for the values of MDPs~\cite{chatterjee2025FixedPointCertificates}.
  We use them as certificates in witnesses for concurrent stochastic games.
  These witnesses are verified in polynomial time and replace the NP oracle of previous approaches.
\end{remark}

\subparagraph{Witness.}
Consider a game $G$, a reward function $r$ whose values are rational numbers defined using at most $\Binp$ bits, a succinct discount factor $\lambda = 1 - 2^{-2^{\Blam}}$, and an additive error $\eps=2^{-\Beps}$.
A \emph{witness} is a tuple $(\strategyone, \strategytwo, I_{\strategyone}, I_{\strategytwo})$ such that
\begin{compactenum}
    \item \emph{(Witness strategies)} $\strategyone$ and $\strategytwo$ are power-series representable stationary strategies for Player~1 and Player~2, respectively.
    \item \emph{($\frac{\eps}{2}$-close shrinking intervals)} $I_\strategyone$ (resp. $I_\strategytwo$) is a power-series representable $\frac{\eps}{2}$-close shrinking interval bound of $G_\strategyone$ (resp. $G_\strategytwo$).
    \item \emph{(Interval intersection)} $I_\strategyone \cap I_\strategytwo \neq \varnothing$, i.e., $[\low_\strategyone(\state), \up_\strategyone(\state)] \cap [\low_\strategytwo(\state), \up_\strategytwo(\state)] \neq \varnothing$ for all states $\state \in \States$.
\end{compactenum}
The following result shows the \emph{soundness} of witnesses, i.e., every witness yields an $\eps$-approximation of the value and $\eps$-optimal strategies.

\begin{lemma}[Soundness of witnesses]
\label{Result: Soundness of witnesses}
  Consider a game $G$ with a reward function $r$, a succinct discount factor $\lambda$, and additive error $\eps = 2^{-\Beps}$, and let $(\strategyone, \strategytwo, I_{\strategyone}, I_{\strategytwo})$ be a witness.
  Then, $(\low_\strategyone, \up_\strategytwo)$ is an $\eps$-approximation of $\val_\lambda$,
  i.e., $\val_\lambda(\state) \in [\low_\strategyone(\state), \up_\strategytwo(\state)]$ and $\up_\strategytwo(\state) - \low_\strategyone(\state) \le \eps$ for all states $\state \in \States$.
  Moreover, $\strategyone$ and $\strategytwo$ are $\eps$-optimal.
\end{lemma}

\begin{proof}
  The intervals are shrinking. Hence, by \Cref{Result: Properties of Bellman operator}, we have $\low_\strategyone \preceq \val_\lambda^{G_\strategyone} \preceq \up_\strategyone$ and $\low_\strategytwo \preceq \val_\lambda^{G_\strategytwo} \preceq \up_\strategytwo$.
  In particular, $\low_{\strategyone}$ is a lower bound on the value of $G$, since $\strategyone$ guarantees at least $\low_{\strategyone}$.
  Similarly, $\up_{\strategytwo}$ is an upper bound on the value of $G$, since $\strategytwo$ guarantees at most $\up_{\strategytwo}$.
  Since $I_{\strategyone}$ and $I_{\strategytwo}$ intersect and both intervals have size at most $\eps / 2$, we have $\up_\strategytwo - \low_\strategyone \preceq \eps \cdot \vec{1}$.
  Therefore, $(\low_{\strategyone}, \up_{\strategytwo})$ is an $\eps$-approximation of the value.
  Lastly, for all states $\state \in \States$, we have $\val_\lambda^{G_\strategyone}(\state) \ge \low_\strategyone(\state) \ge \up_\strategytwo(\state) - \eps \ge \val_\lambda(\state) - \eps$, i.e., $\strategyone$ is $\eps$-optimal, and symmetrically, $\strategytwo$ is $\eps$-optimal.
\end{proof}

The \emph{size} of a witness is the number of bits of its power-series representation.
The following result shows the \emph{completeness} of witnesses, i.e., every game has a witness of polynomial size.

\begin{theorem}[Existence of polynomial-size witness]
\label{Result: Completeness of witnesses}
Consider a game $G$ with a reward function $r$, a succinct discount factor $\lambda$,
and additive error $\eps = 2^{-\Beps}$.
Then, there exists a witness $(\strategyone, \strategytwo, I_{\strategyone}, I_{\strategytwo})$ whose size is polynomial in $n$, $m$, $\Binp$, $\Beps$, and $\Blam$.
\end{theorem}

\Cref{Result: Completeness of witnesses} is our main technical result.
We dedicate \Cref{Section: Existence of Witness} to the proof.

\subparagraph{Verifier.}
Given a game $G$, a reward function, a discount factor, additive error $\eps$, and a tuple $(\strategyone, \strategytwo, I_\strategyone, I_\strategytwo)$,
the \emph{verifier} checks that the tuple satisfies all the conditions to be a witness.
\Cref{algorithm: polynomial verifier} describes such a verifier.
\Cref{Result: Polynomial-time verifier} states that \Cref{algorithm: polynomial verifier} accepts a tuple if and only if it is a witness, and that it runs in polynomial time.

\begin{algorithm}[ht]
  \caption{A polynomial verifier for \CSGAV}
  \label{algorithm: polynomial verifier}
  \begin{algorithmic}[1]
    \Require Game $G$, reward function $r$, discount factor $\lambda$, additive error $\eps$, tuple $(\strategyone, \strategytwo, I_{\strategyone}, I_{\strategytwo})$
    \Ensure Returns true if and only if the tuple is a valid witness

    \Procedure{VERIFIER}{$G, r, \lambda, \eps, (\strategyone, \strategytwo, I_{\strategyone}, I_{\strategytwo})$}
      \State Check $\strategyone$ and $\strategytwo$ are stationary strategies (non-negative probabilities that sum to one) and all entries are power-series representable as required
      \State Check $I_\strategyone$ and $I_\strategytwo$ are interval bounds, i.e., $\low_\strategyone \preceq \up_\strategyone$ and $\low_\strategytwo \preceq \up_\strategytwo$
      \State $I'_\strategyone \gets \left (\Psi_{G_\strategyone}(\low_\strategyone), \Psi_{G_\strategyone}(\up_\strategyone) \right )$ \label{line: start computing updates}
      \State $I'_\strategytwo \gets \left (\Psi_{G_\strategytwo}(\low_\strategytwo), \Psi_{G_\strategytwo}(\up_\strategytwo) \right )$ \label{line: end computing updates}
      \If{$I'_\strategyone \not\subseteq I_\strategyone$ or $I'_\strategytwo \not\subseteq I_\strategytwo$ }  \Comment{shrinking property}
        \State \Return \emph{false} \label{line: false if not shrinking}
      \EndIf
      \If{$\up_\strategyone - \low_\strategyone \not\preceq \frac{\eps}{2} \cdot \vec{1}$ or $\up_\strategytwo - \low_\strategytwo \not\preceq \frac{\eps}{2} \cdot \vec{1} $ } \Comment{$\frac{\eps}{2}$-closeness property}
      \State \Return \emph{false}
      \EndIf
      \If{ $I_\strategyone \cap I_\strategytwo \neq \varnothing $} \Comment{interval intersection}
      \State \Return \emph{true}
      \Else
      \State \Return \emph{false}
      \EndIf
    \EndProcedure
  \end{algorithmic}
\end{algorithm}

\begin{theorem}[Polynomial-time verifier]
\label{Result: Polynomial-time verifier}
    Consider a game $G$ with a reward function $r$, a succinct discount factor $\lambda$, additive error $\eps = 2^{-\Beps}$, and a tuple $(\strategyone, \strategytwo, I_{\strategyone}, I_{\strategytwo})$.
    Then, the following assertions hold.
    \begin{compactenum}
      \item \emph{(Correctness)} \Cref{algorithm: polynomial verifier} returns true if and only if the tuple is a witness.
      \item \emph{(Running time)} \Cref{algorithm: polynomial verifier} runs in time polynomial in the size of the input.
    \end{compactenum}
\end{theorem}

\begin{proof}
  We present the proof of the two items.
  \begin{compactitem}
    \item \emph{(Correctness)}
      \Cref{algorithm: polynomial verifier} checks the conditions of the definition of a witness.
      All its computations are exact, since they use exact arithmetic with power-series numbers and rational numbers (\Cref{lemm:power_arithmetic,Result: Rational coefficients}).
      Hence, it returns true if and only if the tuple is a witness.

    \item \emph{(Running time)}
      We show that the verifier performs polynomially many checks, and that every check decides the sign of a linear combination of polynomially many power-series numbers of polynomial size with rational coefficients from the input.
      Then, every check takes polynomial time by \Cref{Result: Rational coefficients}.
      \begin{compactitem}
        \item \emph{(Power-series numbers)}
          The numbers of the tuple are power-series numbers whose size is at most the size of the tuple.
          Moreover, $\lambda$ and $1 - \lambda$ are power-series numbers with at most two terms (\Cref{Section: Discounted-sum}), and $\frac{\eps}{2} = 2^{-(\Beps + 1)}$ is a power-series number with one term.
          By \Cref{lemm:power_arithmetic}, products of a constant number of these numbers are power-series numbers of polynomial size.

        \item \emph{(Basic checks)}
          Checking that the numbers of the tuple are power-series numbers is syntactic.
          The remaining conditions on the strategies, the interval bounds, the $\frac{\eps}{2}$-closeness, and the intersection are $\mathcal{O}(nm)$ comparisons between sums of numbers of the tuple and $\frac{\eps}{2}$.

        \item \emph{(Shrinking property)}
          Consider the player-2 MDP $G_\strategyone$ and a state $\state$.
          By the definitions of $\Psi_{G_\strategyone}$ and $G_\strategyone$, we have $\Psi_{G_\strategyone}(\low_\strategyone)(\state) \ge \low_\strategyone(\state)$ if and only if, for all actions $\baction \in \Bactions$,
          \begin{equation}
            \label{equation: shrinking check}
            \low_\strategyone(\state) \le \sum_{\action \in \Actions} \strategyone(\state)(\action) \Big( (1 - \lambda) \, r(\state, \action, \baction) + \lambda \sum_{\state' \in \States} \prob(\state, \action, \baction)(\state') \, \low_\strategyone(\state') \Big) \,.
          \end{equation}
          Similarly, $\Psi_{G_\strategyone}(\up_\strategyone)(\state) \le \up_\strategyone(\state)$ if and only if the analogous inequality, with $\up_\strategyone$ in place of $\low_\strategyone$ and $\ge$ in place of $\le$, holds for some action $\baction \in \Bactions$.
          For the player-1 MDP $G_\strategytwo$, the roles of ``for all'' and ``for some'' are exchanged.
          Hence, the shrinking property consists of at most $4nm$ inequalities of the form of \Cref{equation: shrinking check}.

        \item \emph{(Linear combinations)}
          After expanding the products, \Cref{equation: shrinking check} is equivalent to
          \begin{align*}
            \sum_{\action \in \Actions} r(\state, \action, \baction) \Big( (1 - \lambda) \, \strategyone(\state)(\action) \Big)
            + \sum_{\action \in \Actions} \sum_{\state' \in \States} \prob(\state, \action, \baction)(\state') \Big( \lambda \, \strategyone(\state)(\action) \, \low_\strategyone(\state') \Big)
            - \low_\strategyone(\state) \ge 0 \,.
          \end{align*}
          The left-hand side is a linear combination of $(n + 1) m + 1$ products of at most three power-series numbers, with the rewards and transition probabilities of the input as coefficients.
          Hence, by the first item, it is a linear combination of polynomially many power-series numbers of polynomial size with rational coefficients from the input, which completes the proof. \qedhere
      \end{compactitem}
  \end{compactitem}
\end{proof}

The soundness and completeness of witnesses (\Cref{Result: Soundness of witnesses} and \Cref{Result: Completeness of witnesses}) and the polynomial-time verifier (\Cref{Result: Polynomial-time verifier}) yield \Cref{Result: Discounted in TFNP}.
Formally, we show that \CSGAV{} reduces in polynomial time to the search problem of finding a witness, which is in TFNP~\cite{megiddo1991TotalFunctions}.
To do so, we compute an $\eps$-approximation $\hat{v}$ of the value from the witness in polynomial time.
\begin{proof}[Proof of \Cref{Result: Discounted in TFNP}]
    Consider a game $G$, a reward function, a succinct discount factor, and an additive error $\eps = 2^{-\Beps}$.
    By \Cref{Result: Completeness of witnesses}, there exists a polynomial $p$ such that a witness for the additive error $\frac{\eps}{2}$ of size at most $p(n, m, \Binp, \Beps, \Blam)$ exists.
    By \Cref{Result: Polynomial-time verifier}, such witnesses are checked in polynomial time, and hence finding one is in TFNP.
    Moreover, by \Cref{Result: Soundness of witnesses}, every such witness satisfies, for all states $\state \in \States$,
    \[
      \low_\strategyone(\state) \le \val_\lambda(\state) \le \up_\strategytwo(\state) \le \low_\strategyone(\state) + \frac{\eps}{2} \,.
    \]

    \subparagraph{Value approximation.}
    For every state $\state$, let $\hat{v}(\state)$ be the largest integer multiple of $2^{-(\Beps+2)}$ in $[-1, 1]$ that is at most $\low_\strategyone(\state)$, which is computed by binary search (\Cref{lemm:power_arithmetic}).
    Then, $|\hat{v}(\state) - \val_\lambda(\state)| \le \frac{\eps}{4} + \frac{\eps}{2} \le \eps$.
    Hence, \CSGAV{} reduces in polynomial time to finding such a witness.

    \subparagraph{Gap problem.}
    Consider a state $\state$ and a threshold $\alpha$.
    If $\val_\lambda(\state) \ge \alpha + \eps$, then $\low_\strategyone(\state) \ge \val_\lambda(\state) - \frac{\eps}{2} > \alpha$, and if $\val_\lambda(\state) \le \alpha - \eps$, then $\up_\strategytwo(\state) \le \val_\lambda(\state) + \frac{\eps}{2} < \alpha$.
    Hence, guessing such a witness and accepting if and only if $\low_\strategyone(\state) \ge \alpha$ (resp. $\up_\strategytwo(\state) \le \alpha$) shows membership in NP (resp. coNP),
    where the comparisons with $\alpha$ are decided in polynomial time by \Cref{Result: Rational coefficients}.
\end{proof}

\section{Existence of Power-Series Representable Witness}
\label{Section: Existence of Witness}

In this section, we prove the completeness of witnesses (\Cref{Result: Completeness of witnesses}).
First, we show that both players have power-series representable $\eps$-optimal stationary strategies (\Cref{Section: Strategies}).
Second, we discuss natural candidates for power-series representable $\eps$-close shrinking interval bounds that fail (\Cref{Section: Failed candidates}).
Third, we show that every MDP has power-series representable $\eps$-close shrinking interval bounds (\Cref{Section: Interval Bounds in MDPs}).
Lastly, we combine these results to prove \Cref{Result: Completeness of witnesses} (\Cref{Section: Proof of Completeness}).

\subsection{Power-Series Representable \texorpdfstring{$\eps$}{Epsilon}-Optimal Strategies}
\label{Section: Strategies}

In this subsection, we show that both players have power-series representable $\eps$-optimal stationary strategies (\Cref{Result: Existence of Floating-Point Strategies}).
We obtain them from optimal stationary strategies in two steps.
\begin{compactitem}
  \item \emph{(Removing small probabilities)}
    We remove the probabilities below a threshold, which bounds the exponents in the power-series representation.
    We bound the resulting error by an absolute continuity property of MDPs (\Cref{Result: Continuity,Result: Bounded patience}).
  \item \emph{(Truncation)}
    We truncate the remaining probabilities with a fixed relative precision, which bounds the number of terms.
    We bound the resulting error, independently of the discount factor, by interpreting discounted-sum objectives as reachability objectives and applying a relative continuity property of Markov chains (\Cref{Result: Stopping MDP,Result: Relative continuity,Result: Existence of Floating-Point Strategies}).
\end{compactitem}

\subparagraph{Markov chains of MDPs.}
For $\mathfrak{p} \in \{1, 2\}$, consider a player-$\mathfrak{p}$ MDP $P$ with a reward function $r$ and a discount factor $\lambda$.
A pure stationary strategy $\pi$ for Player~$\mathfrak{p}$ assigns an action $\pi(\state) \in \Actions$ to every state $\state$.
By fixing $\pi$, we obtain the MC $P_\pi$ with transition function $\prob(\state, \pi(\state))$, reward function $r(\state, \pi(\state))$, and the same discount factor.
Since Player~$\mathfrak{p}$ has an optimal pure stationary strategy in $P$~\cite{shapley1953stochastic, filar2012competitive}, for all states $\state \in \States$, we have
\begin{equation}
  \label{equation: MDP value by pure strategies}
  \val_\lambda^P(\state) = \op_{\pi} \val_\lambda^{P_\pi}(\state) \,,
\end{equation}
where $\op$ is $\max$ (resp. $\min$) for player-1 (resp. player-2) MDPs, and $\pi$ ranges over the pure stationary strategies of Player~$\mathfrak{p}$.

\subparagraph{Absolute continuity.}
The following continuity property of MDPs generalizes~\cite[Eq.~(4.19)]{filar2012competitive} and~\cite[Lemma~36]{asadi2024StatefulDiscounted}.

\begin{lemma}[Continuity]
\label{Result: Continuity}
  For $\mathfrak{p} \in \{1, 2\}$, let $P$ and $\tilde{P}$ be player-$\mathfrak{p}$ MDPs with identical state and action sets and discount factor $\lambda$,
  with transition functions $\prob$ and $\tilde{\prob}$, and reward functions $r$ and $\tilde{r}$, respectively.
  Let
  \[
    \eta \defas \max_{\state \in \States, \action \in \Actions} \max \left\{ |r(\state, \action) - \tilde{r}(\state, \action)| \,,\, \sum_{\state' \in \States} |\prob(\state, \action)(\state') - \tilde{\prob}(\state, \action)(\state')| \right\} \,.
  \]
  Then, $|\val_\lambda^P(\state) - \val_\lambda^{\tilde{P}}(\state)| \le \eta / (1 - \lambda)$ for all states $\state \in \States$.
\end{lemma}

\begin{proof}
  Note that $\val_\lambda^P$ and $\val_\lambda^{\tilde{P}}$ take values in $[0, 1]$.
  For every state $\state \in \States$, we have
  \begin{align*}
    |\val_\lambda^P(\state) - \val_\lambda^{\tilde{P}}(\state)|
      &= \big| \Psi_P(\val_\lambda^P)(\state) - \Psi_{\tilde{P}}(\val_\lambda^{\tilde{P}})(\state) \big| \\
      &\le \max_{\action \in \Actions} \Big| (1 - \lambda) \big( r(\state, \action) - \tilde{r}(\state, \action) \big) \\
      &\qquad + \lambda \sum_{\state' \in \States} \big( \prob(\state, \action)(\state') \, \val_\lambda^P(\state') - \tilde{\prob}(\state, \action)(\state') \, \val_\lambda^{\tilde{P}}(\state') \big) \Big| \\
      &\le \max_{\action \in \Actions} \Big( (1 - \lambda) \big| r(\state, \action) - \tilde{r}(\state, \action) \big| \\
      &\qquad + \lambda \sum_{\state' \in \States} \big| \prob(\state, \action)(\state') - \tilde{\prob}(\state, \action)(\state') \big| \, \val_\lambda^P(\state') \\
      &\qquad + \lambda \sum_{\state' \in \States} \tilde{\prob}(\state, \action)(\state') \, \big| \val_\lambda^P(\state') - \val_\lambda^{\tilde{P}}(\state') \big| \Big) \\
      &\le (1 - \lambda) \eta + \lambda \eta + \lambda \| \val_\lambda^P - \val_\lambda^{\tilde{P}} \|_\infty \\
      &= \eta + \lambda \| \val_\lambda^P - \val_\lambda^{\tilde{P}} \|_\infty \,,
  \end{align*}
  where the first equality follows from $\val_\lambda^P = \Psi_P(\val_\lambda^P)$ and $\val_\lambda^{\tilde{P}} = \Psi_{\tilde{P}}(\val_\lambda^{\tilde{P}})$, the first inequality follows from the fact that $\op$ is $1$-Lipschitz, the second inequality follows from the triangle inequality, and the third inequality follows from the definition of $\eta$, $\val_\lambda^P \preceq \vec{1}$, and $\tilde{\prob}(\state, \action) \in \Delta(\States)$.
  Taking the maximum over all states $\state$ yields $\| \val_\lambda^P - \val_\lambda^{\tilde{P}} \|_\infty \le \eta + \lambda \| \val_\lambda^P - \val_\lambda^{\tilde{P}} \|_\infty$, i.e., $\| \val_\lambda^P - \val_\lambda^{\tilde{P}} \|_\infty \le \eta / (1 - \lambda)$.
\end{proof}

\begin{lemma}[Bounded patience]
\label{Result: Bounded patience}
  Consider a game $G$ with a reward function $r$ and a discount factor $\lambda$.
  Let $\eps \in (0, 1]$ and $p_{\min} \defas (1 - \lambda) \eps / (2m)$.
  Then, there exist $\eps$-optimal stationary strategies $\strategyone$ and $\strategytwo$ such that all their non-zero probabilities are at least $p_{\min}$.
  In particular, their patience is at most $1 / p_{\min}$.
\end{lemma}

\begin{proof}
  We present the proof for Player~1, and the proof for Player~2 is symmetric.
  Let $\strategyone^*$ be an optimal stationary strategy~\cite{shapley1953stochastic}.
  For every state $\state$, let $\action^*_\state \in \Actions$ be such that $\strategyone^*(\state)(\action^*_\state)$ is maximal.
  Therefore, we have $\strategyone^*(\state)(\action^*_\state) \ge 1/m$.
  For all states $\state$ and actions $\action$, we define
  \[
    \strategyone(\state)(\action) \defas
      \begin{cases}
        1 - \sum_{\action' \neq \action^*_\state} \strategyone(\state)(\action') & \action = \action^*_\state \,, \\
        \strategyone^*(\state)(\action) & \action \neq \action^*_\state \text{ and } \strategyone^*(\state)(\action) \ge p_{\min} \,, \\
        0 & \text{otherwise.}
      \end{cases}
  \]
  First, we show that $\strategyone$ is a stationary strategy whose non-zero probabilities are at least $p_{\min}$.
  For all states $\state$ and actions $\action \neq \action^*_\state$, by definition, we have $\strategyone(\state)(\action) \le \strategyone^*(\state)(\action)$.
  Therefore, for every state $\state$, we have
  \[
    \strategyone(\state)(\action^*_\state)
      = 1 - \sum_{\action \neq \action^*_\state} \strategyone(\state)(\action)
      \ge 1 - \sum_{\action \neq \action^*_\state} \strategyone^*(\state)(\action)
      = \strategyone^*(\state)(\action^*_\state)
      \ge \frac{1}{m}
      \ge p_{\min} \,.
  \]
  Moreover, by definition, we have $\sum_{\action \in \Actions} \strategyone(\state)(\action) = 1$, and the other non-zero probabilities of $\strategyone(\state)$ are at least $p_{\min}$.
  Second, we bound the $L_1$ distance between $\strategyone(\state)$ and $\strategyone^*(\state)$.
  For every state $\state$, we have
  \[
    \sum_{\action \in \Actions} |\strategyone(\state)(\action) - \strategyone^*(\state)(\action)|
      = 2 \sum_{\action \neq \action^*_\state} \big( \strategyone^*(\state)(\action) - \strategyone(\state)(\action) \big)
      \le 2 m \, p_{\min}
      = (1 - \lambda) \eps \,,
  \]
  where the equality follows from $\strategyone(\state)(\action) \le \strategyone^*(\state)(\action)$ for all actions $\action \neq \action^*_\state$ and $\sum_{\action \in \Actions} \strategyone(\state)(\action) = \sum_{\action \in \Actions} \strategyone^*(\state)(\action) = 1$,
  and the inequality follows from the fact that every non-zero term of the sum is $\strategyone^*(\state)(\action) < p_{\min}$.
  Lastly, we show that $\strategyone$ is $\eps$-optimal.
  For all states $\state$ and actions $\baction \in \Bactions$, we have
  \begin{align*}
    |r_\strategyone(\state, \baction) - r_{\strategyone^*}(\state, \baction)|
      &\le \sum_{\action \in \Actions} |\strategyone(\state)(\action) - \strategyone^*(\state)(\action)| \, r(\state, \action, \baction)
      \le (1 - \lambda) \eps \,, \\
    \sum_{\state' \in \States} |\prob_\strategyone(\state, \baction)(\state') - \prob_{\strategyone^*}(\state, \baction)(\state')|
      &\le \sum_{\action \in \Actions} |\strategyone(\state)(\action) - \strategyone^*(\state)(\action)| \sum_{\state' \in \States} \prob(\state, \action, \baction)(\state')
      \le (1 - \lambda) \eps \,,
  \end{align*}
  where the first inequalities follow from the definition of $G_\strategyone$ and the triangle inequality,
  and the second inequalities follow from $r(\state, \action, \baction) \le 1$, $\prob(\state, \action, \baction) \in \Delta(\States)$, and the previous step.
  Hence, $G_\strategyone$ and $G_{\strategyone^*}$ satisfy the condition of \Cref{Result: Continuity} with $\eta \le (1 - \lambda) \eps$.
  Therefore, for all states $\state$, we have $\val_\lambda^{G_\strategyone}(\state) \ge \val_\lambda^{G_{\strategyone^*}}(\state) - \eps = \val_\lambda(\state) - \eps$, i.e., $\strategyone$ is $\eps$-optimal, which completes the proof.
\end{proof}

\subparagraph{Relative continuity.}
To bound the error of truncation with a fixed relative precision, we interpret discounted-sum objectives as reachability objectives in Markov chains (\Cref{Result: Stopping MDP}).
An MC is \emph{absorbing} if, from every state, it reaches an absorbing state with probability one.

\begin{lemma}[Discounted-sum as reachability]
\label{Result: Stopping MDP}
  Consider an MC $C = (\States, \prob)$ with a reward function $r \colon \States \to [0, 1]$ and a discount factor $\lambda$.
  Let $C^{\mathrm{abs}}$ be the MC with states $\States \cup \{\top, \bot\}$, where $\top$ and $\bot$ are absorbing, and, for all states $\state, \state' \in \States$,
  \[
    \prob^{\mathrm{abs}}(\state)(\state') \defas \lambda \prob(\state)(\state') \,, \qquad
    \prob^{\mathrm{abs}}(\state)(\top) \defas (1 - \lambda) r(\state) \,, \qquad
    \prob^{\mathrm{abs}}(\state)(\bot) \defas (1 - \lambda) (1 - r(\state)) \,.
  \]
  Then, $C^{\mathrm{abs}}$ is absorbing, and, for all states $\state \in \States$, we have $\val_\lambda^C(\state) = \val_\top^{C^{\mathrm{abs}}}(\state)$.
\end{lemma}

\begin{proof}
  First, from every state $\state \in \States$, the MC $C^{\mathrm{abs}}$ moves to $\{\top, \bot\}$ with probability $1 - \lambda > 0$.
  Hence, it reaches an absorbing state with probability one, i.e., $C^{\mathrm{abs}}$ is absorbing.
  Second, by conditioning on the first step, we have $\val_\top^{C^{\mathrm{abs}}}(\state) = (1 - \lambda) r(\state) + \lambda \sum_{\state' \in \States} \prob(\state)(\state') \val_\top^{C^{\mathrm{abs}}}(\state')$ for all states $\state \in \States$.
  Hence, the restriction of $\val_\top^{C^{\mathrm{abs}}}$ to $\States$ is a fixed point of the Bellman operator of $C$, whose unique fixed point is $\val_\lambda^C$.
\end{proof}

\begin{lemma}[{\cite[Theorem~6]{solan2003ContinuityValueCompetitive}, see also~\cite[Lemma~27]{asadi2024StatefulDiscounted}}]
\label{Result: Relative continuity}
  Consider absorbing MCs $C$ and $\tilde{C}$ with an identical finite set of states $\States'$, transition functions $\prob$ and $\tilde{\prob}$, and an absorbing state $\top$.
  Let $\eta \defas \max_{\state, \state' \in \States'} \rel(\prob(\state)(\state'), \tilde{\prob}(\state)(\state'))$.
  Then, for all states $\state \in \States'$, we have $|\val_\top^C(\state) - \val_\top^{\tilde{C}}(\state)| \le 4 |\States'| \eta$.
\end{lemma}

\begin{lemma}[Polynomial-size $\eps$-optimal strategies]
\label{Result: Existence of Floating-Point Strategies}
	Consider a game $G$ with a reward function $r$ and a succinct discount factor $\lambda = 1 - 2^{-2^{\Blam}}$.
	Let $\eps = 2^{-\Beps}$, and let
	\[
	  \ell_{\mathrm{str}} \defas \Beps + \lceil \log (n + 2) \rceil + \lceil \log m \rceil + 5
	  \qquad \text{and} \qquad
	  B_{\mathrm{str}} \defas \Blam + \Beps + \lceil \log m \rceil + \ell_{\mathrm{str}} + 3 \,.
	\]
	Then, there exist $\eps$-optimal stationary strategies $\strategyone$ and $\strategytwo$ such that $\strategyone(\state), \strategytwo(\state) \in \D(m \ell_{\mathrm{str}}, B_{\mathrm{str}})$ for all states $\state \in \States$.
\end{lemma}

\begin{proof}
    We present the proof for Player~1, and the proof for Player~2 is symmetric.
    \begin{compactitem}
      \item \emph{(Bounded patience)}
        Let $\strategyone$ be an $\frac{\eps}{2}$-optimal stationary strategy given by \Cref{Result: Bounded patience}.
        Then, we have
        \begin{align*}
          \pati(\strategyone)
            &\le \frac{4m}{(1 - \lambda) \eps}
              && (\text{\Cref{Result: Bounded patience} for } \tfrac{\eps}{2}) \\
            &= m \cdot 2^{2^{\Blam} + \Beps + 2}
              && (1 - \lambda = 2^{-2^{\Blam}},\ \eps = 2^{-\Beps}) \\
            &\le 2^{2^{\Blam} + \Beps + \lceil \log m \rceil + 2}
              && (m \le 2^{\lceil \log m \rceil}) \\
            &= 2^{2^{\Blam} + (B_{\mathrm{str}} - \Blam) - \ell_{\mathrm{str}} - 1}
              && (\text{definition of } B_{\mathrm{str}}) \\
            &\le 2^{2^{B_{\mathrm{str}}} - \ell_{\mathrm{str}} - 1}
              && (2^{a} + b \le 2^{a + b} \text{ for } a, b \in \NN_0) \\
            &< 2^{2^{B_{\mathrm{str}}} - \ell_{\mathrm{str}}} \,.
        \end{align*}
      \item \emph{(Truncation)}
        By \Cref{Result: Distribution representation in power series} applied to every distribution $\strategyone(\state)$, there exists a stationary strategy $\strategyone'$ such that, for all states $\state \in \States$, we have $\strategyone'(\state) \in \D(m \ell_{\mathrm{str}}, B_{\mathrm{str}})$ and
        \begin{align*}
          \rel(\strategyone(\state), \strategyone'(\state))
            &\le m \cdot 2^{1 - \ell_{\mathrm{str}}}
              && (\text{\Cref{Result: Distribution representation in power series}}) \\
            &\le 2^{\lceil \log m \rceil + 1 - \ell_{\mathrm{str}}}
              && (m \le 2^{\lceil \log m \rceil}) \\
            &= 2^{-(\Beps + \lceil \log (n + 2) \rceil + 4)}
              && (\text{definition of } \ell_{\mathrm{str}}) \\
            &= \frac{\eps}{16 \cdot 2^{\lceil \log (n + 2) \rceil}}
              && (\eps = 2^{-\Beps}) \\
            &\le \frac{\eps}{16 (n + 2)}
              && (n + 2 \le 2^{\lceil \log (n + 2) \rceil}) \,.
        \end{align*}
      \item \emph{(Error of the truncation)}
        Consider a pure stationary strategy $\pi$ for Player~2, and the absorbing MCs
        \[
          C \defas ((G_\strategyone)_\pi)^{\mathrm{abs}}
          \quad \text{and} \quad
          \tilde{C} \defas ((G_{\strategyone'})_\pi)^{\mathrm{abs}} \,,
        \]
        which have the $n + 2$ states $\States \cup \{\top, \bot\}$, and denote their transition functions by $\prob_C$ and $\prob_{\tilde{C}}$, respectively.
        Then, for all states $\state \in \States$ and $\state' \in \States \cup \{\top, \bot\}$, we have
        \[
          \prob_C(\state)(\state') =
            \begin{cases}
              \lambda \sum_{\action \in \Actions} \strategyone(\state)(\action) \, \prob(\state, \action, \pi(\state))(\state') & \state' \in \States \,, \\
              (1 - \lambda) \sum_{\action \in \Actions} \strategyone(\state)(\action) \, r(\state, \action, \pi(\state)) & \state' = \top \,, \\
              (1 - \lambda) \sum_{\action \in \Actions} \strategyone(\state)(\action) \big( 1 - r(\state, \action, \pi(\state)) \big) & \state' = \bot \,,
            \end{cases}
        \]
        and $\prob_{\tilde{C}}(\state)(\state')$ is obtained by replacing $\strategyone$ with $\strategyone'$.
        Therefore, for all states $\state, \state' \in \States$, we have
        \begin{align*}
          \prob_{\tilde{C}}(\state)(\state')
            &= \lambda \sum_{\action \in \Actions} \strategyone'(\state)(\action) \, \prob(\state, \action, \pi(\state))(\state')
              && (\text{definition of } \prob_{\tilde{C}}) \\
            &\le \lambda \sum_{\action \in \Actions} \Big( 1 + \frac{\eps}{16 (n + 2)} \Big) \strategyone(\state)(\action) \, \prob(\state, \action, \pi(\state))(\state')
              && (\rel(\strategyone(\state), \strategyone'(\state)) \le \tfrac{\eps}{16 (n + 2)}) \\
            &= \Big( 1 + \frac{\eps}{16 (n + 2)} \Big) \prob_C(\state)(\state')
              && (\text{definition of } \prob_C) \,.
        \end{align*}
        The cases $\state' = \top$ and $\state' = \bot$ are analogous, since $r(\state, \action, \pi(\state))$ and $1 - r(\state, \action, \pi(\state))$ are non-negative.
        Symmetrically, we have $\prob_C(\state)(\state') \le (1 + \frac{\eps}{16 (n + 2)}) \prob_{\tilde{C}}(\state)(\state')$.
        Hence, by the definition of $\rel$, for all states $\state \in \States$, we have
        \begin{equation}
          \label{equation: relative distance of absorbing MCs}
          \rel(\prob_C(\state), \prob_{\tilde{C}}(\state)) \le \frac{\eps}{16 (n + 2)} \,.
        \end{equation}
        Moreover, the transitions from $\top$ and $\bot$ are identical in $C$ and $\tilde{C}$.
        Therefore, for all states $\state \in \States$, we have
        \begin{align}
          \big| \val_\lambda^{(G_\strategyone)_\pi}(\state) - \val_\lambda^{(G_{\strategyone'})_\pi}(\state) \big|
            &= \big| \val_\top^{C}(\state) - \val_\top^{\tilde{C}}(\state) \big|
              && (\text{\Cref{Result: Stopping MDP}}) \nonumber \\
            &\le 4 (n + 2) \cdot \frac{\eps}{16 (n + 2)}
              && (\text{\Cref{Result: Relative continuity}, \Cref{equation: relative distance of absorbing MCs}}) \nonumber \\
            &= \frac{\eps}{4} \,. \label{equation: truncation error for pure strategies}
        \end{align}
        Lastly, for all states $\state \in \States$, we have
        \begin{align*}
          \big| \val_\lambda^{G_\strategyone}(\state) - \val_\lambda^{G_{\strategyone'}}(\state) \big|
            &= \big| \min_\pi \val_\lambda^{(G_\strategyone)_\pi}(\state) - \min_\pi \val_\lambda^{(G_{\strategyone'})_\pi}(\state) \big|
              && (\text{\Cref{equation: MDP value by pure strategies}}) \\
            &\le \max_\pi \big| \val_\lambda^{(G_\strategyone)_\pi}(\state) - \val_\lambda^{(G_{\strategyone'})_\pi}(\state) \big|
              && (\min \text{ is } 1\text{-Lipschitz}) \\
            &\le \frac{\eps}{4}
              && (\text{\Cref{equation: truncation error for pure strategies}}) \,,
        \end{align*}
        where $\pi$ ranges over the pure stationary strategies of Player~2.
    \end{compactitem}
    Therefore, since $\strategyone$ is $\frac{\eps}{2}$-optimal, $\strategyone'$ is $\eps$-optimal, which completes the proof.
\end{proof}

It remains to show that the MDPs $G_\strategyone$ and $G_\strategytwo$ have power-series representable $\eps$-close shrinking interval bounds.
We first discuss natural candidates for such bounds that fail in \Cref{Section: Failed candidates}, and then we prove the existence of such bounds for every MDP in \Cref{Section: Interval Bounds in MDPs}.

\subsection{Failed Natural Candidates for Interval Bounds}
\label{Section: Failed candidates}

In this subsection, we discuss three natural candidates for power-series representable $\eps$-close shrinking interval bounds, namely, (1)~independent truncation, (2)~shifts by order, and (3)~a greedy construction, and we show by examples why they fail.

\subparagraph{Setting.}
Throughout this subsection, we consider a player-2 MDP $P$ with a reward function $r$, a succinct discount factor $\lambda = 1 - 2^{-2^{\Blam}}$, and an additive error $\eps \in (0, \frac{1}{4})$.
We focus on the lower bounds of $\eps$-close shrinking interval bounds, i.e., on functions $\low \colon \States \to \RR$ such that $\low \preceq \Psi_P(\low)$ and $\val_\lambda - \eps \cdot \vec{1} \preceq \low \preceq \val_\lambda$, since shrinking interval bounds contain the value (\Cref{Result: Properties of Bellman operator}).

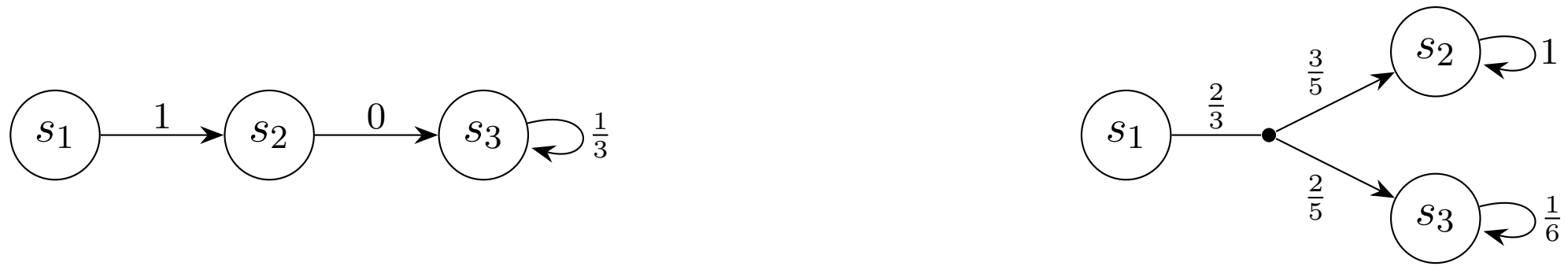
\begin{figure}[t]
  \centering
  \begin{tikzpicture}[
      st/.style={draw, circle, minimum size=0.75cm, inner sep=1pt},
      >={Stealth[length=2mm]},
      lab/.style={font=\footnotesize, inner sep=1.5pt}]
    \begin{scope}
      \node[st] (z) at (-1.8, 0) {$\state_1$};
      \node[st] (a0) at (0, 0) {$\state_2$};
      \node[st] (a1) at (1.8, 0) {$\state_3$};
      \draw[->] (z) -- node[lab, above] {$1$} (a0);
      \draw[->] (a0) -- node[lab, above] {$0$} (a1);
      \draw[->] (a1) edge[loop right] node[lab, right] {$\frac{1}{3}$} (a1);
      \node at (0.2, -1.6) {(a)};
    \end{scope}
    \begin{scope}[xshift=7.2cm]
      \node[st] (b0) at (0, 0) {$\state_1$};
      \node[fill, circle, inner sep=1.2pt] (d) at (1.2, 0) {};
      \node[st] (b1) at (2.6, 0.7) {$\state_2$};
      \node[st] (b2) at (2.6, -0.7) {$\state_3$};
      \draw (b0) -- node[lab, above] {$\frac{2}{3}$} (d);
      \draw[->] (d) -- node[lab, above left] {$\frac{3}{5}$} (b1);
      \draw[->] (d) -- node[lab, below left] {$\frac{2}{5}$} (b2);
      \draw[->] (b1) edge[loop right] node[lab, right] {$1$} (b1);
      \draw[->] (b2) edge[loop right] node[lab, right] {$\frac{1}{6}$} (b2);
      \node at (1.3, -1.6) {(b)};
    \end{scope}
  \end{tikzpicture}
  \caption{The MDPs of \Cref{Example: Deterministic transitions,Example: Probabilistic transitions}.
    Every action is labeled by its reward.
    In (b), the action of $\state_1$ moves from the black dot to $\state_2$ and $\state_3$ with the given probabilities.}
  \label{Figure: Failed candidates}
\end{figure}

\subparagraph{Challenges.}
The existence of power-series representable lower bounds is challenging.
On the one hand, the values are in general not power-series numbers, e.g., $\frac{1}{3}$, and hence the lower bounds must deviate from the values.
On the other hand, the lower bounds must approximate the values, and hence they must be close to the values.
Moreover, distinct values may be double exponentially close to each other.
Since the inequalities of lower bounds compare the bound of every state with a $\lambda$-weighted average of the bounds of its successors, where $\lambda$ is double exponentially close to one, these challenges make many natural candidates fail.
We illustrate this with the following two examples, shown in \Cref{Figure: Failed candidates}.

\begin{example}
  \label{Example: Deterministic transitions}
  Consider the player-2 MDP in \Cref{Figure: Failed candidates}~(a) with three states $\state_1$, $\state_2$, and $\state_3$, where every state has a single action.
  The state $\state_1$ moves to $\state_2$ with the reward $1$, and the state $\state_2$ moves to $\state_3$ with the reward $0$.
  The state $\state_3$ is absorbing, and its self-loop has the reward $\frac{1}{3}$.
  The values of the states are
  \begin{align*}
    \val_\lambda(\state_1) &= (1 - \lambda) + \lambda \val_\lambda(\state_2) = (1 - \lambda) + \frac{\lambda^2}{3} \,, \\
    \val_\lambda(\state_2) &= \lambda \val_\lambda(\state_3) = \frac{\lambda}{3} \,, \\
    \val_\lambda(\state_3) &= \frac{1}{3} \,.
  \end{align*}
\end{example}

\begin{example}
  \label{Example: Probabilistic transitions}
  Consider the player-2 MDP in \Cref{Figure: Failed candidates}~(b) with three states $\state_1$, $\state_2$, and $\state_3$, where every state has a single action.
  The state $\state_1$ moves to $\state_2$ and $\state_3$ with probabilities $\frac{3}{5}$ and $\frac{2}{5}$, respectively, and with the reward $\frac{2}{3}$.
  The states $\state_2$ and $\state_3$ are absorbing, and their self-loops have the rewards $1$ and $\frac{1}{6}$, respectively.
  The values of the states are
  \begin{align*}
    \val_\lambda(\state_1) &= (1 - \lambda) \frac{2}{3} + \lambda \Big( \frac{3}{5} \val_\lambda(\state_2) + \frac{2}{5} \val_\lambda(\state_3) \Big) = \frac{2}{3} \,, \\
    \val_\lambda(\state_2) &= 1 \,, \\
    \val_\lambda(\state_3) &= \frac{1}{6} \,.
  \end{align*}
\end{example}

\subparagraph{Candidate 1: Independent truncation.}
A natural candidate is to truncate every value independently to a power-series number with few terms, and to shift all values by a common amount to obtain a lower bound.
Formally, for $\ell \in \NN$ and $c \in \RR$, the candidate lower bound is $\low \defas \trunc_\ell \circ \val_\lambda - c \cdot \vec{1}$, where $\trunc_\ell$ is the truncation of \Cref{Result: Number representation in power series}.
This candidate fails in \Cref{Example: Deterministic transitions}, because it truncates the double exponentially close values of $\state_2$ and $\state_3$ to the same number, whereas the inequalities of shrinking bounds require a positive difference between their bounds.
Indeed, every function $\low$ with $\low \preceq \Psi_P(\low)$ satisfies $\low(\state_2) \le \Psi_P(\low)(\state_2) = \lambda \low(\state_3)$, i.e.,
\[
  \low(\state_3) - \low(\state_2) \ge (1 - \lambda) \, \low(\state_3) \,.
\]
However, since $1 - \lambda = 2^{-2^{\Blam}}$, the binary expansions of $\val_\lambda(\state_2)$ and $\val_\lambda(\state_3)$ coincide in their first $2^{\Blam} - 2$ digits.
Moreover, the closeness at $\state_3$, i.e., $\low(\state_3) \ge \val_\lambda(\state_3) - \eps$, implies $c \le \eps < \frac{1}{4}$.
Hence, for all $\ell \le 2^{\Blam} - 3$, the candidate satisfies $\low(\state_2) = \low(\state_3) > 0$, and therefore it violates the inequality above.
This suggests that the bounds of different states should be chosen jointly.

\subparagraph{Candidate 2: Shifts by order.}
To choose the bounds of different states jointly, a natural candidate is to shift the values down by amounts that increase as the values decrease, which preserves the order of the values, and to truncate the shifted values within the slack $\Psi_P(\low) - \low$ that the shifts create.
Formally, for $\eta > 0$ and $c \colon \States \to \RR_{> 0}$ such that $c(\state') \ge (1 + \eta) \, c(\state)$ for all states $\state, \state'$ with $\val_\lambda(\state') < \val_\lambda(\state)$, the candidate lower bound is obtained by truncating $\val_\lambda - c$.
This candidate fails in \Cref{Example: Deterministic transitions}, because the slack at $\state_1$ is double exponentially small compared to the shifts, and hence the candidate either violates $\low \preceq \Psi_P(\low)$ or requires truncation with an absolute precision.
Indeed, consider $\low \defas \val_\lambda - c$ before truncation.
Since $\val_\lambda(\state_2) < \val_\lambda(\state_1)$, we have $c(\state_2) \ge (1 + \eta) \, c(\state_1)$, and therefore
\begin{align*}
  \Psi_P(\low)(\state_1) - \low(\state_1)
    &= (1 - \lambda) + \lambda \low(\state_2) - \low(\state_1)
      && (\text{definition of } \Psi_P) \\
    &= c(\state_1) - \lambda \, c(\state_2)
      && (\low = \val_\lambda - c,\ \val_\lambda = \Psi_P(\val_\lambda)) \\
    &\le \big( 1 - \lambda (1 + \eta) \big) \, c(\state_1)
      && (c(\state_2) \ge (1 + \eta) \, c(\state_1)) \,.
\end{align*}
Hence, if $\eta > \frac{1 - \lambda}{\lambda}$, e.g., if $\eta > 2^{1 - 2^{\Blam}}$, then $\low$ violates $\low \preceq \Psi_P(\low)$ already before truncation.
Otherwise, the slack at $\state_1$ is at most $(1 - \lambda) \, c(\state_1) = 2^{-2^{\Blam}} c(\state_1)$, and hence the truncation requires an absolute precision of $2^{-2^{\Blam}} c(\state_1)$, whereas power-series numbers with few terms only guarantee a relative precision (\Cref{Result: Number representation in power series}).
This suggests that the bounds should be built from the gaps between consecutive values, truncated with a relative precision.

\subparagraph{Candidate 3: Greedy construction.}
Following these observations, a natural candidate is to process the distinct values in decreasing order, and to obtain the bound of every value from the bound of the previous value by subtracting the gap between the two values, truncated with a fixed relative precision.
Formally, let $v_1 > v_2 > \cdots > v_J$ be the distinct values of $P$, and, for $j \ge 2$, let $g_j \defas v_{j-1} - v_j$ be the gap between consecutive values and $d_j \defas \trunc_\ell(g_j)$ its truncation, where $\trunc_\ell$ is the truncation of \Cref{Result: Number representation in power series}.
Then, for $c \in \RR$, the candidate lower bound is $\low(\state) \defas v_1 - c - \sum_{i = 2}^{j} d_i$ for all states $\state$ with $\val_\lambda(\state) = v_j$.
These bounds have polynomial size, since every gap is truncated with a relative precision and the gaps are at least double exponentially small (\Cref{Result: Value separation}).
This candidate fails in \Cref{Example: Probabilistic transitions}, because the gaps are truncated with different relative errors, and hence the lower bound is not a convex function of the values.
Indeed, the values are $v_1 = 1$, $v_2 = \frac{2}{3}$, and $v_3 = \frac{1}{6}$.
The first gap $g_2 = \frac{1}{3}$ is not a power-series number, and hence $d_2 = \frac{1}{3} - \rho$ for some $\rho \ge \frac{1}{3} 2^{-(\ell + 1)}$, whereas the second gap $g_3 = \frac{1}{2}$ is truncated exactly, i.e., $d_3 = \frac{1}{2}$.
Then, we have $\low(\state_2) = 1 - c$, $\low(\state_1) = \frac{2}{3} - c + \rho$, and $\low(\state_3) = \frac{1}{6} - c + \rho$, and therefore
\[
  \low(\state_1) - \Psi_P(\low)(\state_1)
    = \frac{2}{3} - c + \rho - (1 - \lambda) \frac{2}{3} - \lambda \Big( \frac{3}{5} (1 - c) + \frac{2}{5} \Big( \frac{1}{6} - c + \rho \Big) \Big)
    = \Big( 1 - \frac{2}{5} \lambda \Big) \rho - (1 - \lambda) \, c \,.
\]
Moreover, the closeness at $\state_2$, i.e., $1 - \eps \le \low(\state_2) \le 1$, implies $0 \le c \le \eps < 1$.
Hence, since $1 - \frac{2}{5} \lambda \ge \frac{3}{5}$, for all $\ell \le 2^{\Blam} - 5$, the candidate violates $\low \preceq \Psi_P(\low)$.
Note that, as a function of the values, the bounds have the slope $1$ between $\frac{1}{6}$ and $\frac{2}{3}$, and the smaller slope $1 - 3\rho$ between $\frac{2}{3}$ and $1$.

\subparagraph{Our construction.}
Our construction in \Cref{Section: Interval Bounds in MDPs} also builds the lower bound from the truncated gaps between consecutive values, but it multiplies the gaps by increasing factors before truncation.
Hence, the slopes increase, i.e., the lower bound is a convex function of the values.
Then, Jensen's inequality yields $\low \preceq \Psi_P(\low)$ (\Cref{Result: Convex concave transformations}).

\subsection{Power-Series Representable \texorpdfstring{$\eps$}{Epsilon}-Close Shrinking Interval Bounds in MDPs}
\label{Section: Interval Bounds in MDPs}

In this subsection, we show that every MDP with a discounted-sum objective has power-series representable $\eps$-close shrinking interval bounds (\Cref{Result: Shrinking Bounds in MDP}).
To do so, we apply a convex piecewise-linear function to the value function to obtain a lower bound, and we obtain the upper bound symmetrically.
\begin{compactitem}
  \item \emph{(Convex transformations)}
    We show that, for every non-decreasing convex function $f$ with $f(x) \le x$, the function $\low \defas f \circ \val_\lambda$ satisfies $\low \preceq \Psi_P(\low)$ (\Cref{Result: Convex concave transformations}).
  \item \emph{(Separation of values)}
    We show that two distinct values of an MDP differ by at least a double exponentially small amount (\Cref{Result: Value separation}).
  \item \emph{(Power-series representable interval bounds)}
    We choose $f$ piecewise linear with breakpoints at the values.
    Its increments between consecutive values are the gaps between these values, multiplied by increasing factors and truncated with a fixed relative precision, which makes $f$ convex (\Cref{Result: Shrinking Bounds in MDP}).
    The truncation requires few terms regardless of how small the gaps are, and \Cref{Result: Value separation} bounds the exponents, so the lower bound is power-series representable.
\end{compactitem}
In contrast to the candidates of \Cref{Section: Failed candidates}, these bounds satisfy the inequalities of shrinking bounds exactly, regardless of how close the values are.

\begin{lemma}[Convex transformations]
  \label{Result: Convex concave transformations}
  For $\mathfrak{p} \in \{1, 2\}$, let $P$ be a player-$\mathfrak{p}$ MDP with a reward function and a discount factor $\lambda$, and let $f \colon [0, 1] \to \RR$ be a non-decreasing convex function such that $f(x) \le x$ for all $x \in [0, 1]$.
  Then, $f \circ \val_\lambda \preceq \Psi_P(f \circ \val_\lambda)$.
\end{lemma}

\begin{proof}
  Note that $\val_\lambda$ takes values in $[0, 1]$, since rewards are in $[0, 1]$.
  Consider a state $\state \in \States$.
  If $\mathfrak{p} = 1$, let $\action \in \Actions$ be an action that attains the maximum in the definition of $\Psi_P(\val_\lambda)(\state)$, and if $\mathfrak{p} = 2$, let $\action \in \Actions$ be an arbitrary action.
  Since $\val_\lambda = \Psi_P(\val_\lambda)$, in both cases, we have
  \[
    \val_\lambda(\state) \le (1 - \lambda) \, r(\state, \action) + \lambda \sum_{\state' \in \States} \prob(\state, \action)(\state') \, \val_\lambda(\state') \,.
  \]
  Therefore, we have
  \begin{align}
    f(\val_\lambda(\state))
      &\le f \Big( (1 - \lambda) \, r(\state, \action) + \lambda \sum_{\state' \in \States} \prob(\state, \action)(\state') \, \val_\lambda(\state') \Big)
        && (f \text{ is non-decreasing}) \nonumber \\
      &\le (1 - \lambda) \, f(r(\state, \action)) + \lambda \sum_{\state' \in \States} \prob(\state, \action)(\state') \, f(\val_\lambda(\state'))
        && (f \text{ is convex},\ \prob(\state, \action) \in \Delta(\States)) \nonumber \\
      &\le (1 - \lambda) \, r(\state, \action) + \lambda \sum_{\state' \in \States} \prob(\state, \action)(\state') \, f(\val_\lambda(\state'))
        && (f(x) \le x) \,. \label{equation: convex transformation}
  \end{align}
  It remains to compare the right-hand side of \Cref{equation: convex transformation} with $\Psi_P(f \circ \val_\lambda)(\state)$.
  If $\mathfrak{p} = 1$, then we have
  \begin{align*}
    f(\val_\lambda(\state))
      &\le (1 - \lambda) \, r(\state, \action) + \lambda \sum_{\state' \in \States} \prob(\state, \action)(\state') \, f(\val_\lambda(\state'))
        && (\text{\Cref{equation: convex transformation}}) \\
      &\le \max_{\action' \in \Actions} \Big( (1 - \lambda) \, r(\state, \action') + \lambda \sum_{\state' \in \States} \prob(\state, \action')(\state') \, f(\val_\lambda(\state')) \Big)
        && (\action \in \Actions) \\
      &= \Psi_P(f \circ \val_\lambda)(\state)
        && (\text{definition of } \Psi_P,\ \op = \max) \,.
  \end{align*}
  If $\mathfrak{p} = 2$, then \Cref{equation: convex transformation} holds for every action $\action \in \Actions$, since $\action$ is arbitrary, and therefore we have
  \begin{align*}
    f(\val_\lambda(\state))
      &\le \min_{\action \in \Actions} \Big( (1 - \lambda) \, r(\state, \action) + \lambda \sum_{\state' \in \States} \prob(\state, \action)(\state') \, f(\val_\lambda(\state')) \Big)
        && (\text{\Cref{equation: convex transformation}}) \\
      &= \Psi_P(f \circ \val_\lambda)(\state)
        && (\text{definition of } \Psi_P,\ \op = \min) \,,
  \end{align*}
  which completes the proof.
\end{proof}

To bound the size of the representation of the interval bounds, we use that distinct values of an MDP cannot be arbitrarily close.
Bounds of this kind are standard, see, e.g.,~\cite{condon1992complexity} and~\cite[Theorem~23]{auger2021GenericStrategyImprovement}.
For completeness, we present the proof with an explicit bound in terms of the succinct discount factor.

\begin{lemma}[Separation of values]
  \label{Result: Value separation}
  For $\mathfrak{p} \in \{1, 2\}$, let $P$ be a player-$\mathfrak{p}$ MDP with a reward function $r$ and a succinct discount factor $\lambda = 1 - 2^{-2^{\Blam}}$.
  Let $B_P \in \NN_0$ be such that all rewards and transition probabilities of $P$ have a common denominator at most $2^{2^{B_P}}$, and let
  \[
    \Bsep \defas \Blam + B_P + 2 \lceil \log n \rceil + 1 \,.
  \]
  Then, two distinct elements of $\{\val_\lambda(\state) \mid \state \in \States\} \cup \{0, 1\}$ differ by at least $2^{-2^{\Bsep}}$.
\end{lemma}

\begin{proof}
  Let $\pi$ be an optimal pure stationary strategy in $P$~\cite{shapley1953stochastic, filar2012competitive}, i.e., $\val_\lambda^P = \val_\lambda^{P_\pi}$.
  Let $Q \le 2^{2^{B_P}}$ be a common denominator of all rewards and transition probabilities of $P$.
  First, we write $\val_\lambda$ as the solution of a linear system with integer coefficients.
  For every state $\state \in \States$, we have
  \begin{align*}
    2^{2^{\Blam}} Q \, \val_\lambda(\state)
      &= 2^{2^{\Blam}} Q \Big( (1 - \lambda) \, r(\state, \pi(\state)) + \lambda \sum_{\state' \in \States} \prob(\state, \pi(\state))(\state') \, \val_\lambda(\state') \Big)
        && (\val_\lambda = \Psi_{P_\pi}(\val_\lambda)) \\
      &= Q \, r(\state, \pi(\state)) + (2^{2^{\Blam}} - 1) \sum_{\state' \in \States} Q \, \prob(\state, \pi(\state))(\state') \, \val_\lambda(\state')
        && (\lambda = 1 - 2^{-2^{\Blam}}) \,.
  \end{align*}
  Therefore, the vector $x = \val_\lambda$ is a solution of the linear system $A x = c$ with the following equation for every state $\state \in \States$
  \begin{equation}
    \label{equation: integer linear system}
    2^{2^{\Blam}} Q \, x_\state - (2^{2^{\Blam}} - 1) \sum_{\state' \in \States} Q \, \prob(\state, \pi(\state))(\state') \, x_{\state'} = Q \, r(\state, \pi(\state)) \,.
  \end{equation}
  By the choice of $Q$, the matrix $A$ and the vector $c$ have integer entries.
  Moreover, dividing \Cref{equation: integer linear system} by $2^{2^{\Blam}} Q$ yields the fixed-point equations of the Bellman operator of $P_\pi$, whose unique solution is $\val_\lambda$.
  Hence, $\val_\lambda$ is the unique solution of $A x = c$, and therefore $A$ is invertible.
  Moreover, since $Q \, r(\state, \pi(\state))$ and $Q \, \prob(\state, \pi(\state))(\state')$ are in $[0, Q]$ and $Q \le 2^{2^{B_P}}$, all entries of $A$ and $c$ are at most $2^{2^{\Blam} + 2^{B_P}}$ in absolute value.
  Second, we bound the common denominator of the solution.
  By Cramer's rule, for all states $\state \in \States$, we have $x_\state = \det(A_\state) / \det(A)$, where $A_\state$ is obtained from $A$ by replacing the column of $\state$ with $c$.
  Both determinants are integers.
  Since every row of $A$ has $n$ entries, we have
  \begin{align*}
    |\det(A)|
      &\le \left( \sqrt{n} \cdot 2^{2^{\Blam} + 2^{B_P}} \right)^{n}
        && (\text{Hadamard's inequality}) \\
      &\le 2^{n(2^{\Blam} + 2^{B_P} + \lceil \log n \rceil)}
        && (\sqrt{n} \le 2^{\lceil \log n \rceil}) \\
      &\le 2^{2^{\lceil \log n \rceil} (2^{\Blam + B_P + 1} + \lceil \log n \rceil)}
        && (n \le 2^{\lceil \log n \rceil},\ 2^{a} + 2^{b} \le 2^{a + b + 1}) \\
      &\le 2^{2^{\lceil \log n \rceil} \cdot 2^{\Blam + B_P + 1 + \lceil \log n \rceil}}
        && (2^{a} + b \le 2^{a + b}) \\
      &= 2^{2^{\Bsep}}
        && (\text{definition of } \Bsep) \,.
  \end{align*}
  Lastly, define $q \defas |\det(A)| \le 2^{2^{\Bsep}}$.
  Then, $q \cdot \val_\lambda(\state) = \pm \det(A_\state)$ is an integer for all states $\state \in \States$.
  For two distinct elements $v, v'$ of $\{\val_\lambda(\state) \mid \state \in \States\} \cup \{0, 1\}$, the numbers $q \cdot v$ and $q \cdot v'$ are distinct integers, and therefore $|v - v'| \ge 1/q \ge 2^{-2^{\Bsep}}$, which completes the proof.
\end{proof}

We are now ready to state the main result of this subsection.

\begin{lemma}[Power-series representable interval bounds in MDPs]
\label{Result: Shrinking Bounds in MDP}
    For $\mathfrak{p} \in \{1, 2\}$, let $P$ be a player-$\mathfrak{p}$ MDP with a reward function and a succinct discount factor $\lambda = 1 - 2^{-2^{\Blam}}$, and let $B_P \in \NN_0$ and $\Bsep$ be as defined in \Cref{Result: Value separation}.
    Let $\eps = 2^{-\Beps} \in (0, 1)$, and let
    \[
      \ell_{\mathrm{bnd}} \defas (n + 1)(\Beps + \lceil \log (n + 1) \rceil + 8) + 1
      \qquad \text{and} \qquad
      B_{\mathrm{bnd}} \defas \Bsep + \Beps + \lceil \log (n + 1) \rceil + 10 \,.
    \]
    Then, there exists an $\eps$-close shrinking interval bound $I = (\low, \up)$ such that $\low(\state), \up(\state) \in \F(\ell_{\mathrm{bnd}}, B_{\mathrm{bnd}})$ for all states $\state \in \States$.
\end{lemma}

\begin{proof}
  We construct the lower bound $\low$ by applying a convex function $f$ to the value function, and the upper bound $\up$ by applying the same construction to a dual MDP.
  The function $f$ is piecewise linear, and its breakpoints are the values of $P$, i.e., $f$ is linear between every two consecutive values.
  Hence, $f$ is determined by $f(0) = 0$ and its \emph{increments}, i.e., the differences between the values of $f$ at consecutive values of $P$.
  The proof consists of the following four steps.
  \begin{compactitem}
    \item \emph{(Increments)}
      We choose every increment by truncating the gap between the corresponding consecutive values, scaled by a factor, to a power-series number.
      For $f$ to be convex, its slopes must increase from left to right.
      Truncation to power-series numbers decreases every slope slightly.
      Therefore, before truncation, we multiply every gap by a factor slightly smaller than one, and these factors increase from left to right by more than the truncation decreases the slopes.
      Hence, the slopes of $f$ are close to one, at most one, and increasing, and therefore $f$ is convex.
    \item \emph{(Lower bound)}
      We show that $\low \defas f \circ \val_\lambda$ satisfies $\low \preceq \Psi_P(\low)$, approximates $\val_\lambda$ up to $\frac{\eps}{2}$, and is power-series representable.
    \item \emph{(Upper bound)}
      We apply the construction to the MDP with rewards $1 - r$ and the opposite player, whose value function is $\vec{1} - \val_\lambda$, and we obtain the upper bound $\up$.
    \item \emph{(Interval bound)}
      We show that $(\low, \up)$ is an $\eps$-close shrinking interval bound.
  \end{compactitem}
  We now prove each step.
  \begin{compactitem}
    \item \emph{(Increments)}
      Let $0 = v_0 < v_1 < \dots < v_J = 1$ be the distinct elements of $\{\val_\lambda(\state) \mid \state \in \States\} \cup \{0, 1\}$, and, for $j \in [J]$, let $g_j \defas v_j - v_{j-1}$ be the gap between consecutive values.
      Note that $J \le n + 1$ and, by \Cref{Result: Value separation}, $g_j \ge 2^{-2^{\Bsep}}$ for all $j \in [J]$.
      Let $\ell_{\mathrm{inc}} \defas \Beps + \lceil \log (n + 1) \rceil + 8$ be the number of terms of every increment in power-series representation.
      Then, we have
      \begin{align}
        2^{1 - \ell_{\mathrm{inc}}}
          &= \eps \cdot 2^{-\lceil \log (n + 1) \rceil - 7}
            && (\text{definition of } \ell_{\mathrm{inc}},\ \eps = 2^{-\Beps}) \nonumber \\
          &\le \frac{\eps}{128 (n + 1)}
            && (n + 1 \le 2^{\lceil \log (n + 1) \rceil}) \nonumber \\
          &\le \frac{\eps}{128 J}
            && (J \le n + 1) \,. \label{equation: precision of increments}
      \end{align}
      For $j \in [J]$, define the factor $\gamma_j \defas 1 - \frac{\eps}{4} + 2^{3 - \ell_{\mathrm{inc}}} j$ by which we multiply the gap $g_j$ before truncation.
      Then, we have $\gamma_j \ge 1 - \frac{\eps}{4} > \frac{1}{2}$ and
      \begin{align*}
        \gamma_j
          &\le \gamma_J
            && (j \le J) \\
          &= 1 - \frac{\eps}{4} + 4 \cdot 2^{1 - \ell_{\mathrm{inc}}} J
            && (\text{definition of } \gamma_J) \\
          &\le 1 - \frac{\eps}{4} + \frac{\eps}{32}
            && (\text{\Cref{equation: precision of increments}}) \\
          &< 1 \,.
      \end{align*}
      Since $\frac{1}{2} < \gamma_j < 1$ and $2^{-2^{\Bsep}} \le g_j \le 1$, we have $\gamma_j g_j \in (2^{-2^{\Bsep} - 1}, 1]$.
      Moreover, since $B_{\mathrm{bnd}} = \Bsep + \ell_{\mathrm{inc}} + 2$ and $2^{a + b} \ge 2^{a} + b$ for all $a, b \in \NN_0$, we have $2^{B_{\mathrm{bnd}}} \ge 2^{\Bsep} + \ell_{\mathrm{inc}} + 2$, and therefore $\gamma_j g_j \in (2^{\ell_{\mathrm{inc}} - 2^{B_{\mathrm{bnd}}}}, 1]$.
      By \Cref{Result: Number representation in power series}, the increment $d_j \defas \trunc_{\ell_{\mathrm{inc}}}(\gamma_j g_j)$ satisfies $d_j \in \F_+(\ell_{\mathrm{inc}}, B_{\mathrm{bnd}})$ and
      \begin{equation}
        \label{equation: truncated increments}
        (1 - 2^{1 - \ell_{\mathrm{inc}}}) \, \gamma_j g_j \le d_j \le \gamma_j g_j \,.
      \end{equation}
      We now show that the ratios $d_j / g_j$ are close to one and increasing in $j$.
      On the one hand, for all $j \in [J]$, we have $d_j / g_j \le \gamma_j < 1$ and
      \begin{align*}
        \frac{d_j}{g_j}
          &\ge (1 - 2^{1 - \ell_{\mathrm{inc}}}) \, \gamma_j
            && (\text{\Cref{equation: truncated increments}}) \\
          &\ge (1 - 2^{1 - \ell_{\mathrm{inc}}}) \Big( 1 - \frac{\eps}{4} \Big)
            && (\gamma_j \ge 1 - \tfrac{\eps}{4}) \\
          &\ge 1 - \frac{\eps}{4} - 2^{1 - \ell_{\mathrm{inc}}} \\
          &\ge 1 - \frac{\eps}{2}
            && (\text{\Cref{equation: precision of increments}}) \,.
      \end{align*}
      On the other hand, for all $j \in [J - 1]$, we have
      \begin{align*}
        \frac{d_j}{g_j}
          &\le \gamma_j
            && (\text{\Cref{equation: truncated increments}}) \\
          &= (1 - 2^{1 - \ell_{\mathrm{inc}}}) \, \gamma_{j+1} - 2^{1 - \ell_{\mathrm{inc}}} (4 - \gamma_{j+1})
            && (\gamma_{j+1} - \gamma_j = 4 \cdot 2^{1 - \ell_{\mathrm{inc}}}) \\
          &< (1 - 2^{1 - \ell_{\mathrm{inc}}}) \, \gamma_{j+1}
            && (\gamma_{j+1} < 1) \\
          &\le \frac{d_{j+1}}{g_{j+1}}
            && (\text{\Cref{equation: truncated increments}}) \,.
      \end{align*}
      Therefore, we have
      \begin{equation}
        \label{equation: slopes}
        1 - \frac{\eps}{2} \le \frac{d_1}{g_1} < \frac{d_2}{g_2} < \dots < \frac{d_J}{g_J} \le 1 \,.
      \end{equation}

    \item \emph{(Lower bound)}
      Let $f \colon [0, 1] \to \RR$ be the continuous piecewise-linear function that is linear on $[v_{j-1}, v_j]$ for all $j \in [J]$ and satisfies, for all $j \in \{0, 1, \ldots, J\}$,
      \[
        f(v_j) = \sum_{l = 1}^{j} d_l \,.
      \]
      Then, the slope of $f$ on $[v_{j-1}, v_j]$ is $d_j / g_j$.
      By \Cref{equation: slopes}, the slopes are positive and increasing, and therefore $f$ is non-decreasing and convex.
      Moreover, since $f(0) = 0$ and all slopes are at most $1$, we have $f(x) \le x$ for all $x \in [0, 1]$.
      Define $\low \defas f \circ \val_\lambda$.
      First, by \Cref{Result: Convex concave transformations}, we have $\low \preceq \Psi_P(\low)$.
      Second, for every state $\state$, let $j$ be such that $\val_\lambda(\state) = v_j$.
      Then, we have
      \begin{align*}
        \val_\lambda(\state) - \low(\state)
          &= \sum_{l = 1}^{j} g_l - \sum_{l = 1}^{j} d_l
            && (v_j = \textstyle\sum_{l = 1}^{j} g_l,\ \text{definition of } f) \\
          &= \sum_{l = 1}^{j} \Big( 1 - \frac{d_l}{g_l} \Big) g_l \\
          &\le \frac{\eps}{2} \sum_{l = 1}^{j} g_l
            && (\text{\Cref{equation: slopes}}) \\
          &= \frac{\eps}{2} \, v_j \\
          &\le \frac{\eps}{2}
            && (v_j \le 1) \,.
      \end{align*}
      Moreover, by \Cref{equation: slopes}, we have $\sum_{l = 1}^{j} (1 - d_l / g_l) \, g_l \ge 0$, i.e., $\low(\state) \le \val_\lambda(\state)$.
      Lastly, $\low(\state)$ is the sum of at most $J \le n + 1$ elements of $\F_+(\ell_{\mathrm{inc}}, B_{\mathrm{bnd}})$.
      Therefore, by \Cref{lemm:power_arithmetic}, we have $\low(\state) \in \F((n + 1) \ell_{\mathrm{inc}}, B_{\mathrm{bnd}}) \subseteq \F(\ell_{\mathrm{bnd}}, B_{\mathrm{bnd}})$.

    \item \emph{(Upper bound)}
      Consider the player-$(3 - \mathfrak{p})$ MDP $P'$ with the same states, actions, transition function, and discount factor as $P$, and with the reward function $1 - r$.
      Denote by $\overline{\op}$ the operator of $P'$, i.e., $\overline{\op} = \min$ if $\op = \max$, and $\overline{\op} = \max$ if $\op = \min$.
      First, we relate the Bellman operators of $P$ and $P'$.
      For every function $\rho \colon \States \to \RR$ and state $\state \in \States$, we have
      \begin{align*}
        \Psi_{P'}(\rho)(\state)
          &= \mathop{\overline{\mathrm{op}}}\limits_{\action \in \Actions} \Big( (1 - \lambda) \big( 1 - r(\state, \action) \big) + \lambda \sum_{\state' \in \States} \prob(\state, \action)(\state') \, \rho(\state') \Big)
            && (\text{definition of } \Psi_{P'}) \\
          &= \mathop{\overline{\mathrm{op}}}\limits_{\action \in \Actions} \Big( 1 - (1 - \lambda) \, r(\state, \action) - \lambda \sum_{\state' \in \States} \prob(\state, \action)(\state') \big( 1 - \rho(\state') \big) \Big)
            && (\prob(\state, \action) \in \Delta(\States)) \\
          &= 1 - \op_{\action \in \Actions} \Big( (1 - \lambda) \, r(\state, \action) + \lambda \sum_{\state' \in \States} \prob(\state, \action)(\state') \big( 1 - \rho(\state') \big) \Big)
            && (\text{definition of } \overline{\op}) \\
          &= 1 - \Psi_P(\vec{1} - \rho)(\state)
            && (\text{definition of } \Psi_P) \,.
      \end{align*}
      Therefore, for every function $\rho \colon \States \to \RR$, we have
      \begin{equation}
        \label{equation: dual Bellman operator}
        \Psi_{P'}(\rho) = \vec{1} - \Psi_P(\vec{1} - \rho) \,.
      \end{equation}
      In particular, $\vec{1} - \val_\lambda^P$ is a fixed point of $\Psi_{P'}$, i.e., $\val_\lambda^{P'} = \vec{1} - \val_\lambda^P$.
      Moreover, a common denominator of all rewards and transition probabilities of $P$ is also a common denominator of all rewards and transition probabilities of $P'$, and therefore $P'$ satisfies the assumptions of the lemma with the same $B_P$.
      Second, by the previous items applied to $P'$, there exists a function $\low' \colon \States \to \RR$ such that $\low' \preceq \Psi_{P'}(\low')$, $\val_\lambda^{P'} - \frac{\eps}{2} \cdot \vec{1} \preceq \low' \preceq \val_\lambda^{P'}$, and $\low'(\state) \in \F((n + 1) \ell_{\mathrm{inc}}, B_{\mathrm{bnd}})$ for all states $\state$.
      Define $\up \defas \vec{1} - \low'$.
      Then, we have
      \begin{align*}
        \Psi_P(\up)
          &= \Psi_P(\vec{1} - \low')
            && (\text{definition of } \up) \\
          &= \vec{1} - \Psi_{P'}(\low')
            && (\text{\Cref{equation: dual Bellman operator}}) \\
          &\preceq \vec{1} - \low'
            && (\low' \preceq \Psi_{P'}(\low')) \\
          &= \up \,.
      \end{align*}
      Moreover, since $\up - \val_\lambda^P = \val_\lambda^{P'} - \low'$, we have $\val_\lambda^P \preceq \up \preceq \val_\lambda^P + \frac{\eps}{2} \cdot \vec{1}$.
      Lastly, for every state $\state$, since $\up(\state) = 2^{-0} - \low'(\state)$, by \Cref{lemm:power_arithmetic}, we have $\up(\state) \in \F((n + 1) \ell_{\mathrm{inc}} + 1, B_{\mathrm{bnd}}) = \F(\ell_{\mathrm{bnd}}, B_{\mathrm{bnd}})$.

    \item \emph{(Interval bound)}
      By the previous items, we have $\low \preceq \val_\lambda \preceq \up$, and therefore $I = (\low, \up)$ is an interval bound.
      Moreover, $I$ is shrinking, since $\low \preceq \Psi_P(\low)$ and $\Psi_P(\up) \preceq \up$.
      Lastly, $I$ is $\eps$-close, since $\up(\state) - \low(\state) \le \frac{\eps}{2} + \frac{\eps}{2} = \eps$ for all states $\state \in \States$, which completes the proof. \qedhere
  \end{compactitem}
\end{proof}

\subsection{\texorpdfstring{Proof of Theorem~\ref{Result: Completeness of witnesses}}{Proof of the Existence of a Witness}}
\label{Section: Proof of Completeness}

In this subsection, we prove \Cref{Result: Completeness of witnesses} by combining \Cref{Result: Existence of Floating-Point Strategies,Result: Value separation,Result: Shrinking Bounds in MDP}.

\begin{proof}[Proof of {\Cref{Result: Completeness of witnesses}}]
  We construct the witness in four steps.
  \begin{compactitem}
    \item \emph{(Strategies)}
      Let $\ell_{\mathrm{str}}$ and $B_{\mathrm{str}}$ be the constants of \Cref{Result: Existence of Floating-Point Strategies} for the additive error $\frac{\eps}{4} = 2^{-(\Beps + 2)}$, i.e.,
      \[
        \ell_{\mathrm{str}} \defas \Beps + \lceil \log (n + 2) \rceil + \lceil \log m \rceil + 7
        \qquad \text{and} \qquad
        B_{\mathrm{str}} \defas \Blam + \Beps + \lceil \log m \rceil + \ell_{\mathrm{str}} + 5 \,.
      \]
      By \Cref{Result: Existence of Floating-Point Strategies}, there exist $\frac{\eps}{4}$-optimal stationary strategies $\strategyone$ and $\strategytwo$ such that $\strategyone(\state), \strategytwo(\state) \in \D(m \ell_{\mathrm{str}}, B_{\mathrm{str}})$ for all states $\state \in \States$.
      Since $\strategyone$ and $\strategytwo$ are $\frac{\eps}{4}$-optimal, for all states $\state \in \States$, we have
      \begin{equation}
        \label{equation: values of the witness strategies}
        \val_\lambda(\state) - \frac{\eps}{4} \le \val_\lambda^{G_\strategyone}(\state) \le \val_\lambda(\state) \le \val_\lambda^{G_\strategytwo}(\state) \le \val_\lambda(\state) + \frac{\eps}{4} \,.
      \end{equation}

    \item \emph{(Common denominator)}
      Let $Q \le 2^{\Binp n m^2 (n + 1)}$ be the product of the denominators of all rewards and transition probabilities of $G$, and let $B_P \defas B_{\mathrm{str}} + \Binp n m^2 (n + 1)$.
      For all states $\state$ and actions $\action$, the exponents of $\strategyone(\state)(\action) \in \F_+(m \ell_{\mathrm{str}}, B_{\mathrm{str}})$ are less than $2^{B_{\mathrm{str}}}$, and therefore $\strategyone(\state)(\action)$ is an integer multiple of $2^{-2^{B_{\mathrm{str}}}}$.
      Moreover, all rewards and transition probabilities of $G$ are integer multiples of $1 / Q$.
      By the definition of $G_\strategyone$, its rewards and transition probabilities are sums of products of these numbers, e.g., $r_\strategyone(\state, \baction) = \sum_{\action \in \Actions} \strategyone(\state)(\action) \, r(\state, \action, \baction)$.
      Therefore, all rewards and transition probabilities of $G_\strategyone$ are integer multiples of $1 / (2^{2^{B_{\mathrm{str}}}} Q)$, where
      \begin{align*}
        2^{2^{B_{\mathrm{str}}}} Q
          &\le 2^{2^{B_{\mathrm{str}}} + \Binp n m^2 (n + 1)}
            && (Q \le 2^{\Binp n m^2 (n + 1)}) \\
          &\le 2^{2^{B_{\mathrm{str}} + \Binp n m^2 (n + 1)}}
            && (2^{a} + b \le 2^{a + b}) \\
          &= 2^{2^{B_P}}
            && (\text{definition of } B_P) \,.
      \end{align*}
      The same holds for $G_\strategytwo$.
      Hence, $G_\strategyone$ and $G_\strategytwo$ satisfy the assumptions of \Cref{Result: Value separation} with $B_P$, and $\Bsep = \Blam + B_P + 2 \lceil \log n \rceil + 1$ as defined there.

    \item \emph{(Interval bounds)}
      Let $\ell_{\mathrm{bnd}}$ and $B_{\mathrm{bnd}}$ be the constants of \Cref{Result: Shrinking Bounds in MDP} for the additive error $\frac{\eps}{4}$, i.e.,
      \[
        \ell_{\mathrm{bnd}} \defas (n + 1)(\Beps + \lceil \log (n + 1) \rceil + 10) + 1
        \quad \text{and} \quad
        B_{\mathrm{bnd}} \defas \Bsep + \Beps + \lceil \log (n + 1) \rceil + 12 \,.
      \]
      By \Cref{Result: Shrinking Bounds in MDP}, there exist $\frac{\eps}{4}$-close shrinking interval bounds $(\low_\strategyone, \up_\strategyone)$ of $G_\strategyone$ and $(\low_\strategytwo, \up_\strategytwo)$ of $G_\strategytwo$ whose entries are in $\F(\ell_{\mathrm{bnd}}, B_{\mathrm{bnd}})$.
      Since these interval bounds are shrinking, by \Cref{Result: Properties of Bellman operator}, for all states $\state \in \States$, we have
      \begin{equation}
        \label{equation: interval bounds contain the values}
        \low_\strategyone(\state) \le \val_\lambda^{G_\strategyone}(\state) \le \up_\strategyone(\state)
        \qquad \text{and} \qquad
        \low_\strategytwo(\state) \le \val_\lambda^{G_\strategytwo}(\state) \le \up_\strategytwo(\state) \,.
      \end{equation}

    \item \emph{(Witness)}
      We widen the interval bounds by $\frac{\eps}{4}$, i.e., we define
      \[
        I_\strategyone \defas \Big( \low_\strategyone, \, \up_\strategyone + \frac{\eps}{4} \cdot \vec{1} \Big)
        \qquad \text{and} \qquad
        I_\strategytwo \defas \Big( \low_\strategytwo - \frac{\eps}{4} \cdot \vec{1}, \, \up_\strategytwo \Big) \,.
      \]
      Since $\low_\strategyone \preceq \up_\strategyone \preceq \up_\strategyone + \frac{\eps}{4} \cdot \vec{1}$ and $\low_\strategytwo - \frac{\eps}{4} \cdot \vec{1} \preceq \low_\strategytwo \preceq \up_\strategytwo$, the pairs $I_\strategyone$ and $I_\strategytwo$ are interval bounds.
      We show that $(\strategyone, \strategytwo, I_\strategyone, I_\strategytwo)$ is a witness of polynomial size.
      First, we show that $I_\strategyone$ and $I_\strategytwo$ are shrinking.
      For every MDP $P$, function $\rho \colon \States \to \RR$, and constant $c \in \RR$, since the transition probabilities from every state and action sum to one, we have
      \begin{equation}
        \label{equation: Bellman operator of shifted functions}
        \Psi_P(\rho + c \cdot \vec{1}) = \Psi_P(\rho) + \lambda c \cdot \vec{1} \,.
      \end{equation}
      Therefore, we have
      \begin{align*}
        \Psi_{G_\strategyone}\Big( \up_\strategyone + \frac{\eps}{4} \cdot \vec{1} \Big)
          &= \Psi_{G_\strategyone}(\up_\strategyone) + \lambda \frac{\eps}{4} \cdot \vec{1}
            && (\text{\Cref{equation: Bellman operator of shifted functions}}) \\
          &\preceq \up_\strategyone + \lambda \frac{\eps}{4} \cdot \vec{1}
            && ((\low_\strategyone, \up_\strategyone) \text{ is shrinking}) \\
          &\preceq \up_\strategyone + \frac{\eps}{4} \cdot \vec{1}
            && (\lambda < 1) \,.
      \end{align*}
      Moreover, since $(\low_\strategyone, \up_\strategyone)$ is shrinking, we have $\low_\strategyone \preceq \Psi_{G_\strategyone}(\low_\strategyone)$, and hence $I_\strategyone$ is shrinking.
      Symmetrically, $I_\strategytwo$ is shrinking.
      Second, $I_\strategyone$ and $I_\strategytwo$ are $\frac{\eps}{2}$-close, since $(\low_\strategyone, \up_\strategyone)$ and $(\low_\strategytwo, \up_\strategytwo)$ are $\frac{\eps}{4}$-close, and the widening adds $\frac{\eps}{4}$.
      Third, we show that $I_\strategyone$ and $I_\strategytwo$ intersect.
      For every state $\state \in \States$, we have
      \begin{align*}
        \low_\strategyone(\state)
          &\le \val_\lambda^{G_\strategyone}(\state)
            && (\text{\Cref{equation: interval bounds contain the values}}) \\
          &\le \val_\lambda(\state)
            && (\text{\Cref{equation: values of the witness strategies}}) \\
          &\le \val_\lambda^{G_\strategyone}(\state) + \frac{\eps}{4}
            && (\text{\Cref{equation: values of the witness strategies}}) \\
          &\le \up_\strategyone(\state) + \frac{\eps}{4}
            && (\text{\Cref{equation: interval bounds contain the values}}) \,,
      \end{align*}
      i.e., $\val_\lambda(\state)$ is in the interval of $I_\strategyone$ at $\state$.
      Symmetrically, $\val_\lambda(\state)$ is in the interval of $I_\strategytwo$ at $\state$, and hence $I_\strategyone$ and $I_\strategytwo$ intersect.
      Lastly, we bound the size of the witness.
      Since $\frac{\eps}{4} = 2^{-(\Beps + 2)}$ and $2^{B_{\mathrm{bnd}}} > \Beps + 2$, by \Cref{lemm:power_arithmetic}, the entries of $I_\strategyone$ and $I_\strategytwo$ are in $\F(\ell_{\mathrm{bnd}} + 1, B_{\mathrm{bnd}})$.
      The strategies consist of at most $2nm$ probabilities in $\F(m \ell_{\mathrm{str}}, B_{\mathrm{str}})$, and the interval bounds consist of $4n$ numbers in $\F(\ell_{\mathrm{bnd}} + 1, B_{\mathrm{bnd}})$.
      Therefore, the witness has size $\mathcal{O}(n m (m \ell_{\mathrm{str}} B_{\mathrm{str}} + \ell_{\mathrm{bnd}} B_{\mathrm{bnd}}))$.
      This size is polynomial in $n$, $m$, $\Binp$, $\Beps$, and $\Blam$, since $\ell_{\mathrm{str}}$, $B_{\mathrm{str}}$, $B_P$, $\Bsep$, $\ell_{\mathrm{bnd}}$, and $B_{\mathrm{bnd}}$ are polynomial in these parameters, which completes the proof. \qedhere
  \end{compactitem}
\end{proof}

\section{Limit-Average and Reachability Objectives}
\label{Section: Liminf-average}

In this section, we prove \Cref{Result: CSG in TFNP}.
To do so, we first reduce reachability objectives to limit-average objectives.
Then, we approximate the values of limit-average objectives by the values of discounted-sum objectives with a succinct discount factor, and we apply \Cref{Result: Discounted in TFNP}.

\begin{lemma}[Reachability as limit-average]
\label{Result: Reachability as limit-average}
  Consider a game $G$ with transition function $\prob$ and a target set $\Target$.
  Let $G_\Target$ be the game obtained from $G$ by making every target state absorbing, i.e., the game with the transition function $\prob_\Target$, and let $r_\Target$ be the reward function defined, for all states $\state$ and actions $\action, \baction$, by
  \[
    \prob_\Target(\state, \action, \baction) \defas
      \begin{cases}
        \1[\state] & \state \in \Target \,, \\
        \prob(\state, \action, \baction) & \text{otherwise,}
      \end{cases}
    \qquad \text{and} \qquad
    r_\Target(\state, \action, \baction) \defas
      \begin{cases}
        1 & \state \in \Target \,, \\
        0 & \text{otherwise,}
      \end{cases}
  \]
  where $\1[\state]$ is the Dirac distribution on $\state$.
  Then, for all states $\state \in \States$, we have $\val_\Target^G(\state) = \val^{G_\Target}(\state)$, where $\val^{G_\Target}$ is the value of $\liminfavg(r_\Target)$ in $G_\Target$.
  In particular, if every target state is absorbing in $G$, then $G_\Target = G$.
\end{lemma}

\begin{proof}
  Every play of $G_\Target$ either never visits $\Target$, and then all its rewards are $0$, or it visits $\Target$ and stays there forever, and then its limit-average is $1$.
  Moreover, the plays of $G$ and $G_\Target$ coincide until the first visit to $\Target$.
  Therefore, for all states $\state \in \States$, we have $\val_\Target^G(\state) = \val^{G_\Target}(\state)$.
  Lastly, if every target state is absorbing in $G$, then $\prob_\Target = \prob$, i.e., $G_\Target = G$.
\end{proof}

The value of a limit-average objective is the limit of the values of the discounted-sum objectives as the discount factor tends to one~\cite{mertens1981StochasticGames}.
The following result bounds how close to one the discount factor needs to be so that the two values are close.
It follows from~\cite[Theorem~3]{oliubarton2021NewAlgorithms}, where we replace the minimal number $\min\{ |\Actions|^n, |\Bactions|^n \}$ of pure stationary strategies of a player by its upper bound $m^n$, and the numbers of bits of $n$, $m^n$, and $Q$ by their upper bounds $\lceil \log n \rceil + 1$, $n \lceil \log m \rceil + 1$, and $B_Q + 1$, respectively, which only increases $\Blam$.

\begin{lemma}[{\cite[Theorem~3]{oliubarton2021NewAlgorithms}}]
\label{Result: Liminf-average approximation}
  Let $G$ be a game and $r$ a reward function such that all transition probabilities and rewards are integer multiples of $1/Q$, for a positive integer $Q \le 2^{B_Q}$.
  Let $\eps = 2^{-\Beps}$, and let
  \[
    \Blam \defas \left\lceil \log \left( 4 n m^n \left( \lceil \log n \rceil + n \lceil \log m \rceil + B_Q + 3 \right) + \Beps \, n \, m^n \right) \right\rceil \,.
  \]
  Then, for every discount factor $\lambda \in [1 - 2^{-2^{\Blam}}, 1)$ and state $\state \in \States$, we have $|\val_\lambda(\state) - \val(\state)| \le \eps$.
\end{lemma}

We are now ready to prove our main result.

\begin{proof}[Proof of \Cref{Result: CSG in TFNP}]
  By \Cref{Result: Reachability as limit-average}, since $G_\Target$ and $r_\Target$ are computed in polynomial time, it suffices to prove the result for limit-average objectives.
  Consider an instance $(G, \phi)$ with a limit-average objective $\phi = \liminfavg(r)$, and an additive error $\eps = 2^{-\Beps}$.
  We proceed in three steps.
  \begin{compactitem}
    \item \emph{(Discount factor)}
      We construct, in polynomial time, a succinct discount factor $\lambda$ such that the values of $\disc(r, \lambda)$ approximate the values of $\phi$ up to $\frac{\eps}{4}$.
      Let $Q$ be the product of the denominators of all transition probabilities and rewards of $G$.
      Since there are $n m^2 (n + 1)$ such numbers, each with a denominator at most $2^{\Binp}$, we have $Q \le 2^{B_Q}$ for $B_Q \defas \Binp n m^2 (n + 1)$.
      Let $\Blam$ be the constant of \Cref{Result: Liminf-average approximation} for the additive error $\frac{\eps}{4} = 2^{-(\Beps + 2)}$, i.e.,
      \[
        \Blam \defas \left\lceil \log \left( 4 n m^n \left( \lceil \log n \rceil + n \lceil \log m \rceil + B_Q + 3 \right) + (\Beps + 2) \, n \, m^n \right) \right\rceil \,,
      \]
      and let $\lambda \defas 1 - 2^{-2^{\Blam}}$.
      Since $\log(n m^n) = \log n + n \log m$, and $B_Q$ is polynomial in $n$, $m$, and $\Binp$, the bit-size $\Blam$ is polynomial in $n$, $m$, $\Binp$, and $\Beps$, and it is computed in polynomial time.
      By \Cref{Result: Liminf-average approximation}, for all states $\state \in \States$, we have
      \begin{equation}
        \label{equation: discounted approximation}
        |\val_\lambda(\state) - \val(\state)| \le \frac{\eps}{4} \,.
      \end{equation}

    \item \emph{(Value approximation)}
      By \Cref{Result: Discounted in TFNP}, the \CSGAV{} problem for the discounted-sum objective $\disc(r, \lambda)$ and the additive error $\frac{\eps}{2}$ reduces in polynomial time to a search problem in TFNP.
      Let $\hat{v}$ be a solution of this problem, i.e., $|\hat{v}(\state) - \val_\lambda(\state)| \le \frac{\eps}{2}$ for all states $\state \in \States$.
      Then, for all states $\state \in \States$, we have
      \begin{align*}
        |\hat{v}(\state) - \val(\state)|
          &\le |\hat{v}(\state) - \val_\lambda(\state)| + |\val_\lambda(\state) - \val(\state)|
            && (\text{triangle inequality}) \\
          &\le \frac{\eps}{2} + \frac{\eps}{4}
            && (\text{choice of } \hat{v},\ \text{\Cref{equation: discounted approximation}}) \\
          &\le \eps \,.
      \end{align*}
      Therefore, the \CSGAV{} problem for $\phi$ is in TFNP.

    \item \emph{(Gap problem)}
      Consider a state $\state$ and a threshold $\alpha$, and suppose that $\val(\state) \ge \alpha + \eps$ or $\val(\state) \le \alpha - \eps$, since otherwise every answer is correct.
      If $\val(\state) \ge \alpha + \eps$, then we have
      \begin{align*}
        \val_\lambda(\state)
          &\ge \val(\state) - \frac{\eps}{4}
            && (\text{\Cref{equation: discounted approximation}}) \\
          &\ge \alpha + \frac{\eps}{2}
            && (\val(\state) \ge \alpha + \eps) \,.
      \end{align*}
      Symmetrically, if $\val(\state) \le \alpha - \eps$, then we have $\val_\lambda(\state) \le \alpha - \frac{\eps}{2}$.
      Therefore, the instance has the same answer as the instance of the \CSGGV{} problem for the discounted-sum objective $\disc(r, \lambda)$ with the same state, the same threshold, and the additive error $\frac{\eps}{2}$, which is computed in polynomial time.
      By \Cref{Result: Discounted in TFNP}, the \CSGGV{} problem for $\phi$ is in NP $\cap$ coNP, which completes the proof. \qedhere
  \end{compactitem}
\end{proof}

\section{Value Decision}
\label{Section: Value decision}

In this section, we prove \Cref{Result: Value decision}, i.e., that the \CSGD{} problem is PosSLP-hard and in $\ER \cap \coER$ for limit-average and reachability objectives.
We prove the lower bound in \Cref{Section: Decision lower bound} and the upper bound in \Cref{Section: Decision upper bound}.

\subsection{Lower Bound}
\label{Section: Decision lower bound}

In this subsection, we prove the lower bound of \Cref{Result: Value decision}.
To do so, we simulate straight-line programs by games whose centered values encode the integers computed by the programs.

\begin{proof}[Proof of the lower bound of \Cref{Result: Value decision}]
  We reduce PosSLP to the \CSGD{} problem for reachability objectives.
  By \Cref{Result: Reachability as limit-average}, reachability objectives reduce in polynomial time to limit-average objectives with the same values, and the reduction preserves the transitions of games in which the target states are absorbing.
  Hence, the same holds for limit-average objectives.
  We construct a game $G$ with the action sets $\Actions = \{\action_1, \action_2\}$ and $\Bactions = \{\baction_1, \baction_2\}$ and two absorbing states $\top$ and $\bot$, and we consider the target set $\Target \defas \{\top\}$.
  We add the states one by one, and every new state moves only to previously added states.
  Hence, every play reaches $\top$ or $\bot$ after at most $n$ steps.
  For every state $\state \notin \{\top, \bot\}$, let $M_\state$ be the $2 \times 2$ matrix with entries $M_\state(\action, \baction) \defas \sum_{\state' \in \States} \prob(\state, \action, \baction)(\state') \, \val_\Target(\state')$.
  By backward induction, $\val_\Target(\state)$ is the value $\val(M_\state)$ of the matrix game $M_\state$, where Player~1 chooses a row and Player~2 chooses a column.
  For every state $\state$, we denote by $\chi(\state) \defas 2 \val_\Target(\state) - 1 \in [-1, 1]$ its \emph{centered value}.
  The proof consists of the following steps.
  \begin{compactitem}
    \item \emph{(Gadgets)}
      We construct states that negate, average, and multiply centered values.
    \item \emph{(Straight-line programs)}
      We represent the integer of every gate as the ratio of the centered values of two states.
    \item \emph{(Reduction)}
      We compare the value of the state of the output gate with the threshold $\frac{1}{2}$.
    \item \emph{(Deterministic transitions)}
      We replace every stochastic transition by deterministic transitions that preserve the values.
  \end{compactitem}
  We now prove each step.
  \begin{compactitem}
    \item \emph{(Gadgets)}
      First, for every new state $\state$, we add a \emph{dual} state $\bar{\state}$ that swaps the roles of the players and replaces every successor by its dual, i.e., the action pair $(\action_i, \baction_j)$ at $\bar{\state}$ leads to the duals of the successors of $(\action_j, \baction_i)$ at $\state$.
      Formally, for all $i, j \in \{1, 2\}$ and states $\state' \in \States$, we define
      \[
        \prob(\bar{\state}, \action_i, \baction_j)(\overline{\state'}) \defas \prob(\state, \action_j, \baction_i)(\state') \,,
      \]
      where $\bar{\top} \defas \bot$ and $\bar{\bot} \defas \top$, and the dual of $\bar{\state}$ is $\state$.
      By induction over the construction, $\val_\Target(\overline{\state'}) = 1 - \val_\Target(\state')$ for all successors $\state'$ of $\state$, and therefore, for all $i, j \in \{1, 2\}$, we have
      \begin{align*}
        M_{\bar{\state}}(\action_i, \baction_j)
          &= \sum_{\state' \in \States} \prob(\bar{\state}, \action_i, \baction_j)(\overline{\state'}) \, \val_\Target(\overline{\state'})
            && (\text{definition of } M_{\bar{\state}}) \\
          &= \sum_{\state' \in \States} \prob(\state, \action_j, \baction_i)(\state') \, \big( 1 - \val_\Target(\state') \big)
            && (\text{definition of } \bar{\state},\ \text{induction hypothesis}) \\
          &= 1 - M_\state(\action_j, \baction_i)
            && (\text{definition of } M_\state) \,.
      \end{align*}
      In the matrix game $M_{\bar{\state}}$, Player~1 plays the role of Player~2 in $M_\state$, and hence
      \begin{equation}
        \label{equation: dual state}
        \val_\Target(\bar{\state}) = 1 - \val_\Target(\state) \,,
        \qquad \text{i.e.,} \qquad
        \chi(\bar{\state}) = -\chi(\state) \,.
      \end{equation}
      Second, for all states $\state_1$ and $\state_2$, let $\mathrm{Avg}(\state_1, \state_2)$ be the state that moves to $\state_1$ and $\state_2$ with probability $\frac{1}{2}$ each.
      Then, we have
      \begin{equation}
        \label{equation: averaging gadget}
        \chi(\mathrm{Avg}(\state_1, \state_2)) = \frac{\chi(\state_1) + \chi(\state_2)}{2} \,.
      \end{equation}
      In particular, since $\chi(\top) = 1$ and $\chi(\bot) = -1$, we have $\chi(\mathrm{Avg}(\top, \bot)) = 0$.
      Third, for all states $\state_1$ and $\state_2$, let $\mathrm{Mul}(\state_1, \state_2)$ be the state $\state$ with the transitions
      \begin{align*}
        \prob(\state, \action_1, \baction_1) &\defas \tfrac{1}{2} \1[\top] + \tfrac{1}{4} \1[\state_1] + \tfrac{1}{4} \1[\bar{\state}_2] \,, &
        \prob(\state, \action_1, \baction_2) &\defas \tfrac{1}{2} \1[\bot] + \tfrac{1}{4} \1[\state_1] + \tfrac{1}{4} \1[\state_2] \,, \\
        \prob(\state, \action_2, \baction_1) &\defas \tfrac{1}{2} \1[\bot] + \tfrac{1}{4} \1[\bar{\state}_1] + \tfrac{1}{4} \1[\bar{\state}_2] \,, &
        \prob(\state, \action_2, \baction_2) &\defas \tfrac{1}{2} \1[\top] + \tfrac{1}{4} \1[\bar{\state}_1] + \tfrac{1}{4} \1[\state_2] \,.
      \end{align*}
      These transitions are depicted in the following figure, where every cell is an action pair, and $\top$ and $\bot$ are drawn twice to avoid crossing arrows.
      
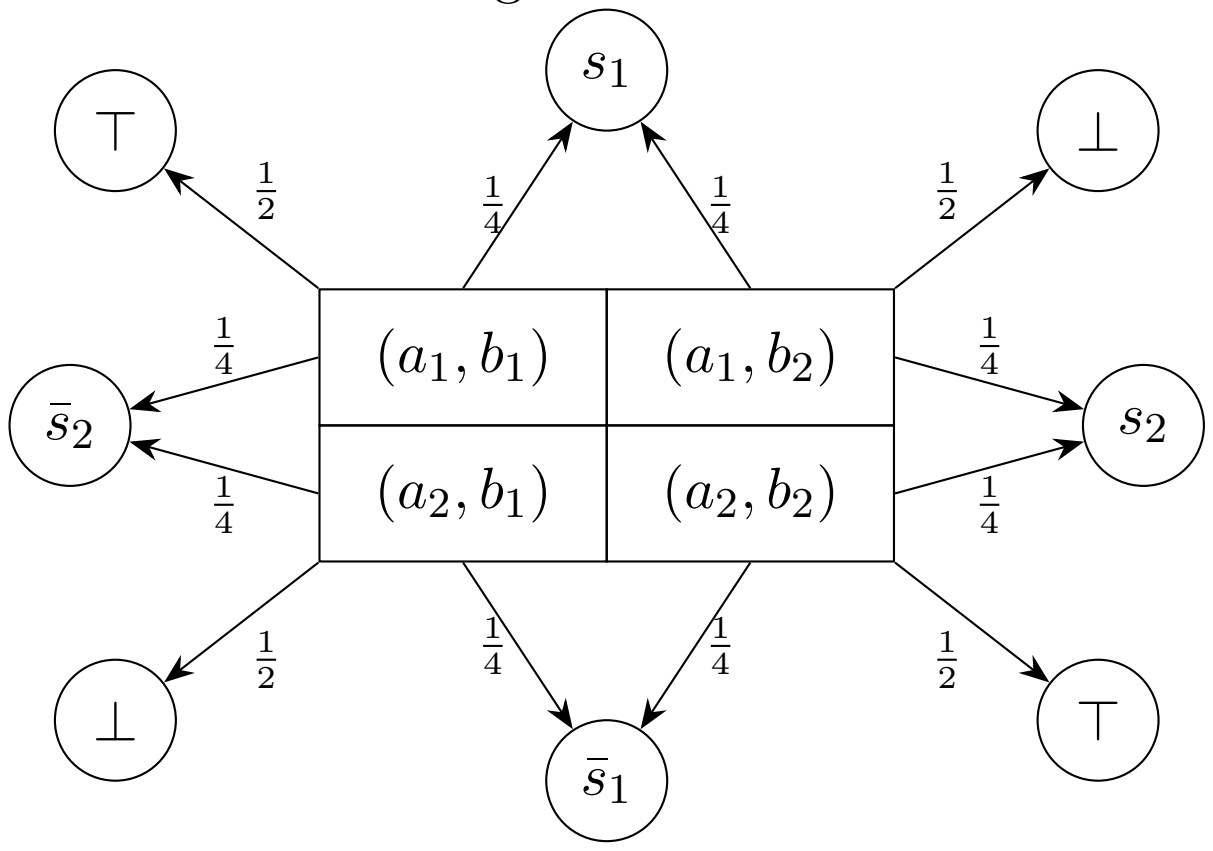
\begin{center}
        \begin{tikzpicture}[
            cell/.style={draw, minimum width=1.9cm, minimum height=0.9cm, inner sep=1pt},
            st/.style={draw, circle, minimum size=0.8cm, inner sep=1pt},
            >={Stealth[length=2mm]},
            lab/.style={font=\footnotesize, inner sep=1.5pt}]
          \node[cell] (c11) at (0, 0) {$(\action_1, \baction_1)$};
          \node[cell] (c12) at (1.9, 0) {$(\action_1, \baction_2)$};
          \node[cell] (c21) at (0, -0.9) {$(\action_2, \baction_1)$};
          \node[cell] (c22) at (1.9, -0.9) {$(\action_2, \baction_2)$};
          \node[st] (s1) at (0.95, 1.9) {$\state_1$};
          \node[st] (bs1) at (0.95, -2.8) {$\bar{\state}_1$};
          \node[st] (bs2) at (-2.6, -0.45) {$\bar{\state}_2$};
          \node[st] (s2) at (4.5, -0.45) {$\state_2$};
          \node[st] (t1) at (-2.3, 1.5) {$\top$};
          \node[st] (f1) at (4.2, 1.5) {$\bot$};
          \node[st] (f2) at (-2.3, -2.4) {$\bot$};
          \node[st] (t2) at (4.2, -2.4) {$\top$};
          \draw[->] (c11.north) -- node[lab, left] {$\frac{1}{4}$} (s1);
          \draw[->] (c12.north) -- node[lab, right] {$\frac{1}{4}$} (s1);
          \draw[->] (c21.south) -- node[lab, left] {$\frac{1}{4}$} (bs1);
          \draw[->] (c22.south) -- node[lab, right] {$\frac{1}{4}$} (bs1);
          \draw[->] (c11.west) -- node[lab, above] {$\frac{1}{4}$} (bs2);
          \draw[->] (c21.west) -- node[lab, below] {$\frac{1}{4}$} (bs2);
          \draw[->] (c12.east) -- node[lab, above] {$\frac{1}{4}$} (s2);
          \draw[->] (c22.east) -- node[lab, below] {$\frac{1}{4}$} (s2);
          \draw[->] (c11.north west) -- node[lab, above right] {$\frac{1}{2}$} (t1);
          \draw[->] (c12.north east) -- node[lab, above left] {$\frac{1}{2}$} (f1);
          \draw[->] (c21.south west) -- node[lab, below right] {$\frac{1}{2}$} (f2);
          \draw[->] (c22.south east) -- node[lab, below left] {$\frac{1}{2}$} (t2);
        \end{tikzpicture}
      \end{center}
      Let $x \defas \chi(\state_1)$ and $y \defas \chi(\state_2)$.
      By \Cref{equation: dual state}, we have
      \[
        M_{\mathrm{Mul}(\state_1, \state_2)} =
          \begin{pmatrix}
            \frac{3}{4} + \frac{x - y}{8} & \frac{1}{4} + \frac{x + y}{8} \\[2pt]
            \frac{1}{4} - \frac{x + y}{8} & \frac{3}{4} - \frac{x - y}{8}
          \end{pmatrix} .
      \]
      If Player~1 chooses $\action_1$ with probability $\frac{1}{2} + \frac{y}{4}$, then the expected payoff is $\frac{1}{2} + \frac{xy}{16}$ against both $\baction_1$ and $\baction_2$.
      Similarly, if Player~2 chooses $\baction_1$ with probability $\frac{1}{2} - \frac{x}{4}$, then the expected payoff is $\frac{1}{2} + \frac{xy}{16}$ against both $\action_1$ and $\action_2$.
      Since $x, y \in [-1, 1]$, both numbers are in $[0, 1]$, and hence these are mixed strategies that guarantee the expected payoff $\frac{1}{2} + \frac{xy}{16}$ for Player~1 and Player~2, respectively.
      Therefore, we have $\val_\Target(\mathrm{Mul}(\state_1, \state_2)) = \val(M_{\mathrm{Mul}(\state_1, \state_2)}) = \frac{1}{2} + \frac{xy}{16}$, and hence
      \begin{equation}
        \label{equation: multiplication gadget}
        \chi(\mathrm{Mul}(\state_1, \state_2)) = \frac{\chi(\state_1) \, \chi(\state_2)}{8} \,.
      \end{equation}

    \item \emph{(Straight-line programs)}
      Consider a straight-line program with gates $1, \ldots, k$, and recall that $z_i$ is the integer computed by gate $i$.
      The integer $z_i$ may be double exponentially large, whereas centered values are in $[-1, 1]$.
      Hence, we represent $z_i$ as the ratio of the centered values of two states, i.e., for every gate $i$, we construct a \emph{numerator} state $\state_i^{\mathrm{num}}$ and a \emph{denominator} state $\state_i^{\mathrm{den}}$ such that
      \begin{equation}
        \label{equation: program invariant}
        \chi(\state_i^{\mathrm{den}}) > 0
        \qquad \text{and} \qquad
        \chi(\state_i^{\mathrm{num}}) = z_i \, \chi(\state_i^{\mathrm{den}}) \,.
      \end{equation}
      We construct these states by induction over the gates, and we show \Cref{equation: program invariant} for every type of gate.
      \begin{compactitem}
        \item \emph{(Constants)}
          If $z_i = 0$, then let $\state_i^{\mathrm{num}} \defas \mathrm{Avg}(\top, \bot)$ and $\state_i^{\mathrm{den}} \defas \top$.
          If $z_i = 1$, then let $\state_i^{\mathrm{num}} \defas \top$ and $\state_i^{\mathrm{den}} \defas \top$.
          Since $\chi(\mathrm{Avg}(\top, \bot)) = 0$ and $\chi(\top) = 1$, \Cref{equation: program invariant} holds.
        \item \emph{(Multiplication)}
          If $z_i = z_j \times z_{j'}$, then let
          \[
            \state_i^{\mathrm{num}} \defas \mathrm{Mul}(\state_j^{\mathrm{num}}, \state_{j'}^{\mathrm{num}})
            \qquad \text{and} \qquad
            \state_i^{\mathrm{den}} \defas \mathrm{Mul}(\state_j^{\mathrm{den}}, \state_{j'}^{\mathrm{den}}) \,.
          \]
          By \Cref{equation: multiplication gadget} and the induction hypothesis, we have $\chi(\state_i^{\mathrm{den}}) = \frac{1}{8} \chi(\state_j^{\mathrm{den}}) \, \chi(\state_{j'}^{\mathrm{den}}) > 0$ and
          \begin{align*}
            \chi(\state_i^{\mathrm{num}})
              &= \frac{\chi(\state_j^{\mathrm{num}}) \, \chi(\state_{j'}^{\mathrm{num}})}{8}
                && (\text{\Cref{equation: multiplication gadget}}) \\
              &= z_j z_{j'} \, \frac{\chi(\state_j^{\mathrm{den}}) \, \chi(\state_{j'}^{\mathrm{den}})}{8}
                && (\text{induction hypothesis}) \\
              &= z_i \, \chi(\state_i^{\mathrm{den}})
                && (\text{\Cref{equation: multiplication gadget}}) \,.
          \end{align*}
        \item \emph{(Addition)}
          If $z_i = z_j + z_{j'}$, then let
          \begin{align*}
            \state_i^{\mathrm{num}} &\defas \mathrm{Avg}\big( \mathrm{Mul}(\state_j^{\mathrm{num}}, \state_{j'}^{\mathrm{den}}), \, \mathrm{Mul}(\state_{j'}^{\mathrm{num}}, \state_j^{\mathrm{den}}) \big) \,, \\
            \state_i^{\mathrm{den}} &\defas \mathrm{Avg}\big( \mathrm{Mul}(\state_j^{\mathrm{den}}, \state_{j'}^{\mathrm{den}}), \, \mathrm{Avg}(\top, \bot) \big) \,.
          \end{align*}
          By \Cref{equation: averaging gadget,equation: multiplication gadget} and the induction hypothesis, we have $\chi(\state_i^{\mathrm{den}}) = \frac{1}{16} \chi(\state_j^{\mathrm{den}}) \, \chi(\state_{j'}^{\mathrm{den}}) > 0$ and
          \begin{align*}
            \chi(\state_i^{\mathrm{num}})
              &= \frac{\chi(\state_j^{\mathrm{num}}) \, \chi(\state_{j'}^{\mathrm{den}}) + \chi(\state_{j'}^{\mathrm{num}}) \, \chi(\state_j^{\mathrm{den}})}{16}
                && (\text{\Cref{equation: averaging gadget,equation: multiplication gadget}}) \\
              &= (z_j + z_{j'}) \, \frac{\chi(\state_j^{\mathrm{den}}) \, \chi(\state_{j'}^{\mathrm{den}})}{16}
                && (\text{induction hypothesis}) \\
              &= z_i \, \chi(\state_i^{\mathrm{den}})
                && (\text{\Cref{equation: averaging gadget,equation: multiplication gadget}}) \,.
          \end{align*}
        \item \emph{(Subtraction)}
          If $z_i = z_j - z_{j'}$, then let $\state_i^{\mathrm{den}}$ be defined as for addition, and let
          \[
            \state_i^{\mathrm{num}} \defas \mathrm{Avg}\big( \mathrm{Mul}(\state_j^{\mathrm{num}}, \state_{j'}^{\mathrm{den}}), \, \overline{\mathrm{Mul}(\state_{j'}^{\mathrm{num}}, \state_j^{\mathrm{den}})} \big) \,.
          \]
          As for addition, we have $\chi(\state_i^{\mathrm{den}}) = \frac{1}{16} \chi(\state_j^{\mathrm{den}}) \, \chi(\state_{j'}^{\mathrm{den}}) > 0$ and
          \begin{align*}
            \chi(\state_i^{\mathrm{num}})
              &= \frac{\chi(\state_j^{\mathrm{num}}) \, \chi(\state_{j'}^{\mathrm{den}}) - \chi(\state_{j'}^{\mathrm{num}}) \, \chi(\state_j^{\mathrm{den}})}{16}
                && (\text{\Cref{equation: dual state,equation: averaging gadget,equation: multiplication gadget}}) \\
              &= (z_j - z_{j'}) \, \frac{\chi(\state_j^{\mathrm{den}}) \, \chi(\state_{j'}^{\mathrm{den}})}{16}
                && (\text{induction hypothesis}) \\
              &= z_i \, \chi(\state_i^{\mathrm{den}})
                && (\text{\Cref{equation: averaging gadget,equation: multiplication gadget}}) \,.
          \end{align*}
      \end{compactitem}
      The construction adds a constant number of states for every gate, and it does not compute the integers $z_i$.

    \item \emph{(Reduction)}
      We append a constant number of gates to the program that compute $z^\star \defas 2 z_k - 1$.
      Since $z^\star$ is an odd integer, we have $z^\star \neq 0$, and $z_k > 0$ if and only if $z^\star > 0$.
      Let $\state_\star^{\mathrm{num}}$ and $\state_\star^{\mathrm{den}}$ be the states of the output gate.
      By \Cref{equation: program invariant}, we have
      \[
        \val_\Target(\state_\star^{\mathrm{num}}) = \frac{1 + \chi(\state_\star^{\mathrm{num}})}{2} = \frac{1}{2} + \frac{z^\star \chi(\state_\star^{\mathrm{den}})}{2} \,,
      \]
      where $\chi(\state_\star^{\mathrm{den}}) > 0$.
      Therefore, we have $z_k > 0$ if and only if $\val_\Target(\state_\star^{\mathrm{num}}) \ge \frac{1}{2}$.

    \item \emph{(Deterministic transitions)}
      Stochastic transitions can be simulated by deterministic transitions in concurrent games, i.e., there is a polynomial-time reduction to concurrent games with deterministic transitions that preserves the values of reachability objectives~\cite[Proposition~2.1]{etessami2008RecursiveConcurrentStochastic}, see also~\cite{chatterjee2015RandomnessForFree}.
      The reduction replaces every probabilistic choice by new states in which both players have two actions and the successor depends only on whether their actions match.
      Since all transition probabilities above are integer multiples of $\frac{1}{4}$, these new states do not introduce cycles, and hence every play still reaches $\top$ or $\bot$, which completes the proof. \qedhere
  \end{compactitem}
\end{proof}

\begin{remark}[Double exponentially close values]
  \label{Remark: Close values}
  The values in the constructed games may be double exponentially close to the threshold $\frac{1}{2}$.
  For example, if every gate $i \ge 2$ squares the previous gate, starting from the constant $1$, then $\chi(\state_i^{\mathrm{num}}) = 2^{-3(2^{i - 1} - 1)}$.
  Hence, the lower bound of \Cref{Result: Value decision} does not contradict \Cref{Result: CSG in TFNP}, where the additive error is given in binary.
\end{remark}

\subsection{Upper Bound}
\label{Section: Decision upper bound}

In this subsection, we prove the upper bound of \Cref{Result: Value decision}.
First, we show that strict comparisons between the values of discounted-sum objectives and real thresholds are witnessed by stationary strategies together with bounds on the value function (\Cref{Result: Real witnesses}).
Second, we show that the value of a limit-average objective differs from a rational threshold by at least a double exponentially small amount, unless they are equal (\Cref{Result: Algebraic values,Result: Separation from thresholds}).
Lastly, we combine these results to construct, in polynomial time, existential sentences over the reals that express $\val(\state) \ge \alpha$ and $\val(\state) < \alpha$, which shows that the \CSGD{} problem is in $\ER \cap \coER$.

\subparagraph{Real witnesses.}
The following result shows that strict comparisons between the value of a discounted-sum objective and a real threshold are witnessed by a stationary strategy together with a lower or an upper bound on the value function.
Similar witnesses are used in~\cite[Theorem~39]{chatterjee2007Thesis} and~\cite{chatterjee2008LimitAverageEXPTIME}, where a single existential sentence contains strategies for both players and certifies both a lower and an upper bound on the value.
Our witnesses are similar to those of \Cref{Section: Witnesses and Verifier}, but their strategies and bounds are arbitrary real numbers instead of power-series numbers.
Since their conditions are polynomial inequalities, we use them to construct existential sentences over the reals.
For completeness, we present the proof.

\begin{lemma}[Real witnesses, cf.~{\cite[Theorem~39]{chatterjee2007Thesis}}]
\label{Result: Real witnesses}
  Let $G$ be a game with a reward function and a discount factor $\lambda$, let $\state \in \States$ be a state, and let $c \in \RR$.
  Then, the following assertions hold.
  \begin{compactenum}
    \item \emph{(Lower witness)} We have $\val_\lambda(\state) > c$ if and only if there exist a stationary strategy $\strategyone$ for Player~1 and a function $\low \colon \States \to \RR$ such that $\low \preceq \Psi_{G_\strategyone}(\low)$ and $\low(\state) > c$.
    \item \emph{(Upper witness)} We have $\val_\lambda(\state) < c$ if and only if there exist a stationary strategy $\strategytwo$ for Player~2 and a function $\up \colon \States \to \RR$ such that $\Psi_{G_\strategytwo}(\up) \preceq \up$ and $\up(\state) < c$.
  \end{compactenum}
\end{lemma}

\begin{proof}
  We present the proof for lower witnesses, and the proof for upper witnesses is symmetric.
  First, let $\strategyone$ and $\low$ be such that $\low \preceq \Psi_{G_\strategyone}(\low)$ and $\low(\state) > c$.
  Then, we have
  \begin{align*}
    \val_\lambda(\state)
      &\ge \val_\lambda^{G_\strategyone}(\state)
        && (\strategyone \text{ is a strategy for Player~1}) \\
      &\ge \low(\state)
        && (\text{\Cref{Result: Properties of Bellman operator}}) \\
      &> c \,.
  \end{align*}
  Second, suppose that $\val_\lambda(\state) > c$.
  Let $\strategyone$ be an optimal stationary strategy for Player~1~\cite{shapley1953stochastic}, and let $\low \defas \val_\lambda^{G_\strategyone}$.
  Since $\low$ is the fixed point of $\Psi_{G_\strategyone}$, we have $\low = \Psi_{G_\strategyone}(\low)$.
  Moreover, since $\strategyone$ is optimal, we have $\low(\state) = \val_\lambda(\state) > c$, which completes the proof.
\end{proof}

\subparagraph{Separation from thresholds.}
To compare the value of a limit-average objective with a rational threshold, we show that, if they differ, then they differ by at least a double exponentially small amount.
To do so, we use that the value is an algebraic number, i.e., a root of a nonzero polynomial with integer coefficients, whose degree and coefficients are bounded as follows.
The result follows from~\cite[Proposition~3.3]{oliubarton2021NewAlgorithms}, where we replace the minimal number of pure stationary strategies of a player by its upper bound $m^n$, and the numbers of bits of $n$, $m^n$, and $Q$ by their upper bounds $\lceil \log n \rceil + 1$, $n \lceil \log m \rceil + 1$, and $B_Q + 1$, respectively, which only increases the bounds.

\begin{lemma}[{\cite[Proposition~3.3]{oliubarton2021NewAlgorithms}}]
\label{Result: Algebraic values}
  Let $G$ be a game and $r$ a reward function such that all transition probabilities and rewards are integer multiples of $1/Q$, for a positive integer $Q \le 2^{B_Q}$, and let
  \[
    B_{\mathrm{coef}} \defas \left\lceil \log \left( 4 n m^n \left( \lceil \log n \rceil + n \lceil \log m \rceil + B_Q + 3 \right) \right) \right\rceil \,.
  \]
  Then, for every state $\state \in \States$, there exist integers $e_0, \ldots, e_d$ with $e_d \neq 0$ such that $\sum_{i = 0}^{d} e_i \val(\state)^i = 0$, $d \le m^n$, and $|e_i| < 2^{2^{B_{\mathrm{coef}}}}$ for all $i \in \{0, \ldots, d\}$.
\end{lemma}

\begin{lemma}[Separation from thresholds]
\label{Result: Separation from thresholds}
  Let $G$ be a game and $r$ a reward function such that all transition probabilities and rewards are integer multiples of $1/Q$, for a positive integer $Q \le 2^{B_Q}$, and let $\alpha = p/q \in [0, 1]$ be a rational number defined using at most $\Binp$ bits, for integers $0 \le p \le q$ with $q \ge 1$.
  Let $B_{\mathrm{coef}}$ be as defined in \Cref{Result: Algebraic values}, and let
  \[
    B_\alpha \defas B_{\mathrm{coef}} + \left\lceil \log \left( m^n \left( \Binp + 3 \right) \right) \right\rceil + 1 \,.
  \]
  Then, for every state $\state \in \States$ such that $\val(\state) \neq \alpha$, we have $|\val(\state) - \alpha| \ge 2^{-2^{B_\alpha}}$.
\end{lemma}

\begin{proof}
  Consider a state $\state \in \States$ such that $\val(\state) \neq \alpha$, and let $\beta \defas \val(\state) - \alpha \neq 0$.
  First, we transfer the algebraic bound to $\beta$.
  By \Cref{Result: Algebraic values}, there exist integers $e_0, \ldots, e_d$ with $e_d \neq 0$ such that $\sum_{i = 0}^{d} e_i \val(\state)^i = 0$, $d \le m^n$, and $|e_i| < 2^{2^{B_{\mathrm{coef}}}}$ for all $i \in \{0, \ldots, d\}$.
  Substituting $\val(\state) = \frac{p}{q} + \beta$ and multiplying by $q^d$, we obtain
  \begin{equation}
    \label{equation: shifted algebraic equation}
    \sum_{j = 0}^{d} e'_j \beta^j = 0 \,,
    \qquad \text{where} \qquad
    e'_j \defas \sum_{i = j}^{d} e_i \, q^{d - i} \binom{i}{j} p^{i - j} q^j \,.
  \end{equation}
  The numbers $e'_j$ are integers, and $e'_d = e_d q^d \neq 0$.
  Moreover, for all $j \in \{0, \ldots, d\}$, we have
  \begin{align*}
    |e'_j|
      &\le \sum_{i = j}^{d} 2^{2^{B_{\mathrm{coef}}}} q^d \binom{i}{j}
        && (|e_i| < 2^{2^{B_{\mathrm{coef}}}},\ p \le q) \\
      &\le (d + 1) \, 2^{2^{B_{\mathrm{coef}}} + d} q^d
        && \Big( \tbinom{i}{j} \le 2^d \Big) \\
      &\le 2^{2^{B_{\mathrm{coef}}} + 2 m^n + m^n \Binp}
        && (d + 1 \le 2^d,\ d \le m^n,\ q \le 2^{\Binp}) \\
      &\le 2^{2^{B_{\mathrm{coef}}} + m^n (\Binp + 3) - 1}
        && (m^n \ge 1) \\
      &\le 2^{2^{B_\alpha} - 1}
        && (\text{definition of } B_\alpha,\ 2^a + 2^b \le 2^{a + b + 1}) \,.
  \end{align*}
  Second, we bound $|\beta|$ from below.
  Let $j_0$ be the smallest index such that $e'_{j_0} \neq 0$, which exists since $e'_d \neq 0$.
  Suppose towards a contradiction that $|\beta| < 2^{-2^{B_\alpha}}$.
  Then, we have
  \begin{align*}
    1
      &\le |e'_{j_0}|
        && (e'_{j_0} \text{ is a nonzero integer}) \\
      &= \Big| \sum_{j = j_0 + 1}^{d} e'_j \beta^{j - j_0} \Big|
        && (\text{\Cref{equation: shifted algebraic equation}},\ \beta \neq 0) \\
      &\le 2^{2^{B_\alpha} - 1} \sum_{j \ge 1} |\beta|^j
        && (|e'_j| \le 2^{2^{B_\alpha} - 1}) \\
      &= 2^{2^{B_\alpha} - 1} \frac{|\beta|}{1 - |\beta|}
        && (|\beta| < 1) \\
      &< 2^{2^{B_\alpha}} |\beta|
        && \Big( |\beta| < \tfrac{1}{2} \Big) \\
      &< 1
        && \big( |\beta| < 2^{-2^{B_\alpha}} \big) \,,
  \end{align*}
  which is a contradiction.
  Therefore, we have $|\val(\state) - \alpha| = |\beta| \ge 2^{-2^{B_\alpha}}$, which completes the proof.
\end{proof}

We are now ready to prove the upper bound of \Cref{Result: Value decision}.

\begin{proof}[Proof of the upper bound of \Cref{Result: Value decision}]
  By \Cref{Result: Reachability as limit-average}, reachability objectives reduce in polynomial time to limit-average objectives with the same values.
  Hence, it suffices to consider an instance $(G, \phi)$, where $\phi = \liminfavg(r)$ is a limit-average objective, a state $\state$, and a rational threshold $\alpha = p/q$.
  Since $\val(\state) \in [0, 1]$, the answer is positive if $\alpha \le 0$ and negative if $\alpha > 1$, so we assume that $\alpha \in (0, 1]$.
  We proceed in five steps.
  \begin{compactitem}
    \item \emph{(Separation)}
      Let $Q$ be the product of the denominators of all transition probabilities and rewards of $G$.
      As in the proof of \Cref{Result: CSG in TFNP}, we have $Q \le 2^{B_Q}$ for $B_Q \defas \Binp n m^2 (n + 1)$.
      Let $B_\alpha$ be as defined in \Cref{Result: Separation from thresholds}, and let $\eta \defas 2^{-2^{B_\alpha}}$.
      By \Cref{Result: Separation from thresholds}, if $\val(\state) \neq \alpha$, then $|\val(\state) - \alpha| \ge \eta$.
      Since $\log(m^n) = n \log m$, and $B_Q$ is polynomial in the input size, $B_\alpha$ is polynomial in the input size, and it is computed in polynomial time.

    \item \emph{(Discount factor)}
      Let $\Blam$ be the constant of \Cref{Result: Liminf-average approximation} for the additive error $\frac{\eta}{4} = 2^{-(2^{B_\alpha} + 2)}$, i.e.,
      \[
        \Blam \defas \left\lceil \log \left( 4 n m^n \left( \lceil \log n \rceil + n \lceil \log m \rceil + B_Q + 3 \right) + (2^{B_\alpha} + 2) \, n \, m^n \right) \right\rceil \,,
      \]
      and let $\lambda \defas 1 - 2^{-2^{\Blam}}$.
      Since $B_\alpha$ is polynomial in the input size, $\Blam$ is polynomial in the input size, and it is computed in polynomial time.
      By \Cref{Result: Liminf-average approximation}, we have
      \begin{equation}
        \label{equation: decision approximation}
        |\val_\lambda(\state) - \val(\state)| \le \frac{\eta}{4} \,.
      \end{equation}

    \item \emph{(Shifted threshold)}
      Let $c \defas \alpha - \frac{\eta}{2}$.
      If $\val(\state) \ge \alpha$, then we have
      \begin{align*}
        \val_\lambda(\state)
          &\ge \val(\state) - \frac{\eta}{4}
            && (\text{\Cref{equation: decision approximation}}) \\
          &\ge \alpha - \frac{\eta}{4}
            && (\val(\state) \ge \alpha) \\
          &> c
            && (\text{definition of } c) \,.
      \end{align*}
      If $\val(\state) < \alpha$, then we have
      \begin{align*}
        \val_\lambda(\state)
          &\le \val(\state) + \frac{\eta}{4}
            && (\text{\Cref{equation: decision approximation}}) \\
          &\le \alpha - \eta + \frac{\eta}{4}
            && (\text{\Cref{Result: Separation from thresholds}}) \\
          &< c
            && (\text{definition of } c) \,.
      \end{align*}
      Therefore, we have
      \begin{equation}
        \label{equation: shifted threshold}
        \val(\state) \ge \alpha \iff \val_\lambda(\state) > c
        \qquad \text{and} \qquad
        \val(\state) < \alpha \iff \val_\lambda(\state) < c \,.
      \end{equation}

    \item \emph{(Existential sentence)}
      We construct the sentence from the following three formulas.
      \begin{compactitem}
        \item \emph{(Constants)}
          Let $B_{\max} \defas \max\{\Blam, B_\alpha\}$, and define the following formula over the real variables $\lambda$, $c$, and $x = (x_0, \ldots, x_{B_{\max}})$:
          \[
            \mathrm{Const}(\lambda, c, x) \defas
            2 x_0 = 1
            \;\wedge\; \bigwedge_{i = 0}^{B_{\max} - 1} x_{i + 1} = x_i^2
            \;\wedge\; \lambda = 1 - x_{\Blam}
            \;\wedge\; 2 q c = 2 p - q x_{B_\alpha} \,.
          \]
          By induction over $i$, the first two conjuncts hold if and only if $x_i = 2^{-2^i}$ for all $i \in \{0, \ldots, B_{\max}\}$.
          Hence, $\mathrm{Const}(\lambda, c, x)$ holds if and only if, in addition, $\lambda = 1 - 2^{-2^{\Blam}}$ and $c = \alpha - \frac{\eta}{2}$, i.e., $\lambda$ and $c$ are the constants of the previous steps.
        \item \emph{(Strategy)}
          We define the following formula over the real variables $\strategyone(\state')(\action)$, for $\state' \in \States$ and $\action \in \Actions$:
          \[
            \mathrm{Strat}(\strategyone) \defas
            \bigwedge_{\state' \in \States} \Big( \bigwedge_{\action \in \Actions} \strategyone(\state')(\action) \ge 0 \;\wedge\; \sum_{\action \in \Actions} \strategyone(\state')(\action) = 1 \Big) \,.
          \]
          It holds if and only if $\strategyone$ is a stationary strategy for Player~1.
        \item \emph{(Lower certificate)}
          We define the following formula, where $\low(\state')$, for $\state' \in \States$, are additional real variables:
          \begin{align*}
            & \mathrm{LowerCert}(\strategyone, \low, \lambda, c) \defas \\
            & \qquad \bigwedge_{\state' \in \States} \bigwedge_{\baction \in \Bactions} \low(\state') \le \sum_{\action \in \Actions} \strategyone(\state')(\action) \Big( (1 - \lambda) \, r(\state', \action, \baction) + \lambda \sum_{\state'' \in \States} \prob(\state', \action, \baction)(\state'') \, \low(\state'') \Big) \\
            & \qquad \wedge\; \low(\state) > c \,.
          \end{align*}
          It holds if and only if $\low \preceq \Psi_{G_\strategyone}(\low)$ and $\low(\state) > c$.
          Multiplying its inequalities by $Q$ yields integer coefficients, since all rewards and transition probabilities are integer multiples of $1/Q$.
      \end{compactitem}
      By \Cref{equation: shifted threshold} and the lower witness of \Cref{Result: Real witnesses}, we have $\val(\state) \ge \alpha$ if and only if
      \begin{equation}
        \label{equation: existential sentence}
        \exists \strategyone \; \exists \low \; \exists \lambda \; \exists c \; \exists x \colon \;
        \mathrm{Const}(\lambda, c, x) \;\wedge\; \mathrm{Strat}(\strategyone) \;\wedge\; \mathrm{LowerCert}(\strategyone, \low, \lambda, c) \,,
      \end{equation}
      where $\exists \strategyone$, $\exists \low$, and $\exists x$ quantify over the real variables $\strategyone(\state')(\action)$, $\low(\state')$, and $x_i$, for $\state' \in \States$, $\action \in \Actions$, and $i \in \{0, \ldots, B_{\max}\}$.
      Since $B_{\max}$ is polynomial in the input size, \Cref{equation: existential sentence} is an existential sentence over the reals of polynomial size that is computed in polynomial time, and hence the \CSGD{} problem is in $\ER$.

    \item \emph{(Universal sentence)}
      Symmetrically, let $\mathrm{Strat}(\strategytwo)$ be the formula obtained from $\mathrm{Strat}(\strategyone)$ by replacing $\strategyone$ and $\Actions$ with $\strategytwo$ and $\Bactions$, and define
      \begin{align*}
        & \mathrm{UpperCert}(\strategytwo, \up, \lambda, c) \defas \\
        & \qquad \bigwedge_{\state' \in \States} \bigwedge_{\action \in \Actions} \up(\state') \ge \sum_{\baction \in \Bactions} \strategytwo(\state')(\baction) \Big( (1 - \lambda) \, r(\state', \action, \baction) + \lambda \sum_{\state'' \in \States} \prob(\state', \action, \baction)(\state'') \, \up(\state'') \Big) \\
        & \qquad \wedge\; \up(\state) < c \,,
      \end{align*}
      which holds if and only if $\Psi_{G_\strategytwo}(\up) \preceq \up$ and $\up(\state) < c$.
      By \Cref{equation: shifted threshold} and the upper witness of \Cref{Result: Real witnesses}, we have $\val(\state) < \alpha$ if and only if $\exists \strategytwo \; \exists \up \; \exists \lambda \; \exists c \; \exists x \colon \mathrm{Const}(\lambda, c, x) \wedge \mathrm{Strat}(\strategytwo) \wedge \mathrm{UpperCert}(\strategytwo, \up, \lambda, c)$.
      Hence, we have $\val(\state) \ge \alpha$ if and only if
      \begin{equation}
        \label{equation: universal sentence}
        \forall \strategytwo \; \forall \up \; \forall \lambda \; \forall c \; \forall x \colon \;
        \neg \big( \mathrm{Const}(\lambda, c, x) \;\wedge\; \mathrm{Strat}(\strategytwo) \;\wedge\; \mathrm{UpperCert}(\strategytwo, \up, \lambda, c) \big) \,.
      \end{equation}
      As before, \Cref{equation: universal sentence} is a universal sentence over the reals of polynomial size that is computed in polynomial time, and hence the \CSGD{} problem is in $\coER$, which completes the proof. \qedhere
  \end{compactitem}
\end{proof}

\begin{remark}[Implications]
  \label{Remark: Implications}
  Two very recent preprints, which have not been verified yet, claim that $\ER$ is contained in the counting hierarchy~\cite{meiburg2026ExistentialRealsCH, openai2026ExistentialUniversalCH}.
  If these claims hold, then our result implies that the \CSGD{} problem is in the counting hierarchy for limit-average and reachability objectives.
\end{remark}

\section{Extensions}
\label{Section: Extensions}

In this section, we discuss extensions of our results.

\subparagraph{Parity objectives.}
A parity objective is given by a priority function $p \colon \States \to \{0, 1, \ldots, d\}$, and Player~1 wins a play if the minimal priority that is visited infinitely often is even.
All $\omega$-regular objectives that arise in the formal analysis of systems can be expressed by deterministic parity automata, and temporal logics such as LTL (linear-time temporal logic) can be translated to deterministic parity automata.
The product of a game graph with a deterministic parity automaton is a game with a parity objective, and hence parity objectives are a canonical representation of $\omega$-regular objectives.
The value of a parity objective is an iterated limit of the values of stateful-discounted objectives, where the reward is $1$ in states with even priority and $0$ otherwise, every priority $i$ is associated with a discount factor $\lambda_i$, and the discount factors tend to one in the order of the priorities~\cite{dealfaro2003Discounting, gimbert2005Discounting}.
Moreover, the value of a parity objective is approximated up to $\eps$ by the value of such a stateful-discounted objective with succinct discount factors $\lambda_i = 1 - 2^{-2^{B_{\lambda_i}}}$, where every $B_{\lambda_i} \in \NN_0$ is polynomial in the input size, since the proof of~\cite[Theorem~18]{asadi2024StatefulDiscounted} applies to all discount factors that are close enough to one and whose distances to one decrease fast enough with the priority.
Our proof of \Cref{Result: Discounted in TFNP} carries over to such objectives, where the Bellman operator of an MDP $P$ uses the discount factor of the priority of the current state, i.e., for all functions $\rho \colon \States \to \RR$ and states $\state$,
\[
  \Psi_P(\rho)(\state) = \op_{\action \in \Actions} \Big( \big( 1 - \lambda_{p(\state)} \big) \, r(\state, \action) + \lambda_{p(\state)} \sum_{\state' \in \States} \prob(\state, \action)(\state') \, \rho(\state') \Big) \,.
\]
First, the discount factors are power-series numbers with two terms, and this Bellman operator is monotone and a contraction, so shrinking interval bounds contain the value and are verified in polynomial time as before.
Second, the existence of polynomial-size $\eps$-optimal strategies follows as in \Cref{Section: Strategies}, where the smallest probability that we keep depends on $\min_i (1 - \lambda_i)$ instead of $1 - \lambda$, and the relative continuity applies since stateful-discounted objectives are also reachability objectives in a game that stops with probability $1 - \lambda_{p(\state)}$ at every state $\state$.
Third, the existence of polynomial-size $\eps$-close shrinking interval bounds follows as in \Cref{Section: Interval Bounds in MDPs}.
Indeed, the Bellman operator still takes a convex combination of the reward and the values of the successors at every state, so Jensen's inequality yields \Cref{Result: Convex concave transformations}, and Cramer's rule and Hadamard's inequality yield \Cref{Result: Value separation} with $\Blam$ replaced by $\max_i B_{\lambda_i}$.
Therefore, for parity objectives, the \CSGAV{} problem is in TFNP and the \CSGGV{} problem is in NP $\cap$ coNP, which improves the TFNP[NP] upper bound of~\cite{asadi2024StatefulDiscounted}.

\subparagraph{Partial-information games with restricted strategies.}
While our work focuses on perfect-information concurrent stochastic games, where strategies observe the states of the histories, our results also extend to special cases of partial-information games.
For partial-information games, the values that bounded-memory strategies guarantee are approximated by guessing polynomial-size representations of such strategies, whose probabilities may be double exponentially small, and by verifying them with an NP oracle~\cite{bose2024BoundedMemory}.
In general, the NP oracle cannot be avoided, since approximating these values is NP-hard and coNP-hard already for memoryless strategies~\cite{bose2024BoundedMemory}.
However, if the opponent observes both the states and the memory of the guessed strategy, then fixing this strategy yields an MDP on the product of the states and the memory.
We first approximate the limit-average values of this MDP by the values of a discounted-sum objective with a succinct discount factor (\Cref{Result: Liminf-average approximation}), where $\Blam$ is polynomial, since the transition probabilities of this MDP have a common denominator $Q \le 2^{B_Q}$ with $\log B_Q$ polynomial, as in the proof of \Cref{Result: Completeness of witnesses}.
Then, our $\eps$-close shrinking interval bounds verify the guessed strategy in polynomial time.

\subparagraph{Stationary Nash equilibria.}
While our focus is on zero-sum games, our techniques also apply to the approximate constrained stationary Nash equilibrium problem in multi-player turn-based stochastic games with terminal rewards, where plays that never reach a terminal vertex have payoff zero~\cite{asadi2025StationaryNash}.
This is a promise problem: if a stationary Nash equilibrium that satisfies given constraints on the payoffs exists, then the goal is to compute an $\eps$-Nash equilibrium that satisfies the constraints up to $\eps$.
Under this promise, there exists such an equilibrium whose strategies have double-exponential patience~\cite{asadi2025StationaryNash}.
Fixing the strategies of all players except one yields an MDP with double exponentially small transition probabilities.
As in \Cref{Result: Reachability as limit-average}, its terminal-reward objectives are limit-average objectives, which we first approximate by discounted-sum objectives with a succinct discount factor (\Cref{Result: Liminf-average approximation}), as for partial-information games.
Our $\eps$-close shrinking interval bounds for the resulting discounted MDP and for the Markov chain obtained by fixing all strategies certify in polynomial time that, up to the approximation error of the discount factor, this player cannot improve its payoff by more than $\eps$, by checking that an upper bound on its best-response payoff exceeds a lower bound on its payoff under the strategy profile by at most $\eps$.
The same bounds verify the constraints on the payoffs up to $\eps$, and they replace the NP oracle used for this verification in~\cite{asadi2025StationaryNash}.

In summary, MDPs with double exponentially small transition probabilities arise in several settings, and they are captured by MDPs with succinct discount factors.
Previous techniques verify witnesses for such MDPs with an NP oracle, whereas our $\eps$-close shrinking interval bounds verify them in polynomial time.

\section{Conclusion}
\label{Section: Conclusion}

We consider the fundamental problem of value approximation in concurrent stochastic games with limit-average objectives and improve its complexity from TFNP[NP] to TFNP.
As a special case, the same improvement holds for reachability objectives.
Our approach approximates limit-average values by discounted-sum values with a discount factor that is double exponentially close to one and is represented succinctly in power series.
Our approach also extends to parity objectives via stateful-discounted objectives~\cite{dealfaro2003Discounting, asadi2024StatefulDiscounted} (\Cref{Section: Extensions}).
We also show that the value-decision problem for limit-average and reachability objectives is in $\ER \cap \coER$ and is PosSLP-hard.

Further improvements are challenging open problems for future research.
For the approximation problem, improving the TFNP upper bound to polynomial time remains a major long-standing open problem, even for turn-based stochastic games.
For the value-decision problem, $\ER$ has recently been claimed to be in low levels of the counting hierarchy~\cite{meiburg2026ExistentialRealsCH, openai2026ExistentialUniversalCH}, and if these claims hold, then our $\ER \cap \coER$ upper bound places the value-decision problem for concurrent stochastic games in low levels of the counting hierarchy.
Moreover, our PosSLP-hardness brings the upper and lower bounds close to each other, and establishing a tight complexity bound for this problem remains an interesting open problem.

\section*{Acknowledgments}
We thank Raimundo Saona and Jakub Svoboda for helpful discussions at an early stage of this work.
The research was partially supported by the Austrian Science Fund (FWF) 10.55776/COE12 and by the ERC CoG 863818 (ForM-SMArt) grant.

\section*{AI Methodologies}
This project started several years ago, and its main ideas predate our use of AI tools.
In particular, the power-series representation, the witnesses that consist of stationary strategies and $\eps$-close shrinking interval bounds, and the resulting polynomial-time verifier are our own.
We used AI tools as follows: (a)~for calculations related to the $\eps$-close shrinking intervals we used GPT-5.6 (OpenAI); (b)~to edit the text of the paper we used Claude Opus~5.5 (Anthropic) and GPT-6~Astra; and (c)~to check the proofs we used GPT-6~Astra.
The authors verified the correctness and originality of all content, including the references, and take full responsibility for the paper.

\bibliographystyle{alphaurl}
\bibliography{references}

\end{document}